%% file: main.tex
\documentclass[sigconf, acmsmall]{acmart}
\AtBeginDocument{%
  }

\setcopyright{rightsretained}
\acmJournal{TODAES}
\acmYear{2025} \acmVolume{30} \acmNumber{2} \acmArticle{26} \acmMonth{02} 
\acmDOI{10.1145/3711847}

\renewcommand{\shortauthors}{S. Pouget et al.}
\usepackage{hyperref}
\usepackage{tikz}
\usepackage{multirow}
\usetikzlibrary{shapes,arrows,positioning}

\usepackage{listings}
\usepackage{xcolor}
\usepackage{dsfont}
\usepackage{geometry}
\usepackage{changepage}
\usepackage{tikz}
\usetikzlibrary{tikzmark}

\usepackage{multicol}

\definecolor{codegreen}{rgb}{0,0.6,0}
\definecolor{codegray}{rgb}{0.5,0.5,0.5}
\definecolor{codepurple}{rgb}{0.58,0,0.82}
\definecolor{backcolour}{rgb}{0.95,0.95,0.92}

\lstdefinestyle{cppstyle}{
    language=C++,
    commentstyle=\color{codegreen},
    keywordstyle=\color{blue},
    numberstyle=\tiny\color{codegray},
    stringstyle=\color{codepurple},
    basicstyle=\ttfamily\footnotesize,
    breakatwhitespace=false,
    breaklines=true,
    captionpos=b,
    keepspaces=true,
    numbers=left,
    numbersep=5pt,
    showspaces=false,
    showstringspaces=false,
    showtabs=false,
    tabsize=4,
    xleftmargin=0.5cm
}

\usepackage[caption=false, font=normalsize, labelfont=sf, textfont=sf]{subfig}
\usepackage{array}
\newcolumntype{C}[1]{>{\centering\arraybackslash}p{#1}}
\usepackage{adjustbox}
\usepackage{forest}
\usepackage{graphviz}
\usepackage{xspace}
\usepackage{enumitem}
\usepackage{algorithm2e}

\newcommand{\framework}{NLP-DSE\xspace}
\newcommand{\myparagraph}[1]{\par\textit{ #1}~}
\makeatletter


\begin{document}

\title{Automatic Hardware Pragma Insertion in High-Level Synthesis: A Non-Linear Programming Approach}

\author{Stéphane Pouget}
\email{pouget@cs.ucla.edu}
\orcid{0000-0003-3950-5818}
% \affiliation{University of California, Los Angeles}
\affiliation{%
  \institution{University of California, Los Angeles}
  \city{Los Angeles}
  \state{CA}
  \country{USA}
}
% \affiliation{%

\author{Louis-Noël Pouchet}
\email{pouchet@colostate.edu}
\orcid{0000-0001-5103-3097}
% \affiliation{Colorado State University}
\affiliation{%
  \institution{Colorado State University}
  \city{Fort Collins}
  \state{CO}
  \country{USA}
}

\author{Jason Cong}
\email{cong@cs.ucla.edu}
\orcid{0000-0003-2887-6963}
% \affiliation{University of California, Los Angeles}
\affiliation{%
  \institution{University of California, Los Angeles}
  \city{Los Angeles}
  \state{CA}
  \country{USA}
}

\renewcommand{\shortauthors}{S. Pouget et al.}
\renewcommand{\shortauthors}{}

% \input{answer}

% \newpage

\begin{abstract}
\input{sources/abstract}

\end{abstract}





\maketitle

% \begin{abstract}
%     \input{sources/abstract}
% \end{abstract}

% \begin{IEEEkeywords}
% HLS, FPGA, cost model, non linear programming, Program Optimization
% \end{IEEEkeywords}

% \REMARK{The scope of this paper is very narrow from both the problem formulation and its targeted downstream compiler.. 
% From the NLP solving point of view, such constraints are on the simpler side, even though the authors reported up to E+11 
% design points for their tested benchmarks.}

% \FIXME{Emphasize the generalization of the method for various compilers such as those from Intel and Siemens.}

\section{Introduction}
% \label{sec:introduction}
\input{sources/introduction}

% \clearpage

\section{Background and Motivation}
% \label{sec:motivation}
\input{sources/motivation}

\section{Modeling Programs and Their Pragmas}
% \label{sec:model}
\input{sources/model}

% \vspace{-0.2cm}
% \section{Theoretical Latency and Resource Modeling}
% % \label{sec:theoretical}
% % \input{sources/theoretical_and_modeling2}
% \input{sources/th_proof}

\section{Non-Linear Formulation for Pragma Insertion}
% \label{sec:formulation}
\input{sources/NLP_formulation}

\section{Latency and Resource Lower Bound}
\label{sec:theoretical}
% \input{sources/theoretical_and_modeling2}
%\input{sources/th_proof_summary}
\input{sources/th_proof_v2}

% \clearpage

\section{Design Space Exploration}
% \label{sec:implementation}
\input{sources/implementation}

\section{Evaluation}
% \label{sec:evaluation}
\input{sources/evaluation}

\section{Examples}
\input{sources/example}

\section{Related Work}
% \label{sec:related}
\input{sources/related}

\section{Conclusion}
% \label{sec:conclusion}
\input{sources/conclusion}

% \clearpage 

\section*{Acknowledgments}
 This work was supported by the NSF award \#CCF-2211557. It is also supported
by CDSC industrial partners and
the AMD/HACC Program.

\bibliographystyle{ACM-Reference-Format}
\bibliography{bibs/iccad23,bibs/sp,bibs/lnp,bibs/refs,bibs/gabriel,bibs/ierefs,bibs/ics15}

% \clearpage

\appendix

\input{sources/appendix}

\end{document}

%% file: sources/abstract.tex
High-Level Synthesis enables the rapid prototyping of hardware accelerators, by combining a high-level description of the functional behavior of a kernel with a set of micro-architecture optimizations as inputs. Such optimizations can be described by inserting pragmas e.g. pipelining and replication of units, or even higher level transformations for HLS such as automatic data caching using the AMD/Xilinx Merlin compiler. Selecting the best combination of pragmas, even within a restricted set, remains particularly challenging and the typical state-of-practice uses design-space exploration to navigate this space. But due to the highly irregular performance distribution of pragma configurations, typical DSE approaches are either extremely time consuming, or operating on a severely restricted search space.

This work proposes a framework to automatically insert HLS pragmas in regular loop-based programs, supporting pipelining, unit replication, and data caching. We develop an analytical performance and resource model as a function of the input program properties and pragmas inserted, using non-linear constraints and objectives. We prove this model provides a lower bound on the actual performance after HLS. We then encode this model as a Non-Linear Program, by making the pragma configuration unknowns of the system, which is computed  optimally by solving this NLP. This approach can also be used during DSE, to quickly prune points with a (possibly partial) pragma configuration, driven by lower bounds on achievable latency. We extensively evaluate our end-to-end, fully implemented system, showing it can effectively manipulate spaces of billions of designs in seconds to minutes for the kernels evaluated.

%% file: sources/introduction.tex
\label{sec:introduction}

High-level synthesis (HLS) \cite{cong.tcadics.2011, Zhang2008} compilers \cite{vitis, intel_fpga, catapult, legup} and source-to-source compiler for HLS \cite{merlin, pylog, scalehls, heterocl, s2fa, heterohalide} can reduce development time while delivering a good performance for the designs. However, achieving a satisfactory Quality of Results (QoR) often requires design-space exploration (DSE). This is because the design space, including which pragmas to insert and where, can not only contain millions of points, but typically does not present characteristics suitable for fast analytical exploration, such as convexity and regularity.
Although the existing DSE methods \cite{autodse, sohrabizadeh2022gnn, ironman} can find designs with a good QoR, it comes at a high computation cost: for example, hundreds of designs may be concretely instantiated using HLS to compute its estimated QoR during exploration \cite{autodse}. Alternatively, models can be built to estimate performance and resource usage instead of applying HLS \cite{comba,harp}. These approaches typically rely on enumerating the search space and estimating the QoR of candidate designs without deploying HLS, filtering out candidates with lower performance. They can quickly estimate the QoR of thousands of designs in a search space, for example using purely analytical models \cite{comba}, 
or deploying machine learning to learn performance predictors 
% from designs concretely instantiated with HLS \cite{harp}.
from large database \cite{10.5555/3666122.3668084} of designs with HLS synthesis results \cite{harp, 10.1145/3670474.3685952, ding2024efficienttasktransferhls}.
However, these approaches still face challenges in predicting the performance of new designs unseen from the training set, despite the recent success in transfer learning \cite{10.1145/3670474.3685952}.

% lose the accuracy provided by invoking HLS, and do not provide any guarantee the designs pruned out have a lower QoR in practice.

% 

Our main objective in this work is to provide a system to automatically insert a set of hardware pragmas for HLS, that delivers a good QoR and yet significantly reduces the search time needed to obtain the final design. To address this challenge, we propose \framework, a framework built on top of the AMD/Xilinx Merlin source-to-source compiler \cite{merlin}. This framework
 automatically inserts pragmas for unrolling/parallelization, pipelining, tiling and data caching \emph{for affine programs} \cite{mlir}, prior to HLS. These Merlin pragmas can also be inserted using a DSE approach, such as AutoDSE
\cite{autodse} or HARP \cite{harp}, which we use as reference for our evaluations. However, in contrast to AutoDSE \cite{autodse} and HARP \cite{harp}, we specifically restrict the class of programs we optimize to affine programs that are regular loop-based computations. In turn, it enables us to develop a hybrid analytical approach to drive the search, combined with a lightweight DSE to reduce the number of designs actually explored. \framework preserves, and often even improves, the final QoR of designs produced while significantly reducing exploration time.  

To this end, we create a novel Non-Linear Programming approach to automatically insert pragmas in an existing affine program. We develop an analytical model combining latency and resources, targeting regular loop-based kernels \cite{polybench-web}, that is parameterized by the pragma configuration. We can then designate the pragma configuration as the unknowns of this model, solving it by NLP to obtain the set of pragmas that minimizes latency. An important design principle of our approach is to ensure that the latency computed \emph{is  a lower bound on the achievable latency for a given pragma configuration}. This enables efficient pruning during DSE: any design predicted to have a latency lower bound higher than the best latency obtained through exploration so far is necessarily slower and does not need exploration. To overcome the fact that optimizing compilers, and the overall HLS toolchain underneath, may not apply optimizations as expected (e.g., due to insufficient resources, or limitations of the compiler's implementation), we develop a lightweight NLP-based DSE approach, exploring parts of the design space with different types and amounts of hardware parallelism and array partitioning factors. We make the following contributions:
\begin{itemize}[leftmargin=*]
% \begin{itemize}
    \item We present an analytical performance and resource model specifically built for AMD/Xilinx Vitis and Merlin compilers, which is amenable to optimization via non-linear programming. 
    \item We prove our model is  a lower bound on the final latency of the design, under reasonable hypothesis. This enables fast pruning of the design space: it ensures designs which have a higher latency lower bound than the best design found so far can be safely pruned from the search.
    \item We develop an NLP-based DSE approach exploiting this model, targeting regular loop-based kernels, which can significantly outperform DSE-based search approaches such as AutoDSE, delivering equal or better QoR in a significantly less search time.
    % \item Our approach has been successfully applied to Design-Space Exploration (DSE), enabling the efficient exploration of vast search spaces. By leveraging the latency lower bound property, the framework can rapidly eliminate points in the search space during DSE, reducing the exploration time to seconds or minutes, even when dealing with billions of designs.
    \item We implement \framework, an end-to-end, fully automated system and use it to conduct extensive evaluation on 47 benchmarks including kernels from linear algebra, image processing, physics simulation, graph analytics, datamining, etc. \cite{polybench-web}. 
    Our results show the ability of our approach to find in most cases a better QoR than AutoDSE, in significantly less time. Furthermore, in most instances, our approach outperforms HARP in terms of QoR within a comparable timeframe.
\end{itemize}

The paper is organized as follows. Section~\ref{sec:motivation} motivates our approach and solution proposed. Section~\ref{sec:modeling} presents our analytical performance and resource model.  In Section~\ref{sec:formulation}, we introduce a non-linear formulation based on this model to automatically find pragma configurations by NLP optimization. 
Section~\ref{sec:theoretical} delves into proving it is a lower bound on the final QoR.
Section~\ref{sec:implementation} presents our lightweight DSE approach.
Finally, sections~\ref{sec:evaluation} and \ref{sec:conclusion} are devoted to evaluating our method validating the effectiveness of our approach and presenting related work, before concluding.

%% file: sources/motivation.tex
\label{sec:motivation}

%%%%%%%%% Ce qui manque
%
% C'est tjr pas clair pq NLP permet une exploration (th) de l'espace efficace
% Limitation of AutoDSE
% 
%
%
%%%%%%%%

\subsection{Pragma-based Optimizations for HLS}
\label{subsec:pragams-merlin}
This work targets the automatic optimization of FPGA designs using HLS \cite{cong.tcadics.2011, Zhang2008}, 
in particular when using compilers to automatically generate HLS-friendly optimized programs such as the AMD/Xilinx Merlin Compiler \cite{cong.ispled.16, cong.springer.2017, merlin}. 
Falcon Computing Solutions developed this source-to-source automation tool for FPGAs, which was acquired by Xilinx in 2020 and is now open source. 

%in order to quickly obtain a design with a good QoR even for developers who may not possess advanced expertise in the field.
% 

HLS has made FPGA usage more accessible, and many projects are looking to further democratize this field by automating optimizations \cite{autodse, heterocl, pylog, scalehls, lorenzo}. 

Merlin was developed to improve the performance, reduce the development time of HLS-based designs and simplify the search space. 
To achieve this, Merlin automatically generates data transfers between off-chip and on-chip memory based on the specified available on-chip memory and allows overlapping of memory transfers for different arrays when their transfers occur consecutively in the code. 
It also automatically inserts essential Vitis pragmas, such as array partitioning, based on the unrolling factor (pragma \texttt{ACCEL parallel}).
Furthermore, Merlin performs hardware-specific code transformations in line with user-defined directives. For example, it applies strip-mining to loops that are partially unrolled, fully unrolling the newly created inner loop with a trip count equal to the user-specified unroll factor. Additionally, Merlin enables parallel coarse-grained processing by encapsulating inner loops as separate functions.

The pragmas available for use with Merlin are:
\begin{itemize}
    \item \texttt{ACCEL parallel <factor=x>}, which creates \texttt{x} parallel instances, and Merlin restructures the code accordingly if the loop nest has more than \texttt{x} iterations. This pragma can be used for fine-grained and coarse-grained parallelization. Merlin will insert the  array partitioning corresponding to the parallelization and optimize memory coalescing accordingly for the memory transfer;
    \item \texttt{ACCEL pipeline flatten <II=y>} for pipelining;
    \item \texttt{ACCEL tile <factor=z>} for strip-mining a loop by \texttt{z}, enabling Merlin to insert other pragmas such as data caching in a loop with smaller trip count, matching the on-chip resources available and reducing off-chip communications;
    \item \texttt{ACCEL cache <array=a>} which transfers all required elements of array $a$ from off-chip to on-chip to perform computations within the specified sub-region. If the user does not specify this pragma, it can be applied automatically by Merlin;
    \item \texttt{ACCEL pipeline} which creates a double buffer and enables overlapping of computation and communication. 
    % However, this pragma can only be applied in limited scenarios, so we decided not to include it in the design space to maintain a more accurate model.
    This pragma frequently fails to apply as expected in various scenarios, so we opted to exclude it from the design space to preserve a more accurate model.
    Incorporating it into the model requires a minor adjustment to the objective function.
\end{itemize}

In this work we target the automatic generation of pragmas for the Merlin toolchain, to enable the seamless deployment of optimizations such as array partitioning, off-chip data transfers using bursts, coarse-grain and fine-grain replication, etc. These pragma-directed optimizations are implemented by Merlin on loop-based programs, combining source code transformations and the automatic insertion of Vitis pragmas to drive the HLS process.

We note our approach is not restricted to Merlin, nor a particular version of a toolchain: by adjusting the parameters of the performance model, such as operation latency, resource usage per operation(s), etc. one can easily target other toolchains than the one we evaluate here.
In a recent study, we employed a similar NLP-based approach directly on HLS programs \cite{sisyphus}.

%It is important to note that our methodology is adaptable to any HLS compiler, encompassing various versions. The key requirements for adaptation are details regarding operator latency, resource usage, and pragma formats. Nevertheless, it's crucial to acknowledge that compiler constraints may introduce variations in results if anticipated optimizations are not applied by the compilers.

% mouais pas sur pr ca
% We anticipate that our research will serve as a foundation, inspiring other researchers to apply the NPL methodology to diverse HLS compilers. This collaborative exploration of methodologies across different compilers could contribute to a more comprehensive understanding of their strengths and limitations.

% \subsection{Merlin Compiler}

\subsection{DSE for Pragma Insertion}

Design-space exploration techniques typically trade-off coverage for speed \cite{autodse,zuo2015polyhedral}. That is, it may impose restrictions on the input programs supported, on the pragmas/transformations considered, etc. in order to accelerate the search \cite{zuo2015polyhedral,pouchet:fpga13}. To overcome the difficulty of providing accurate performance models for arbitrary programs, HLS may be invoked to obtain a QoR estimate, without imposing any restriction on the input programs and transformations used. However HLS time for highly optimized designs combining various Merlin pragmas (e.g., parallelism and caching) can quickly reach tens of minutes to several hours per design, making the search process particularly time-consuming. One may restrict the space of pragmas considered, and especially their parameter range, to reduce search time. General-purpose DSE approaches, such as AutoDSE \cite{autodse}, are agnostic to the input program features and the search space explored, thereby preserving generality.
But as we demonstrate in this paper, it also misses opportunities for search acceleration that can be provided by careful static analysis, leading to missed performance opportunities.

In this work, we target the \emph{specialization} of the DSE process to \emph{affine programs}, that are programs with a statically analyzable control-flow and dataflow. By restricting the class of programs supported to affine programs, we can deploy \emph{exact} loop and data dependence analysis \cite{Fea92b}. More importantly, as shown in Section~\ref{sec:formulation}, for this class of programs we can model accurately enough the behavior of a design in terms of latency and resource usage by using \emph{non-linear programming}, significantly accelerating the DSE time for such programs by avoiding the need for actual HLS estimation in numerous cases.

%However this comes at the expense of the potentially improved performance that can be reached by considering the complete parameter space.

We illustrate the performance merits and limitations of such general-purpose DSE \cite{autodse} on three important loop-based linear algebra benchmarks. \textit{GEMM}, the classical dense general matrix-multiply, and \textit{2mm} shown in Listing \ref{lst:2mm_motiv} which computes the product of three matrices $D = alpha*A*B*C + beta*D$. Both are key computations in e.g., inference of transformers \cite{devlin2018bert}. \textit{Gramschmidt} computes QR decomposition using the Gram-Schmidt process. In later Section~\ref{sec:evaluation} we evaluate these benchmarks using various problem sizes, ranging from kBs to MBs of footprints for the matrices to demonstrate the robustness of our approach to varying and large problem sizes. Below we use matrices of about 300kB each.

\begin{figure}[!htb]
\begin{lstlisting}[label={lst:2mm_motiv},caption={\vspace{1cm}2mm code: $D=alpha \times A \times B \times C + beta \times D$}]
Loop0: for (i1 = 0; i1 < 180; i1++) 
    Loop1: for (j1 = 0; j1 < 190; j1++) {
        S0: tmp[i1][j1] = 0.0;   
        Loop2: for (k1 = 0; k1 < 210; ++k1) 
            S1: tmp[i1][j1] += alpha * A[i1][k1] * B[k1][j1];
    }
Loop3: for (i2 = 0; i2 < 180; i2++) 
    Loop4: for (j2 = 0; j2 < 220; j2++) {
        S2: D[i2][j2] *= beta;
        Loop5: for (k2 = 0; k2 < 190; ++k2) 
            S3: D[i2][j2] += tmp[i2][k2] * C[k2][j2];
    }
\end{lstlisting}
\end{figure}

The search spaces considered here quickly reach billions of feasible designs, even for kernels containing only a handful of loops and statements. Considering \textit{2mm},  each loop can have a pragma tile and parallel, all with factors that are divisors of the loop trip count and a pragma pipeline. We obtain a space of $1.37 \times 10^{10}$ \textbf{valid} designs. This represents \textbf{432 years} if assuming one design takes a single second to evaluate. Obviously, only a minimal fraction of these spaces is actually explored, making it essential to adequately select the order in which designs are explored.

% \FIXME{LNP to Stephane: put the right version of vitis and board below} DONE!
\paragraph{AutoDSE}
Table~\ref{tab:motiv1}  displays the performance (in GigaFlop/s, GF/s) as reported by Vitis HLS 2021.1 estimation, targeting the AMD/Xilinx Alveo U200 FPGA. We report the performance of the original programs from PolyBench/C \cite{polybench-web} using the medium dataset size for 2mm and Gemm and large dataset for Gramschmidt, when fed to Merlin as-is (column Merlin). The best design found by AutoDSE, given a time budget of 20 hours per benchmark and a timeout of 3 hours per HLS run, is also reported. AutoDSE uses a bottleneck-driven search approach, which targets the improvement of the code section with the lowest throughput \cite{autodse}. It unambiguously achieves particularly solid improvements over a naive design without pragmas. However, we show in Table~\ref{tab:motiv2} below that a carefully built DSE technique, exploiting the regularity of affine programs and leveraging non-linear programming, can provide order(s) of magnitude higher performance for these exact benchmarks, all while using significantly less search time.

%
%using the Medium dataset size and Large for Gramschmidt, without any pragmas, and optimized by Merlin and the final performance achieved by running AutoDSE to insert Merlin pragmas, given a time budget of 20 hours and a timeout of 3 hours per HLS run. AutoDSE uses a bottleneck-driven search approach, which targets the improvement of the code section with the lowest throughput \cite{autodse}.

% DSE time is reported in minutes.

% \vspace{-0.3cm}
\begin{table}[!htb]
\footnotesize
% \begin{table}[]
    \centering
    \begin{tabular}{@{}l rr r rr r rr@{}}
    \toprule
&\multicolumn{2}{l}{\textbf{2mm }(footprint: 773kB)} && \multicolumn{2}{l}{\textbf{Gemm }(footprint: 579kB)} && \multicolumn{2}{l}{\textbf{Gramsch. }(footprint: 15MB)}\\
           \cmidrule{2-3}
           \cmidrule{5-6}
           \cmidrule{8-9}
      &GF/s & Time (min)  && GF/s & Time (min)  && GF/s & Time (min)    \\
       % &  Timeout &  Optimal &  &   \\
      % \hline
      \midrule
      Merlin  & 0.10  & 5
      &&
       0.07 & 5
        &&
       0.14 & 8 \\
       % Expert \\
      AutoDSE & 0.41 & 1,870 
      & & 
      68.91 & 1,345 
      & & 
0.95 & 819  \\
% \framework & 117.48 & 70 
%       & & 
%       105.18 & 185 
%       & & 
%       2.34 & 420  \\

       \bottomrule
\textit{Improvement} & 4.1x & & & 984x & & & 6.8x & \\
\bottomrule
    \end{tabular}
    \caption{Comparison of throughput (GF/s) between the AutoDSE framework and the source-to-source compiler Merlin without pragma insertion for the kernels 2mm, Gemm and Gramschmidt}
    \label{tab:motiv1}
\end{table}

\paragraph{HARP} Machine learning can be deployed to accelerate the search space exploration, substituting for the actual HLS of a particular pragma configuration to estimate its QoR. Table~\ref{tab:motiv-harp} displays the performance achieved by HARP (Hierarchical Augmentation for Representation with Pragma optimization) \cite{harp} on the same setup as in Table~\ref{tab:motiv1}. The HARP model was trained on a database of about 5000 designs, comprising different kernels and problem sizes (including 2mm and gemm, but excluding gramschmidt), as well as different pragma configurations for these, following exactly the methodology in \cite{harp}. We evaluate 75,000 designs using HARP, and run HLS on the top-10 predicted designs. A timeout of 3 hours for each HLS run is implemented. 
We explicitly do not employ re-training or fine-tuning of the model for the 3 benchmarks reported, which requires additional HLS runs, and study the robustness of the model to varying affine benchmarks and problem sizes. We show in Table~\ref{tab:motiv2} below our proposed NLP-DSE approach can match and often outperform HARP, using less HLS runs that would typically be needed to fine-tune the HARP model itself for a benchmark.

\begin{table}[!htb]
\footnotesize
\begin{tabular}{ccccrrr}
\toprule
Benchmark & Pb. size & In training set? & Nb. timeouts & Nb. exceed res. & best GF/s \\ \hline
2mm & small & yes & 0 & 4 & 42.33 \\
2mm & medium & no & 2 & 7 & 96.44 \\
2mm & large & no & 10 & N/A & N/A \\ 

\midrule

gemm & small & yes & 0 & 0 & 27.38 \\
gemm & medium & yes & 0 & 0 & 125.59 \\
gemm & large & no & 3 & 5 & 10.65 \\ 

\midrule

gramsch. & small & no & 0 & 0 & 1.16 \\
gramsch. & medium & no & 10 & N/A & N/A \\
gramsch. & large & no & 10 & N/A & N/A \\ 
\bottomrule
\end{tabular}
\caption{\label{tab:motiv-harp}Performance of HARP on double-precision floating-point data, for 2mm, Gemm and Gramschmidt varying the problem size}

\end{table}

\subsection{Limitations of General-Purpose DSE}

%%%%%%%%%%%%% Limitation on the examples %%%%%%%%%%%%%%%%%$

By analyzing the space explored by the DSE for these three examples, valuable hints can be observed which drive the design of \framework.

\paragraph{Exploration of the space} 

% 2mm 13650482313 1.37E+10
% gemm 2304078 2.30E+06
% gram 122135505 1.22E+08

% The number of valid design for these three examples are $1.37 \times 10^{10}, 2.30 \times 10^{6}$ and $1.22 \times 10^{8}$ for \textit{2mm}, \textit{Gemm} and \textit{Gramschmidt} respectively.

AutoDSE \cite{autodse} 
% and GNN-DSE \cite{sohrabizadeh2022gnn} utilize 
utilizes
the HLS compilers
as a black box, in order to select the configurations that minimize the objective function. 
The tools are agnostic of the input program shape and 
if it detects that Merlin did not apply the pragmas as expected it allows the DSE to prune the design after Merlin has generated the HLS-C code.
These frameworks use incremental DSEs, i.e., having no information on the characteristics of the program, they explore the space by increasing the parallelism in order to respond to a problem, e.g., a bottleneck for AutoDSE.

Table \ref{tab:space_autodse} shows the number of valid designs in each space and the number of synthesized, pruned, timeout designs for each kernel. As we can see with a timeout of 20 hours for the DSE and a timeout of 3 hours for each synthesis, the DSE only allows a tiny part of the space to be explored.

\begin{table}[!htb]
\footnotesize
% \begin{table}[]
    \centering
    \begin{tabular}{@{}l r r r r r@{}}
    \toprule
&\multicolumn{1}{l}{\textbf{2mm }} && \multicolumn{1}{l}{\textbf{Gemm }} && \multicolumn{1}{l}{\textbf{Gramsch. }}\\
           \cmidrule{2-2}
           \cmidrule{4-4}
           \cmidrule{6-6}
           % \cmidrule{4-5}
           % \cmidrule{6-7}
      
      % \midrule
      Nb. valid designs (Space) & $1.37 \times 10^{10}$ && $ 2.30 \times 10^{6}$ && $1.22 \times 10^{8}$   \\
      \midrule
      Nb. design Synthetized (AutoDSE) & 
      % 2mm
        15
        &&
      % gemm
      25
      &&
      % gram 
      15
      \\
      Nb. design pruned (AutoDSE) &
      % 2mm
        49
        &&
      % gemm
      34
      &&
      % gram 
      239
      \\
      Nb. design timeout (AutoDSE) &
      % 2mm
        37
        &&
      % gemm
      27
      &&
      % gram 
      11 
      \\
      \midrule
      Nb. Design explored (AutoDSE) & 
      % 2mm
      101
        &&
      % % gemm
      86
        &&
      % gram 
      265 \\

      % \\
      % Percentage space explored (AutoDSE) & 
      % % 2mm
      %   &&
      % % gemm
      %   &&
      % % gram 

      % \\

       \bottomrule
    \end{tabular}
    \caption{Investigation of design space and exploration coverage for synthesized, pruned, and timeout designs by the AutoDSE framework for the 2mm, Gemm, and Gramschmidt kernels.}
    \label{tab:space_autodse}
\end{table}

\paragraph{Lack of full understanding of pragma impact}
AutoDSE is an incremental method: in order to speed up the search AutoDSE will seek to pipeline certain loops which leads to an unrolling of the innermost loops. Without knowledge of the code, trip counts and resources used this leads to over-use of parallelism, leading to timeouts and/or over-use of resources. For \textit{2mm}, it attempts to pipeline the outermost loops, leading to the above issues.

\paragraph{Parallelism imbalance}
Botlleneck analysis will make it possible to select which part of the code to optimize as a priority \cite{autodse}. However, this priority does not take into account 
%from which level of parallelism the botlleneck changes. 
parallelism (i.e., hardware resources) that shall be deployed for other parts of the code.
This creates code with regions that are extremely/fully parallelized, and others without any parallelism.

For 2mm, the fastest design found by AutoDSE  mainly optimizes one loop body. When AutoDSE tries to optimize the second loop body, it favors the unroll factors to the power of two for the innermost loop and goes directly to the outermost loop. This does not enhance performance or generate configurations pruned by AutoDSE as the pragmas are not applied by AMD Merlin.
The fact that it does not try the other unrolls factors for the innermost loop before optimizing the other pragmas leads a loss of performance.
For Gemm and Gramschmidt, the DSE finds designs with a good 
QoR. However the DSE wastes much time exploring too large unroll factors, which generates ponderously long synthesis times without giving any result as the HLS timeout is reached.
The time spent increasing the unroll factor for certain pragmas without result does not allow the unroll factor of other pragmas to be increased, which results in missed performance.

% \paragraph{Limitations of filtering-based approaches} Approaches that rely on actual HLS runs for exploration have the benefit of being independent of the toolchain being used, as actual synthesis is used to obtain the QoR for designs. Accelerating the search by using a predictor, typically to filter out designs with low predicted QoR can significantly reduce search time and offer solid performance \cite{harp}. However, training requires a significant database of actual designs synthesized by HLS (e.g. 80,000 for HARP \cite{10.5555/3666122.3668084}), itself a particularly time-consuming process. Extending to other data types, other problem sizes, or even other benchmarks may require costly additional retraining or fine-tuning to achieve good performance. 
% In addition, these models are "black-boxes", in that they provide no useful insight to the user regarding the prediction produced. In contrast, our proposed \framework approach provides easily interpretable estimations, similar to more traditional analytical models,
% and \emph{guarantees} designs pruned are necessarily slower than the best design found so far. In contrast to analytical models, we do not rely on using predictors to prune the design space, and instead deploy non-linear programming to directly find relevant candidate designs for HLS. Our approach also requires only minor adjustments to the parameters of the NLP model to support different data types and optimizations for affine programs.

\paragraph{Limitations of filtering-based approaches} Approaches that rely on actual HLS runs for exploration have the benefit of being independent of the toolchain being used, as actual synthesis is used to obtain the QoR for designs. Accelerating the search by using a predictor, typically to filter out designs with low predicted QoR can significantly reduce search time and offer solid performance \cite{harp}. However, training requires a significant database of actual designs synthesized by HLS (e.g. 80,000 for HARP \cite{10.5555/3666122.3668084}), itself a particularly time-consuming process. Extending to other data types, other problem sizes, or even other benchmarks may require costly additional retraining or fine-tuning to achieve good performance. 
In addition, these models are “black-boxes" and do not provide meaningful insight of the predictions they produce. In contrast, our proposed \framework approach offers interpretable estimations, enabling developers to better understand the performance achieved, similar to traditional analytical models. It also \emph{guarantees} that any pruned design is slower than the best design found so far. By leveraging the lower-bound based cost models, our approach estimates latency for specific configurations faster than neural network based approaches for affine programs, allowing for the exploration of many more designs within the same time frame. Moreover, given the explicit latency and resource functions we developed, we can employ existing non-linear programming (NLP) to directly identify the most promising candidate designs for HLS validation, significantly reduce the number of HLS runs.

%%% LNP: para below seems a repeat
%\paragraph{Incremental space exploration}

%The lack of AutoDSE backtracking during bottleneck analysis or the possibilities to change part of the space may be inefficient. Indeed, some workers can be stuck in a sub-optimal local minimal or stuck in the impossibility to synthesize design when there is a over-parallelization like with \textit{2mm}.

%%%%%%%%%%%%% Limitation of current DSEs %%%%%%%%%%%%%%%%%%%%%%%%%

%\subsection{Limitation of the current DSE}

%The limitations can be succinctly outlined as follows:

%\begin{itemize}
%\item Inability to thoroughly explore the design space.
%\item HLS compilers treated as a black box without detailed analysis.
%\item Lack of code analysis, including absence of dependency and array partitioning analysis, posing the risk of designs hitting physical limits (e.g., 1024 for the Vitis compiler) or over-utilizing resources. This complexity can create invalid designs.
%\item Incremental methods, whether too slow (pragma by pragma) or too fast, often lack code awareness. This may lead to pipelining loops without understanding the code, resulting in designs with unrealistic parallelism.
%\item Code exhibiting unbalanced parallelism due to over-optimization when addressing bottlenecks. Highly optimized segments may excessively consume resources, and the addition of new pragmas could render the design invalid.
%\end{itemize}

% % \vspace{-0.01cm}
\section{Overview of \framework}

% \textcolor{red}{reDO}

%In our pursuit of efficiently achieving high-quality designs within a constrained compilation time budget, our primary focus is on addressing the limitations inherent in existing tools. These limitations include restricted space exploration, issues related to over parallelization, parallelism imbalance, and incremental space exploration.

%To overcome these challenges, we present a novel method that enables the theoretical traversal of the entire design space in a matter of seconds or minutes. This is made possible through the implementation of a cost model in the form of a non-linear programming problem, which is then solved using the solver BARON \cite{baron1}. By employing this approach, we can theoretically explore the entire design space, eliminate designs that violate resource constraints, and optimize each loop body uniformly. Additionally, this methodology allows for a faster search by commencing with highly optimized designs tailored to the specific hardware.
%However, it is crucial to note that the practical configuration yielding the best performance may differ. Our cost model is incapable of capturing the intricacies of the downstream toolchain, issues related to resource usage/sharing, and potential performance bugs in optimizing compilers like Merlin and Vitis. As a result, a design space exploration approach remains necessary to achieve optimal performance.

\framework targets the (conservative) modeling of the performance and resources used by a design, such that arbitrary pragma configurations from Section~\ref{subsec:pragams-merlin} are applied on a regular, loop-based affine program. It deploys accurate static analysis to reason on the input program features, and a complex non-linear analytical performance model to drive the design space exploration. That is, \emph{\framework is a method for automatic pragma insertion that is specialized to affine programs}. As demonstrated below, this specialization enables significantly better QoR and DSE time for affine programs than general-purpose DSE approaches such as AutoDSE.

To make our approach feasible and maintain sufficient accuracy in analytical models, we focus on programs with static control flow that can be exactly captured using \emph{polyhedral structures} \cite{Fea92b}. These affine, or polyhedral, programs can be analyzed to obtain the exact information about loop trip counts and dependencies, enabling more accurate performance predictors \cite{zuo2015polyhedral}. The Affine MLIR dialect specializes in modeling such programs \cite{mlir,scalehls}. 

%This restriction aligns with the language affine in MLIR \cite{mlir} and encompasses a significant portion of commonly used programs.
%While our framework primarily targets affine programs, we acknowledge the possibility of supporting non-affine programs by utilizing affine over-approximations \cite{benabderrahmane2010polyhedral}. However, this extension is beyond the scope of our current paper.

%We also propose a novel Design Space Exploration (DSE) framework, exploiting 
Although \framework  can be used to compute pragmas without any DSE, the inherent limits of analytical models persist with \framework: as the implementation details of the back-end toolchain for HLS and synthesis may not be accurately captured by a model. Hence, DSE remains needed for best performance.
\framework enables the exploration of different parallelism and configuration spaces by constraining the level of parallelism, as detailed in Section \ref{sec:implementation}. 
Our model is presented in Section~\ref{sec:modeling}, and we prove it is a performance lower bound in Section~\ref{sec:theoretical}, an important feature to be able to prune designs during the search without the risk of losing performance.
The associated NLP formulation is provided in Section~\ref{sec:formulation}.

% The effectiveness of our framework is demonstrated in Table~\ref{tab:motiv2}, comparing the performance and time-to-solution of \framework. We also display the result of the first synthesizable design produced, \framework-FS. 

The effectiveness of our framework is demonstrated in Table~\ref{tab:motiv2}, which compares the performance and time-to-solution of \framework. Additionally, we present the results of the first synthesizable design produced, referred to as \framework-FS.

\begin{table}[!htb]
% \begin{table}[]
\footnotesize
    \centering
    \begin{tabular}{@{}p{0.0006\linewidth}l
    % @{}\hspace{-0.4cm}
    rrr 
    % @{}\hspace{-0.4cm}
    @{\hspace{-0.4mm}}
    r rrr 
    % @{}\hspace{-0.4cm}
    @{\hspace{-0.4mm}}
    r rrr@{}}
    \toprule

&&\multicolumn{3}{l}{\textbf{2mm }} && \multicolumn{3}{l}{\textbf{Gemm }} && \multicolumn{3}{l}{\textbf{Gramsch. }}\\
           \cmidrule{3-5}
           \cmidrule{7-9}
           \cmidrule{11-13}
    &  &GF/s & Time (mn) & DSP (\%)   && GF/s & Time (mn) & DSP (\%)  && GF/s & Time (mn) & DSP (\%)    \\
       % &  Timeout &  Optimal &  &   \\
      % \hline
      \midrule
\multirow{5}{*}{\rotatebox[origin=c]{90}{\textit{float}}} 
&
      Merlin  & 0.10  & 5 & 0
      &&
       0.07 & 5 & 0
        &&
       0.14 & 8 & 0 \\
       % Expert \\
& AutoDSE & 0.41 & 1,870 & 14
      & & 
      68.91 & 1,345 & 10
      & & 
0.95 & 819 & 1  \\
& \framework-FS & 13.19 & 21  & 24
      & & 
      105.18 & 54 & 26
      & & 
      2.34 & 115 & 2 \\
& \framework & 117.48 & 70 & 39
      & & 
      105.18 & 185 & 26
      & & 
      2.34 & 420 & 2 \\

\midrule

\multirow{4}{*}{\rotatebox[origin=c]{90}{\textit{double}}} &
        Merlin  & 0.12  & 7 & 0
      &&
       0.66 & 5 & 0
        &&
       0.16 & 5 & 0 \\
    & HARP & 96.44  & 420 & 65
      &&
       125.59 & 181 & 80
        &&
       N/A & N/A  & N/A  \\
      &NLP-DSE-FS & 25.49  & 22 & 58
      &&
       29.16 & 9 & 14
        &&
       0.64 & 7 & 1 \\
    &NLP-DSE & 96.44  & 32 & 65
      &&
       120.6 & 66 & 67
        &&
       0.64 & 390 & 1 \\
%        % Expert \\
    
%        % Expert \\

       \bottomrule

& \textit{Imp. vs. AutoDSE} &
286x & 26x &
      & & 
      1.5x & 7.2x &
      & & 
      2.4x & 1.9x & \\
& \textit{Imp. vs. HARP} &
1x & 13x &
      & & 
      0.96x & 2.74x &
      & & 
    N/A & N/A & \\

    \bottomrule
       
    \end{tabular}
    \caption{Comparison of \framework, \framework-FS (which provides the result of the first synthesizable design), the source-to-source compiler Merlin without pragma insertion, AutoDSE and HARP, in terms of throughput (GF/s), DSE time (minutes), and DSP utilization (\%) for the 2mm, Gemm, and Gramschmidt kernels using \textit{float} and \textit{double} data types}
    \label{tab:motiv2}
\end{table}
% \vspace{-0.5cm}

For \textit{Gemm} and \textit{Gramsch} the first design synthetizable has the best QoR of our DSE.
For these two kernels,  \framework implements better parallelism usage compared to AutoDSE. Specifically, our methodology successfully identified configurations with more balanced levels of parallelism.
In contrast, AutoDSE failed to achieve the same level of parallelization within the fixed time for the DSE. 
On one hand, AutoDSE tends to explore first configurations with low levels of parallelism.
%and struggles to reach the same level of parallelism achieved by our approach. 
On the other hand, it concurrently explores design spaces with excessively high levels of parallelism, leading to timeouts and unmet resource constraints. 
%This dual challenge suggests that AutoDSE faces difficulties in striking the right balance and achieving an optimal level of parallelism within the given time constraints, highlighting the advantage of our method in efficiently navigating the design space.
This discrepancy highlights the merits of seeding the DSE with configurations optimized for maximum parallelism, and systematically adjusts this level based on hardware directives and compiler expectations, as implemented in \framework.

For \textit{2mm}, the first design synthetizable allows us to have a better QoR vs. AutoDSE 
but our DSE demonstrates the ability to find a configuration 8.9 times faster than the first configuration found by the DSE. More detail can be found in Section \ref{sec:example}.

It is noteworthy that our Design Space Exploration (DSE) approach, detailed in Section~\ref{sec:implementation}, deviates significantly from AutoDSE. Unlike AutoDSE, which starts with a pragma-free configuration and gradually introduces pragmas, we begin with configurations characterized by the lowest theoretical latency, emphasizing high levels of parallelism. This deliberate departure from the conventional approach is further discussed in Section~\ref{sec:implementation}.

In Section~\ref{sec:evaluation}, we present a comprehensive evaluation of our framework, demonstrating the improvements over AutoDSE that can be achieved by specializing to affine programs, in terms of design throughput and time-to-solutions across various benchmarks. The results indicate an average performance improvement of 5.69x and 17.24x in terms of DSE time and design throughput, respectively, with only a marginal decrease in throughput for 1 out of 47 benchmarks. Importantly, the time-to-solution of \framework it is consistently up to 30x faster than AutoDSE across all benchmarks evaluated.

%% file: sources/model.tex
\label{sec:modeling}

We now present our analytical performance model. We assume the input programs are affine programs \cite{Fea92b,girbal06ijpp}, and therefore exact loop trip counts can be computed by static analysis, similarly for all data dependencies. 
%Further restrictions will be detailed when the models are presented.

\subsection{Program Representation}
\label{subsec:programrepresentation}

We represent programs using a summary of their Abstract syntax tree (AST), with sufficient information to estimate latency and resource consumption by analytical modeling. Intuitively, we can build a constructor-style description of the summary AST, and then directly instantiate the complete formula for estimating e.g., latency, based on loop properties. We first introduce this representation before proving how to compute a latency lower bound with it.

We employ the code below as a running example with the pragma above the loops as AMD/Xilinx Merlin.
For presentation simplicity, we assume each loop iterator in the program region has been renamed to a unique name, so that we can uniquely identify loops by their iterator name.

% \vspace{-0.4cm}
\begin{figure}[H]
\begin{lstlisting}[label={lst:reference}, caption={Running example with the pragma above the loops as AMD/Xilinx Merlin}]
<some-pragmas-for-loop-i>
for (i = lbi; i <= ubi; i++) {
    <some-pragmas-for-loop-j1>
    for (j1 = lbj1(i); j1 <= ubj1(i); j1++) 
        S1(i,j1);
    <some-pragmas-for-loop-j2>
    for(j2=lbj2(i);j2<=ubj2(i);j2++){
        S2(i,j2);
        S3(i,j2);
    }
}
\end{lstlisting}
\end{figure}
%% \vspace{-0.3cm}

The summary AST for this program is simply built by creating one node per \texttt{for} loop and one per statement \texttt{Sx}, the body of a loop is made of loops and/or statements, and their nodes are children of said loop in the tree, listed in their syntactic order. For the example above, using a tree constructor notation, we get:
$
Loop_i(Loop_{j1}(S1), Loop_{j2}(S2,S3))
$.
% $$
% Loop_i(Loop_{j1}(S1), Loop_{j2}(S2,S3))
% $$
Then, a simple rewrite of this formula by substituting the loop and statements by carefully chosen composition operators and descriptors of loop properties will lead to our non-linear analytical model, as outlined in Sec.~\ref{sec:theoretical}.

We first describe the loop properties we associate to each loop. We consider combinations of the following pragmas, based on Merlin's optimizations, for the loop with iterator \texttt{i}:
% 
% \texttt{\#pragma ACCEL parallel <factor=ufi>}, \texttt{\#pragma ACCEL pipeline <II=IIi>}, \texttt{\#pragma ACCEL tile} and  \texttt{\#pragma ACCEL cache <variable=a>}.

\begin{itemize}
    \item \texttt{\#pragma ACCEL parallel <factor=ufi>}
    \item \texttt{\#pragma ACCEL pipeline <II=IIi>}
    \item \texttt{\#pragma ACCEL tile}
    \item \texttt{\#pragma ACCEL cache <variable=a>}
\end{itemize}

We therefore associate to each loop \texttt{i} in the program a \emph{property vector} that informs about the optimizations to be considered. We define $\vec PV_i$ as follows:
% $$
% \vec PV_i : <ispipelined_i,II_i,uf_i,uf_i,TC^{min}_i,TC^{avg}_i,TC^{max}_i>
% $$
% 
% $$
% \vec PV_i : <ispipelined_i,II_i,uf_i,TC^{min}_i,TC^{avg}_i,TC^{max}_i>
% $$
$
\vec PV_i : <ispipelined_i,II_i,uf_i,tile_i,TC^{min}_i,TC^{max}_i>
$
where we have: $ispipelined_i = 1$ if the loop is pipelined, 0 otherwise; $II_i$ is the initiation interval, set to 1 by default; $uf_i$ is the parallelism/unroll factor, set to 1 by default (no \texttt{\#pragma ACCEL parallel} pragma) and set to $TC^{max}_i$ if \texttt{parallel} is defined without a factor $uf_i$ specified. $tile_i$ is the TC of the innermost loop after strip mining.
% Similarly for $uf_i$ for \texttt{unroll}. 
$TC^{min}_i$ is the minimal trip count of loop \texttt{i}, for any of its execution in the program. We also compute the  maximal trip count over all executions. These values are computed using polyhedral analysis on the loops \cite{pouchet:fpga13}. The pragma \texttt{cache} transfers above the loop \texttt{i}  the data needed for the computation of this loop nest for the array \texttt{a}.

%\myparagraph{Modeling the compiler's optimizations}

This vector is built by syntactic analysis on the program, where the default value $PV_i : 
% <0,1,1,1,TC^{min}_i,TC^{avg}_i,TC^{max}_i>$ 
<0,1,1,1,TC^{min}_i,TC^{max}_i>$ 
is used for a loop without any pragma. Once all loops have been annotated by their $\vec PV$ properties, subsequent treatment can be implemented to mirror the optimizations implemented by the back-end tool.

\myparagraph{Modeling Vitis optimizations}
AMD/Xilinx Vitis will apply several optimizations automatically, such as auto-pipeline and auto-loop-flatten, 
some other optimizations when
 the user gives a compilation option such as tree reduction.
 % with the option \textit{-funsafe-math-optimizations}.
% The pragmas unroll and pipeline will respectively allow to unroll a loop and to pipeline a loop. 
Only a loop with a constant trip count ($TC$), i.e., $TC_{max} = TC_{min}$ can be unrolled.
The unroll pragma options allow to specify the unrolling factor, $uf_i$. When the factor is not specified, it implies that the factor is equal to the $TC$ of the loop.
% Another important processing sets the values of $uf_i$ or $uf_i$ to $TC^{max}_i$ when the \texttt{factor=} argument is not specified, if the loop has a constant trip count (i.e.,  $TC^{min}_i = TC^{max}_i$).
%
When a loop is pipelined, all innermost loops are automatically fully unrolled. Hence, we also propagate unrolling information, e.g., to mark a full loop nest for full unrolling if an outer loop is marked with \texttt{\#pragma ACCEL pipeline}. The pipeline pragma options allow to specify the objective II the user wants to achieve. When II is not specified, it is automatically set to 1.
In addition, Vitis  will auto-pipeline  with a target II of 1 the innermost loops which are not fully unrolled for 
each nested loop.
Within Vitis, users can enable optimizations like logarithmic time reduction through tree reduction. This optimization choice will be a global option within our model, applicable across the entire model rather than being limited to a specific loop.

\myparagraph{Modeling Merlin optimizations}
% Finally, Merlin will also add automatic optimizations.
% When a parallel pragma is applied, the loop will be explicitly strip-mined during partial unrolling. The innermost loop will have a trip count (TC) equal to the unrolling factor, with the unroll pragma applied to that loop
% Similarly to Vitis, Merlin will auto-pipeline.
% Merlin also applies a program transformation for certain pragmas. When there are two perfectly nested and partially unrolled loops, Merlin swaps the two loops strip-mined (if legal), unrolled innermost and flatten and pipeline the two outermost loops.
% Further, Merlin will  automatically transfer the data from off-chip to on-chip and cache on-chip with packing, with a maximum packing of 512 bits for our FPGA while computing if the footprint of the data fit on-chip by static analysis. The pragma tile allows Merlin to strip mine a loop and 
% give the compiler the opportunity to transfer 
%  less/more data while respecting resource constraints. For our model we suppose an optimistic data transfer i.e., all memory transfers are done with a packing of 512 bits and each data are transferred once (perfect data reuse).
% 
Merlin applies automatic optimizations to enhance performance. When a parallel pragma is used, the loop undergoes explicit strip-mining during partial unrolling. The innermost loop’s trip count matches the unrolling factor, with the parallel pragma applied at this level. Similar to Vitis, Merlin automatically pipelines loops.
Merlin also performs program transformations for specific pragmas. For example, in the case of two perfectly nested and partially unrolled loops, Merlin (when legal) swaps the strip-mined loops, unrolls the innermost loop, and flattens and pipelines the two outermost loops.
Additionally, Merlin automates data transfers from off-chip to on-chip memory, caching data on-chip with packing. Our FPGA supports a maximum packing size of 512 bits. Merlin uses static analysis to ensure the data footprint fits on-chip. The tile pragma enables loop strip-mining, allowing the compiler to transfer more or less data while adhering to resource constraints.
For our model, we assume an optimistic data transfer scenario—memory transfers occur with 512-bit packing, and each piece of data is transferred only once, reflecting perfect data reuse.

Consequently, the set of possible $\vec{PV_l}$ vectors are adjusted by analyzing the input code, and modifying their initial value, possibly further constraining the set of possible $\vec{PV_l}$ based on which program transformation will be performed, as described above.
Overall, the $\vec{PV_l}$ vectors, along with the summarized AST, contain sufficient information to capture several source-to-source transformations performed by the Merlin compiler for coarse- and fine-grain parallelization, and reason on the likeliness of the optimization to succeed at HLS time (e.g., capturing loops with  non-constant trip count).

%% file: sources/NLP_formulation.tex
\label{sec:formulation}

% \REMARK{The connection between the latency insights of Section 4 and the formulation of Section 5 is unclear. None of the equations in Section 5 includes the latency bound--where exactly is this information used and how? The objective function is not understandable at all; please devote a couple of sentences to at least explain the basic intuition. }
% \FIXME{Explain the objective function \textcolor{red}{DONE}}
% \FIXME{Add a small paragraph to explain how to compute the minimal II \textcolor{red}{DONE}}

We now present the complete set of constraints and variables employed to encode the latency and resource model as a non-linear program.
This section presents a modeling of section \ref{sec:theoretical} in the practical case.

% TODO improve this

% Let $\mathcal{L}$ be the set of loops, $\mathcal{A}$ the set of arrays, $\mathcal{S}$ the set of statements and $\mathcal{O}_s$ the operations of the statements $s$. In order to have an accurate model we distinguish for each statement the operation which can be done in parallel and the operation of reduction. Hence, $\mathcal{O}_{s_{par}}$ which does not have any loop-carried dependence and $\mathcal{O}_{s_{red}}$ which is an associative/commutative operator to reduce one or more values into a single value, leading to loop-carried dependencies, represent respectively the operation of the statement $s$ which can be done in parallel and the operation of reduction. 

Let $\mathcal{L}$ be the set of loops, $\mathcal{A}$ the set of arrays, $\mathcal{S}$ the set of statements and $\mathcal{O}_s$ the operations of the statements $s$. In order to have an accurate model we distinguish for each statement 
the operation which can be done in parallel, i.e., does not have any loop-carried dependence, $\mathcal{O}_{s_{par}}$  
and the reduction operations, i.e., associative/commutative operators to reduce one or more values into a single value, leading to loop-carried dependencies, $\mathcal{O}_{s_{red}}$.

Let $\mathcal{P}$ 
be the set of different possible pipeline configurations.
Let $\forall p \in \mathcal{P}$ define $\mathcal{L}^p_{pip}$ the set of loops pipelined and $\forall l \in \mathcal{L}^p_{pip}$, $\mathcal{L}^p_{under\_pip_l}$ the set of loops within a loop pipelined and $\mathcal{L}^p_{above\_pip_l}$ the set of loops above the loop pipelined $l$.
% Let $s \in \mathcal{S}$ define the set of nested loops which iterate each statement, $ \mathcal{L}_s$ the set of nested loop which iterate the statement $s$.
Let $\forall s \in \mathcal{S}$ define the set of nested loops which iterate the statement $s$, $ \mathcal{L}_s$.
 $\forall a \in \mathcal{A}$ and for $d$ a dimension of the array $a$, let $\mathcal{C}_{a_d}$ be the set of loops which iterates the array $a$ at the dimension $d$. 
$\forall l \in \mathcal{L}$, let designate $d_l$ the maximum dependency distance of the loop $l$.
And let $II_s$ be the II of the loop pipelined for the statement $s$.

The II for each loop, the dependencies, the properties of the loops, the trip count (TC), the iteration latency of the parallel operations and the reduction operations and the number of DSPs per operation per statements are computed at compile time with PolyOpt-HLS \cite{pouchet:fpga13} and used as constants in the NLP problem.

Table \ref{tab:var}, summarize the sets, variable and constants we use.

\subsection{Variables}

% In this subsection, we proceed to define the variables that encompass the design space of the code. Specifically, we consider the possibilities of pipelining $loop\_pip \in \{0,1\}$, unrolling $loop\_UF \in \mathbb{N}^*$, and tiling $loop\_tile \in \mathbb{N}^*$ for each loop. 
% Additionally, we include the possibility of caching an array $a$ that is iterated over by the loop $l$ $loop_l\_cache\_array_a \in \{0,1\}$. 
Variables in the formulation correspond to $PV_l$ attributes.
We consider the possibilities of pipelining (Eq. \ref{eq:pip}), unrolling (Eq. \ref{eq:UF}), and tiling (Eq. \ref{eq:tile}) for each loop. 
Additionally, we include the possibility of caching an array that is iterated over by the loop (Eq. \ref{eq:cache}).

\begin{eqnarray}
\label{eq:UF}
 \forall l \in \mathcal{L}, 1 \leq loop_l\_UF \leq TC_l
\\
% \nonumber % to remove the +1 label without equation
% \end{eqnarray}
% \begin{eqnarray}
 % to remove the +1 label without equation
% 
\label{eq:tile}
\forall l \in \mathcal{L}, 1 \leq loop_l\_tile \leq TC_l
\\
% \nonumber % to remove the +1 label without equation
% \end{eqnarray}
% \begin{eqnarray}
 % to remove the +1 label without equation
% 
\label{eq:pip}
\forall l \in \mathcal{L}, loop_l\_pip \in \{0,1\}
\\
% \nonumber % to remove the +1 label without equation
% \end{eqnarray}
% \begin{eqnarray}
% 
\label{eq:cache}
\forall l \in \mathcal{L}, \forall a \in \mathcal{A}, loop_l\_cache\_array_a \in \{0,1\}
 \end{eqnarray}

\begin{table}[!htb]
% \begin{table*}[]
\footnotesize
\centering

\begin{tabular}{@{}l l@{}}

\toprule
Set & Description \\
\midrule
$\mathcal{L}$ & the set of loops \\
$\mathcal{A}$ & the set of arrays \\
$\mathcal{S}$ & the set of statements \\
$\mathcal{O}_s$ & the list of operations of the statement $s$ \\
$\mathcal{O}_{s_{par}}$ & the operations which can be done in parallel, i.e., do not have any loop-carried dependence \\ 
$\mathcal{P}$ & the set of different possible pipeline configurations \\
$\mathcal{L}^p_{pip}$ & the set of pipelined loops \\
$\mathcal{L}^p_{under\_pip_l}$ & the set of loops under a pipelined loop \\ 
$\mathcal{L}^p_{above\_pip_l}$ & the set of loops above the pipelined loop $l$ \\
$\mathcal{L}_s$ & the set of nested loops which iterate over the statement $s$ \\
$\mathcal{C}_{a_d}$ & the set of loops which iterate over the array $a$ at dimension $d$ \\ 
$d_l$ & the maximum dependency distance of the loop $l$ \\
$AP_{a,d}$ & Array Partition for the array $a$ in dimension $d$ \\
\midrule
Variable & Description \\
\midrule
$loop_l\_UF$ & Unroll factor of the loop $l$\\
$loop_l\_tile$ & Tile size of the innermost loop after strip-mining the loop $l$\\
$loop_l\_pip$  & Boolean to indicate if the loop $l$ is pipelined\\
$loop_l\_cache\_array\_a$ & Boolean to indicate if the array $a$ is transferred on-chip before the loop $l$ \\
\midrule
Constant & Description \\
\midrule
$TC_l$ & Trip Count of the loop $l$ \\
$II_l$ & Initiation Interval (II) of the loop $l$ \\
$IL_{par}$ & Iteration Latency of the operations without dependencies in the statement $s$\\
$IL_{red}$ & Iteration Latency of the operations with dependencies in the statement $s$\\
$DSP_{s_{op}}$ & Number of DSPs used for the statement $s$ for the operation $op$\\
$DSP_{available}$ & Number of DSPs available for the FPGA used \\
$max_{part}$ & Maximum array partitioning defined by the user or the DSE (cf. Section \ref{sec:implementation}) \\
$footprint\_array_a\_loop_l$ & Footprint of the array $a$ if transferred on-chip before the loop $l$ \\
$D_{a}$ & Number (Integer) of dimensions of the array $a$ \\
\bottomrule

\end{tabular}
\caption{Description of the sets, variables, and constants utilized in the formulation of the Nonlinear Problem (NLP) aimed at modeling latency and resource consumption of a design}
\label{tab:var}
% \vspace{-0.8cm}
\end{table}

% \subsection{Rules}
\subsection{Modeling Compiler Transformations}
\label{rules}

Now that we have defined our design space, we need to constrain the space by removing infeasible cases and those that do not comply with the rules of the compilers.

\myparagraph{Pipeline Rules:}
 Vitis HLS  unrolls all loops under  the pipelined loop. This implies that all loop $l$ under the pipelined loop must have $loop_l\_UF == TC_l$ (Eq. \ref{eq:under_pip_fully_unrolled}).
Considering this constraint, it is important to note that for each statement, only one of the loops that iterate the statement can have a pragma pipeline (Eq. \ref{eq:one_pip_per_nesteed_loop}). Thanks to PolyOpt-HLS \cite{pouchet:fpga13}, we have the schedule of the kernel, including the loop order and which loops iterate over other loops.
If multiple pipelined loops were present, the loops beneath the first pipelined loop would be unrolled instead.
For example, in Listing \ref{lst:2mm_motiv}, if Loop4 (line 8) is pipelined, this implies that Loop5 is fully unrolled with an unroll factor of 190.

\begin{eqnarray}
\label{eq:under_pip_fully_unrolled}
% \begin{cases}
\forall p \in \mathcal{P}, \forall lp \in \mathcal{L}^p_{pip}, \forall lbp \in \mathcal{L}_{under\_pip_{lp}},  loop_{lp} \times loop_{lbp}\_UF == loop_{lp} \times TC_{lbp}
\\
\label{eq:one_pip_per_nesteed_loop}
% \begin{cases}
\forall s \in \mathcal{S}, \sum_{l \in \mathcal{L}_s} loop_l\_pip \leq 1
 \end{eqnarray}

\myparagraph{Memory Transfer Rules:}
% Xilinx Merlin
 Merlin  automates the process of transferring data on-chip and applying array partitioning. The tool caches data on-chip and packs it in chunks of up to 512 bits, enabling efficient  transfer speed. When the data is already present on-chip it can be reused provided that resource constraints are satisfied. The compiler caches on-chip the data only above the loop pipelined (Eq. \ref{eq:no_cache_under_pip}).

If Loop4 in Listing \ref{lst:2mm_motiv} is pipelined, we need to cache at least all the data from the second dimension of D and tmp (i.e., the tiles $1 \times 220$ and $1 \times 190$) and all the elements of C (i.e., the tile $190 \times 220$) on board to ensure efficient pipelining and unrolling.

 \begin{eqnarray}
\label{eq:no_cache_under_pip}
\forall p \in \mathcal{P}, \forall lp \in \mathcal{L}^p_{pip}, \forall lbp \in \mathcal{L}_{under\_pip_{lp}}
\forall a \in \mathcal{A}, 
loop_{lbp}\_cache\_array_a == 0
 \end{eqnarray}

\myparagraph{Dependencies:}
Loop-carried dependencies are managed using constraints (Eq \ref{eq:dd}). If a loop has a dependency distance of $n$, this means that if we unroll with an unroll factor $uf > n$ this is equivalent to unrolling the loop with a factor $uf = n$ because the statements corresponding to the iteration $\{uf+1, \dots, n\}$ will be executed only after the first n statements are executed due to the dependency.
Loop-independent data dependencies are managed at the objective function, as elaborated in Section \ref{sec:obj_fct}. 

If we encounter code like $for (j = 2; j < N; j++) y[j] = y[j-2] + 3;$, a straightforward approach to handling this type of dependency is to impose a constraint such as $loop_l\_UF \leq 2$, which is represented by equation \ref{eq:dd}. In this case, due to dependencies an $UF > 2$ is similar 
to 
% have 
$UF=2$.

\begin{eqnarray}
\label{eq:dd}
\forall l \in \mathcal{L}, \text{ if } dd_l > 1,  loop_l\_UF \leq d_l
 \end{eqnarray}

% if the statements are independent then the iteration latency of the loop body corresponds to the longest execution, we take the max. If there are dependencies we add because each statement is executed after the previous one is finished.

\myparagraph{Array Partitioning:}
The array partitioning is constrained (Eq. \ref{eq:max_part1}) to adhere to the maximum array partitioning limit (e.g., 1024 for AMD/Xilinx). Equations \ref{eq:max_part2} and \ref{eq:max_part3} enforce that the array partitioning must be equal to or a multiple of the unroll factor for dimension $d$. Equation \ref{eq:max_part1} constrains the total array partitioning.
% 
% This constraint aims to limit the maximum unrolling factor while providing more flexibility than directly restricting unrolling factors. By constraining the total array partitioning, we limit the unroll factor by limiting the product of unroll factors, rather than directly restricting each unroll factor individually.
% 
This constraint limits the maximum unrolling factor while offering more flexibility than directly restricting individual unroll factors. By constraining total array partitioning, it indirectly controls the unroll factor by limiting the product of unroll factors, rather than imposing direct limits on each one.

For example, if we set \( max_{part} \) to 1024, the array partitioning of array \( D \) in Listing \ref{lst:2mm_motiv} will limit the product of the unroll factors of Loop3 and Loop4 to 1024. Specifically, \( AP_{D,0} = \text{loop}_{Loop3}\_UF \) and \( AP_{D,1} = \text{loop}_{Loop4}\_UF \), and the product across dimensions \( d \in \llbracket 1, D_a \llbracket \) is constrained by \( \prod_{d \in \llbracket 1, D_a \llbracket} AP_{a,d} = \text{loop}_{Loop3}\_UF \times \text{loop}_{Loop4}\_UF \leq max_{part} \).

\begin{eqnarray}
\label{eq:max_part1}
\forall a \in \mathcal{A}, \prod_{d \in \llbracket 1,D_{a} \llbracket} 
AP_{a,d} \leq max_{part} \\
\label{eq:max_part2}
\forall a \in \mathcal{A}, \forall d \in \llbracket 1,D_{a} \llbracket, \forall l \in C_{a_d}, 
AP_{a,d} \% loop_{l}\_UF == 0 \\
\label{eq:max_part3}
\forall a \in \mathcal{A}, \forall d \in \llbracket 1,D_{a} \llbracket, \forall l \in C_{a_d}, 
AP_{a,d} \geq loop_{l}\_UF
\end{eqnarray}

\myparagraph{Supplementary Rules:}
In addition, we add the constraints for the divisibility of the problem size of the unroll factors (Eq. \ref{eq:uf_divide_tc}) and the tile size (Eq. \ref{eq:tile_divide_tc}). For this last, this corresponds to adding an upper bound to the product of the unroll factors of all the loops which iterate the same array on different dimensions.

During our DSE, we can force the solution to be fine-grained. In this case we add a constraint where the loop above the loop pipelined has a UF of 1, i.e., for all loop $l$ above the pipelined loop 
$loop_{l}\_UF == 1$ (Eq. \ref{eq:fine_grained}).
\begin{eqnarray}
\label{eq:uf_divide_tc}
% \begin{cases}
\forall l \in \mathcal{L}, loop_l\_UF \% TC_l == 0
% \end{cases}
\\
\label{eq:tile_divide_tc}
% \begin{cases}
\forall l \in \mathcal{L}, loop_l\_tile \% TC_l == 0
% \end{cases}
\\
\label{eq:fine_grained}
% \begin{cases}
(Opt. )\forall l \in \mathcal{L}, \forall l' \in \mathcal{L}_{above\_l}, loop_l\_pip * loop_{l'}\_UF <= 1
 \end{eqnarray}
%\subsection{Constraints and Resource}
% 
We also constrain the resources, modeling their sharing optimistically. We consider the number of DSPs (Eq. \ref{eq:dsp_optimistic}) and on-chip memory (Eq. \ref{eq:on_chip}) used. 
As the consumption of DSPs can be difficult to estimate due to resource sharing we utilize an optimistic estimate, which considers a perfect reuse/sharing: as soon as a computation unit is free, its resource can be reused. 

\begin{eqnarray}
\label{eq:dsp_optimistic}
% \begin{cases}
% & DSPs\_used_{optimistic} = \\
%     & \sum_{op \in \{+,-,*,/ \}} \max_{s \in \mathcal{S}}(DSP_{s_{op}} / II_s)
% \end{cases}
 DSPs\_used_{optimistic} = 
     \sum_{op \in \{+,-,*,/ \}} \max_{s \in \mathcal{S}}(DSP_{s_{op}} / II_s) \leq DSP_{available}
\\
\label{eq:on_chip}
% \begin{cases}
     \sum_{a \in \mathcal{A}} \sum_{l \in \mathcal{L}} loop_l\_cache\_array_a  \times footprint\_array_a\_loop_l \leq Mem
 \end{eqnarray}

% Xilinx Merlin  automates the process of transferring memory on-chip and conducting array partitioning. The tool caches data on-chip and packs it in chunks of up to 512 bits, allowing for efficient transfer speed. When the data are already present on-chip, they can be reused, provided that resource constraints are satisfied.
% % \vspace{-0.4cm}
% \subsection{Initiation Interval of the Pipeline}

% We compute the minimal II in function of the dependencies of the loop pipelined and the iteration latency of the operations of the statements during the NLP generation. 
% % 
% Let $RecMII$ and $ResMII$ which are respectively the recurrence constrained and the resource constraint of the loop pipelined and we have $II \geq \max{(ResMII, RecMII)}$. 

% $RecMII = \max_i \lceil \frac{ delay(c_i)}{ distance(c_i)} \rceil$ with $delay(c_i)$ the total latency in dependency cycle $c_i$ and $distance(c_i)$ the total distance in dependency cycle $c_i$.
% % As we cannot compute accurately $ResMII$ as we don't know how the resource will be used by the compiler 
% We suppose that $ResMII=1$, as we don't know how the resource will be used by the compiler.
% % 
% Hence, if the loop is a reduction loop then the $II \geq \frac{IL_{reduction}}{1}$ with $IL_{reduction}$ the iteration latency of the operation of reduction. For a kernel like $for (j = 0; j < N; j++) y[j] = y[j-2] + 3;$ the $II \geq \lceil \frac{IL_+}{2} \rceil$.

% \subsection{Constraints}
% % \vspace{-0.2cm}

% \input{sources/eq}

% \vspace{-0.23cm}
\subsection{Objective function}

\label{sec:obj_fct}

Lastly, we need to define the objective function ($obj\_func$) that supports fine-grained and coarse-grained parallelism. Fine-grained parallelism involves duplicating a specific statement(s), while coarse-grained parallelism duplicates modules, including statements and loops. However, it may not always be feasible to achieve parallelism based on the characteristics of the loops and the level of parallelism required.
Therefore, we distinguish between parallel and reduction loops.
A parallel loop can be coarse and fine-grained unrolled, whereas a reduction loop can only be fine-grained unrolled with a tree reduction process that operates in logarithmic time.

As the pragmas \texttt{cache}  are part of the space we compute the communication latency with these pragmas. If more than one array is transferred above the same loop we take the maximum as Merlin transferred them in parallel.
To ensure these properties, we formulate the objective function for each pipeline configuration.
The objective function uses the combined latencies of communication and computation. \textit{When using  Merlin, communication and computation do not overlap, but communication tasks can overlap when they occur consecutively in the code at the same level}. 
Consequently, for each loop, where two arrays are transferred consecutively within the loop, we calculate the sum of the maximum latencies for these transferred arrays ($L_{mem}$).

In every loop nest, there will invariably be a pipeline loop due to either user-inserted or compiler-inserted instructions 
(AMD/Xilinx Merlin and Vitis automatically insert  the pragma pipeline if it is not done by the user or the previous compiler). 
Therefore, the objective function takes the following form: $TC_{ap} \times (IL + II \times (\frac{TC}{UF}-1))$, where $TC_{ap}$ includes the loops situated above the pipeline. Parallel loops above the pipeline can be coarse-grained parallelized. The iteration latency within the unrolled loop body is divided into either reduction operations or non-reduction operations, as reduction operations require logarithmic time for the reduction process.
The variable IL encompasses the latencies of the statements found within the pipelined loop body. Independent statements can be executed in parallel and statements with dependencies are summed. 
% , as detailed in Section \ref{sec:analyticalModelTemplate}.

% \vspace{-0.4cm}

\begin{equation*}
\left.
\begin{cases}
    & TC_{ap} = \prod_{l\in \mathcal{L}^{par}_{above\_pip}} \frac{TC_l}{loop_l\_UF} \times 
\prod_{l\in \mathcal{L}^{red}_{above\_pip}} TC_l 

 \\
& IL = IL_{par} +
IL_{seq} \times \prod_{l\in \mathcal{L}^{red}_{under\_pip}} \frac{TC_l}{loop_l\_UF} \times \log_2(loop_l\_UF)  \\

& L_{mem} = \sum_{l \in \mathcal{L}}   \max_{a\in \mathcal{A}} (loop_l\_cache_a \times footprint_a\_loop_l)
\\
& obj\_func = TC_{ap} \times (IL + II\times (\frac{TC_{lp}}{loop_{lp}\_UF}-1) ) + L_{mem}
    \end{cases}
    \right.
\end{equation*}

% \begin{equation*}
% \left.
% \begin{cases}
%     & obj\_func = \prod_{l\in \mathcal{L}^{par}_{above\_pip}} \frac{TC_l}{loop_l\_UF} \times 
% \prod_{l\in \mathcal{L}^{red}_{above\_pip}} TC_l 
% \times 
% (
%  \\
% & \left( IL_{par} +
% IL_{seq} \times \prod_{l\in \mathcal{L}^{red}_{under\_pip}} \frac{TC_l}{loop_l\_UF} \times \log_2(loop_l\_UF) \right) \\
% & +
% II \times (\frac{TC_{lp}}{loop_{lp}\_UF}-1)
% ) +  \sum_{l \in \mathcal{L}} \\ &  \max_{a\in \mathcal{A}} (loop_l\_cache\_array_a \times footprint\_array_a\_loop_l)
%     \end{case}
%     \right.
% \end{equation*}

\subsection{Example}

% \begin{lstlisting}[label={lst:reference2},caption=Example with dependency]
% for (j = 0; j < N; j++) 
%       S0: y[j] = y[j-2] + 3;
% \end{lstlisting}
% % \vspace{0.5cm}

% \vspace{-0.45cm}
\begin{figure}[!htb]
\begin{lstlisting}[label={lst:reference1},caption={AtAx code for Large problem size: $t = A * x; y = A^t * t$} ]
Loop0: for(i=0; i<2100; i++) 
    S0:y[i] = 0;
Loop1: for(i=0; i<1900; i++) { 
    S1:t[i] = 0;
    Loop2: for(j=0; j<2100; j++) 
        S2:t[i]+=A[i][j]*x[j];
    Loop3: for(j=0; j<2100; j++) 
        S3:y[j]+=A[i][j]*t[i];
}

\end{lstlisting}
\end{figure}
% \vspace{0.5cm}

We now use the AtAx kernel (Listing \ref{lst:reference1}) as an illustration.

$S0$ and $S3$ do not have inter-iteration dependencies within their respective loops, $Loop0$ and $Loop3$. Therefore, it is possible to pipeline $Loop0$ and $Loop3$ with an initiation interval (II) $\geq 1$. In other words, there are no dependencies on previous iterations within the same loop for all iterations in these loops.

Loop1 and Loop2 are reduction loops, where the reduction operation is an addition with a latency of $IL_+$ cycles. Thus, the II for these loops must be greater or equal than $ IL_+$.

If Loop1 is pipelined, there is a dependency between $S1$, $S2$, and $S3$. The dependency distance between $S1$ and $S2$ is 1, so we simply add $IL_{S1}$ and $IL_{S2}$. 
Between $S2$ and $S3$, the array $t$ can be read in statement $S3$ after the reduction of Loop2 is completed in $\log(N)$ cycles due to the tree reduction. Therefore, the final equation for the loop body cycle will be $IL_{S1} + IL_{S2} \times \log(N) + IL_{S3}$.

If statements can be executed simultaneously (i.e., there is no dependency), we use the maximum instead of an addition.

% We now employ the AtAx kernel (Listing \ref{lst:reference1}) as an illustration.
% % 
% $S0$ and $S3$ do not have inter-iteration dependencies within their respective loops, namely $Loop0$ and $Loop3$. Therefore, it is possible to pipeline $Loop0$ and $Loop3$ with an $II \geq 1$. In other words, for all iterations within these loops, there are no dependencies on previous iterations within the same loop.
% % 
% Loop1 and Loop2 are reduction loops, and the reduction operation is an addition which has an IL of $IL_+$ cycles. So the II >= $IL_+$.
% % 
% If loop1 is pipelined, there is a dependency between S1, S2 and S3.
% Between S1 and S2 the distance is 1, so we just add IL\_S1 and IL\_S2.
% Between S2 and S3 the distance is log(N) due to the tree reduction of the unrolled loop Loop2. So the final equation will be IL\_S1 + IL\_S2 * log(N) + IL\_S3 for the loop body cycle.
% If statements can be run at the same time (i.e., there is no dependency) it is a max instead of an addition. 
% If we encounter code resembling Listing \ref{lst:reference2}, a straightforward approach to handle this type of dependency is to impose a constraint such as $loop_l\_UF \leq dd_l$, which is represented by equation \ref{eq:dd}. In this case, for the computation estimation an $UF > 2$ is similar to have $UF=2$.

%% file: sources/th_proof_v2.tex
%\section{Theoretical Latency and Resource Modeling}

% \label{sec:appendix}
%\label{appendix:theoretical}

We now revisit the analytical modeling of programs in \framework, demonstrating that the objective function presented in Section~\ref{sec:formulation} serves as a lower bound on program latency under resource constraints. The proofs for the following theorems are provided in Appendix \ref{appendix:proof}.

\subsection{A Formal Model for Latency}

Our objective is to formulate a lower bound on the latency of a program after HLS. We therefore have put several restrictions: we assume the input program is an affine program, that is the control-flow is statically analyzable; all loops can be recognized and their trip count computed; and all array / memory accesses can be exactly modeled at compile-time. No conditional can occur in the program. While our approach may generalize beyond this specific class, we limit it here to these strict assumptions.

To maintain a lower bound on latency by composition, we operate on a representation of (parts of the) program which is both schedule-independent and storage-independent: indeed, a lower bound on this representation is necessarily valid under any schedule and storage eventually implemented. \emph{We however require HLS to not change the count and type of operations}. Furthermore, for lower bounding purposes, we assume unless stated otherwise $\forall i,~inparallel_i = 1$. We will discuss in the next section a more realistic but compiler-dependent approach to set $inparallel_i$, based on dependence analysis.

We assume programs are made of affine loops, that are loops with statically computable control-flow, with loop bounds made only of intersection of affine expressions of surrounding loop iterators and program constants. We now assume loop bodies (i.e., statements surrounded by loops) have been translated to a list of statements, with at most a single operation (e.g., \texttt{+,-,/,*}) per statement. Operations are n-ary, that is they take $n \ge 0$ input scalar values as operand, and produce 0 or 1 output scalar value. A memory location can be loaded from (resp. stored to) an address stored in a scalar variable. This is often referred to as straight-line code
% \cite{some-good-def}
. 
This normalization of the loop body facilitates the computation of live-in/live-out data for the code block, and the extraction of the computation graph. Note the region can be represented in Static Single Assignment 
% \cite{ssa} 
form, to ensure different storage location for every assignment, facilitating the construction of the operation graph.
In addition we require the input program to not contain useless operations which may be removed by the HLS toolchain e.g. by dead code elimination. 
% , as illustrated in Listing~\ref{lst:dead-code}.

% \begin{lstlisting}[caption={Example of Code Illustrating Dead Code Elimination: As Statement S1 writes to variable x, but the value of x assigned in Statement S0 is never used, the compiler will remove Statement S0.},label={lst:dead-code}]
% int example(int y, int z){
%     int x;
%     S0: x = 12 + y; //  dead-code elimination
%     S1: x = y + z;
%     return x;
% }
% \end{lstlisting}
% \vspace{0.7cm}

The restriction can be summarized as:

\begin{itemize}[leftmargin=*]
    \item The input program is a pure polyhedral program \cite{girbal06ijpp}, and its analysis (loop trip counts for every loop, all data dependences \cite{Fea92b}) is exact.
    \item No HLS optimization shall change the number of operations in the computation: strength reduction, common sub-expression elimination, etc. shall either first be performed in the input program before analysis, or not be performed by the HLS toolchain. The program also does not contain "useless" operations that may be removed by the compiler with a dead-code elimination pass.
    \item We only model DSP and BRAM resources for the considered kernel, ignoring all other resources. We do not model LUT and FF resources, because from experience in the loop-based benchmarks we consider DSP and BRAM resources are the most constraining resources. Moreover, the estimation of LUTs and FFs is more tedious.
    \item We assume that resource (DSP) sharing across different operations executing at the same clock cycle is not possible.
\end{itemize}

An important term is $SL$, a latency lower bound for a region of straight-line code. To maintain a lower bound on latency by composition, we operate on a representation of (parts of the) program which is both schedule-independent and storage-independent: the operation graph, or CDAG \cite{jia1981complexity,elango2015characterizing}. Indeed, a lower bound on this representation is necessarily valid under any schedule and storage eventually implemented, and can be used to prove I/O lower bounds on programs \cite{elango2015characterizing}. The CDAG of a straight-line code region is the directed acyclic graph with one node per operation in the code region, connecting all immediate producer and consumer operations directly. Then, we can easily compute the length of its critical path, which represents the minimal set of operations to execute serially.

% Then, we can build a lower bound on the number of cycles a region $L$ may require to execute, under fixed resources, by simply taking the maximum between the weighted critical path and the work to execute normalized by the resources available.

\begin{definition}[Straight-line code]
\label{def:straightlinecode}
An n-ary operation takes n scalar operands $\vec i$ as input, and produces a single scalar $o$ as output. A statement contains a single n-ary operation, or a load from (resp. store to) a memory location to (resp. from) a scalar. A straight-line code region $L$ is a list of consecutive statements, with a single entry and single exit.
\end{definition}

\begin{definition}[Live-in set] The live-in set $V_I^L$ of region $L$ is the set of scalar values, variables and memory locations that read before being written, under any possible valid execution of $L$.
\end{definition}

\begin{definition}[Live-out set] The live-out set $V_O^L$ of region $L$ is the set of variables and memory locations that written to during any possible valid execution of $L$.
\end{definition}

%%% LNP: I think it's not useful for the reader.
%On the listing \ref{lst:dead-code}, $V_O^L = \{x\}$ and $V_I^L = \{y, z\}$. 
%On the listing \ref{lst:2mm}, let be $L$ the region of the first loop body,
%$V_O^{L} = \{tmp[i_1][j_1], \forall i_1,j_1 \text{ s.t. } 0 \leq i_1 \leq 180 \And 0 \leq j_1 \leq 190  \}$ and $V_I^{L} = \{B[k_1][j_1], \forall k_1,j_1 \text{ s.t. } 0 \leq k_1 \leq 210 \And 0 \leq j_1 \leq 190  \} \cup \{A[i_1][k_1], \forall i_1,k_1 \text{ s.t. } 0 \leq i_1 \leq 180 \And 0 \leq k_1 \leq 210  \} \cup \{alpha\}$. 

We can compute the directed acyclic graph made of all statements (i.e., all n-ary operations), connecting all producer and consumer operations, to build the operation graph: % 
% \cite{someothergoodref}:

\begin{definition}[Operation Graph] 
\label{def:computationgraph}
Given a straight-line code region $R$ made of a list $L$ of statements $S \in L$, the operation graph OG is the directed graph $<\{N,V_I,root,V_O\},E>$ 
such that 
for every statement  $S_i$  in the list  $L$, there exists a corresponding node $N_{S_i}$ in the set of nodes $N$:
$\forall S_i \in L,~N_{S_i} \in N$;
for each statement \( S_i \) in the list \( L \), the statement produces an output \( o_{S_i} \) and takes inputs \( \vec{i}_{S_i} \). $\forall S_i, \forall i_k \in \Vec{i}_{S_i}, e_{i_k, o} \in E$, $\forall S_i:(o_{S_i},{\vec i}_{S_i}) \in L, S_j:(o_{S_j},{\vec i}_{S_j}) \in L$ with $S_i \ne S_j$ then we have $E_{S_i,S_j} \in E$ iff $o_{Si}\cap{\vec i}_{S_j}\ne \emptyset$. 
% 
% for each statement \( S_i \) in the list \( L \), the statement produces an output \( o_{S_i} \) and takes inputs \( \vec{i}_{S_i} \). This creates edges in the graph that represent the flow of data between inputs and outputs. For every input \( i_k \) of an operation \( S_i \), there exists an edge from \( i_k \) to the output \( o_{S_i} \):  
% \[
% \forall S_i \in L, \ \forall i_k \in \vec{i}_{S_i}, \ e_{i_k, o_{S_i}} \in E
% \]
% If two operations \( S_i \) and \( S_j \) exist in the list \( L \), where \( S_i \neq S_j \), an edge exists between them if the output of \( S_i \) is used as input by \( S_j \). Mathematically, this is expressed as:  
% \[
% \forall S_i, S_j \in L, \ S_i \neq S_j, \ E_{S_i, S_j} \in E \iff o_{S_i} \cap \vec{i}_{S_j} \neq \emptyset
% % \]
% This condition ensures that dependencies between operations are captured by edges in the graph. If the output \( o_{S_i} \) of operation \( S_i \) intersects with the input set \( \vec{i}_{S_j} \) of operation \( S_j \), an edge \( E_{S_i, S_j} \) is created, indicating that \( S_j \) relies on the result of \( S_i \). In essence, the graph reflects the flow of data by connecting operations whose outputs and inputs are shared, establishing a clear representation of computational dependencies.
% 
For every input (resp. output) in $S_i$ which is not matched with an output (resp. input) of another $S_j$ in $L$, create a node $V_{val} \in V_I$ (resp. $V_O$) for this input (resp. output) value. If $dim({\vec i}_{S_i}) = 0$ then an edge $e_{root,S_i}$ is added to $E$.
\end{definition}

This formulation ensures that the operation graph effectively represents the flow of data and dependencies between computational statements. By creating edges from inputs to outputs and linking operations with shared data, the graph encodes the structural relationships essential for analyzing or optimizing the execution of straight-line code regions. This representation allows for the straightforward identification of key properties, such as the span or critical path, which are crucial for estimating the latency and area of the code region.

\begin{definition}[Operation Graph critical path]
\label{def:ogcriticalpath}
Given $OG^L$ an operation graph for region $L$. Its critical path $OG_{cp}$ is the longest of all the shortest paths between every pairs $(v_i,v_o) \in <\{V_I,root\},V_O>$. 
% Its length, in number of nodes, is noted $\#OG_{cp}^L$.
Its length is noted $\#OG_{cp}^L$.
\end{definition}

% \FIXME{Link with the other operation graph}

% On the graph on the Figure \ref{fig:tree_diagram} the critical paths from $A[i]$ to 
% $c, \forall i \in$ \([\![0, 7]\!]\), 
% have the same length.

\subsubsection{Latency Lower Bound}

We can build a lower bound on the latency of an operation graph:
\begin{theorem}[Lower bound on latency of an Operation Graph]
\label{th:latlbcp}
Given infinite resources, and assuming no operation nor memory movement can take less than one cycle to complete, the latency $LAT_{cp}^L \ge \#OG_{cp}^L$ is a lower bound on the minimal feasible latency to execute $L$.
\end{theorem}

We can then build a tighter lower bound on the number of cycles a region $L$ may take to execute, under fixed resources, by simply taking the maximum between the weighted span and the work to execute normalized by the resources available.

\begin{theorem}[Latency Lower Bound under Operation Resource Constraints] 
\label{th:latlbresources}
Given $R_{op}$ a count of available resources of type $op$, for each operation type, let $LO(op)$ be the latency function for operation $op$, with $LO(op) \ge 1$. $\#L(op)$ denotes the number of operations of type $op$ in $L$. We define $LO(\#OG_{cp}^L) = \sum_{n \in cp} LO(n)$ the critical path weighted by latency of its operations.
The minimal latency of a region $L$ is bounded by 
$$
Lat_{R_{op}}^L \ge \max(LO(\#OG_{cp}^L),\max_{o \in op} (\lceil  LO(o)  \times \#L(o)/R_{o}\rceil))
$$
\end{theorem}

% \begin{proof}[Proof Th \ref{th:latlbresources}]
% Suppose $\forall o \in op,~R_{o} \ge \#L(o) $. Then there is equal or more resources available than work to execute, this is equivalent to the infinite resource hypothesis of Th.~\ref{th:latlbcp}, the minimal latency is $LO(\#OG_{cp}^L)$.

% Suppose $\exists o \in op,~R_{o} < \#L(o)  $. Then there exists at least one unit in $R_{o}$ that is executing $\lceil \#L(o) /R_{o} \rceil$ operations. As every operation $op$ take at $L(op) \ge 1$ cycle to complete, this unit will execute in at least  $\lceil \#L(o) \times L_{o}/R_{o} \rceil$ cycles. If $\lceil \#L(o) \times L_{o}/R_{o} \rceil \ge LO(\#OG_{cp}^L)$, the computation cannot execute in less than $\lceil \#L(o) \times L_{o}/R_{o} \rceil$ cycles.
% \end{proof}

% \FIXME{May here show an ILP formulation on the CDAG that achieves the lower bound, and reason that all are (over-)approximations of this perfect scheduling+placement solution?}

This theorem provides the building block to our analysis: if reasoning on a straight-line code region, without any loop, then building the operation graph for this region and reasoning on its critical path is sufficient to provide a latency lower bound. 
As a reminder, all proofs are provided in Appendix~\ref{appendix:proof}.

\begin{figure}[!htb]
\begin{lstlisting}[label={lst:reference_bicg},caption={Bicg code: $s=r \times A; q=A \times p$}]
L1: for (i = 0; i < N; i++)  
    S0: s[i] = 0;
L2: for (i = 0; i < M; i++) { 
        S1: q[i] = 0;
    L3: for (j = 0; j < N; j++) {
        S2: s[j]+=r[i]*A[i][j]; 
        S3: q[i]+=A[i][j]*p[j];
    }
}
\end{lstlisting}
\end{figure}
For example, in Listing~\ref{lst:reference_bicg}, if we consider the sub-loop body composed of loops L3 (fully unrolled) and the statements $S2$ and $S3$ as straight-line code regions, we can calculate the critical paths for $S2$ and $S3$ as follows:
For $S2$, the critical path is given by: $cp_{S2} = \max(L(+) + L(*), N \times (DSP_+ + DSP_*) / R_o)$ with $DSP_{o}$ the number of DSP for the operation $o$.
For $S3$, the critical path is determined by: $cp_{S3} = \max(L(+) \times \log(N) + L(*), N \times (DSP_+ + DSP_*) / R_o)$,  considering the possibility of a tree reduction.
In this context, the critical path for the entire sub-loop body is the maximum of these two individual critical paths, expressed as $\max(cp_{S2}, cp_{S3})$.

We now need to integrate loops and enable the composition of latency bounds.

\subsubsection{Loop Unrolling: full unroll}

% Loop unrolling is an HLS optimization that aims to execute multiple iterations of a loop in parallel. Intuitively, for an unroll factor $UF \ge 1$, $UF$ replications of the loop body will be instantiated. If $TC_l \mod UF_l \ne 0$ then an epilogue code to execute the remaining $TC_l \mod UF_l$ iterations is needed.

% We can bound the latency of a loop nest which has been entirely flatten, that is, all loops have been fully unrolled.

We start by reasoning on the bound for latency of a loop nest which has been fully unrolled, e.g. as a result of \texttt{\#pragma ACCEL parallel} or \texttt{\#pragma HLS unroll}. Full unrolling amounts to fully unroll all $TC$ iterations of a loop, replacing the loop by $TC$ replications of its original loop body, where the loop iterator has been updated with the value it takes, for each replication.

It follows a simple corollary:
\begin{corollary}[Equivalence between fully unrolled and straight-line code]
\label{cor:fullflattenstraightlinecode}
Given a loop nest $l$, if full unrolling is applied to $l$ then the code obtained after full unrolling is a straight-line code as per Def.~\ref{def:straightlinecode}.
\end{corollary}

% \begin{proof}[Proof Co \ref{cor:fullflattenstraightlinecode}]
% By construction the process of fully unrolling all the loops creates a straight-line code region without loop control, which therefore fits Def.~\ref{def:straightlinecode}.
% \end{proof}

Consequently, we can bound the latency of a fully unrolled loop nest:
\begin{theorem}[Minimal latency of a fully unrolled loop nest]
\label{th:minlatflattenednest}
Given a loop nest $l$, which is first rewritten by fully unrolling all loops to create a straight-line code region $L$. Given available resources $R_{op}$ and latencies $L(op) \ge 1$. Then its minimal latency is bounded by:
$$
Lat_{R_{op}}^l \ge \max(LO(\#OG_{cp}^L),\max_{o \in op} (\lceil L_{o} \times \#L(o)/R_{o} \rceil))
$$
\end{theorem}

% \begin{proof}[Proof Th \ref{th:minlatflattenednest}]
% By Corollary~\ref{cor:fullflattenstraightlinecode}.
% \end{proof}

\subsubsection{Loop Unrolling: partial unroll} 

Loop unrolling is an HLS optimization that aims to execute multiple iterations of a loop in parallel. Intuitively, for an unroll factor $UF \ge 1$, $UF$ replications of the loop body will be instantiated. If $TC_l \mod UF_l \ne 0$ then an epilogue code to execute the remaining $TC_l \mod UF_l$ iterations is needed.

% \FIXME{NOTE: write here some comment on what happens if you unroll non-rectangular loops, loops where UF does not divide TC, etc. from the Merlin/Vitis implementation point of view.}

% Partial unrolling can be viewed as a two-step transformation: first strip-mine the loop by the unroll factor, then consider the inner loop obtained is fully-unrolled. The latency of the resulting sub-program is determined by how the outer-loop generated will be implemented. We assume without additional explicit information this outer loop will execute in a non-pipelined, non-parallel fashion, to provide the following bound:

% Unrolling can be viewed as a two-step transformation: first, strip-mine the loop by the unroll factor, then consider the inner loop obtained to be fully-unrolled. The latency of the resulting sub-program is determined by how the outer-loop generated will be implemented. 

Unrolling can be viewed as a two-step transformation: first, strip-mine the loop by the unroll factor, then fully unroll the resulting inner loop. The latency of the resulting sub-program is influenced by how the generated outer loop is implemented.
We assume without additional explicit information this unrolled loop will execute in a non-pipelined, non-parallel fashion.
Note this bound requires to build the operation graph for the whole loop body. This is straightforward for inner loops and/or fully unrolled loop nests, but impractical if the loop body contains other loops. We therefore define a weaker, but more practical, bound \emph{that enables composition}:

\begin{theorem}[Minimal latency of a partially unrolled loop with factor UF]
\label{th:minlatunrolled1}
Given a loop $l$ with trip count $TC_l$ and loop body $L$, and unroll factor $UF \le TC$. Given available resources $R_{op}$ and latencies $L(op) \ge 1$. Given $L^{'}$ the loop body obtained by replicating $UF$ times the original loop body $L$. Then the minimal latency of $l$ if executed in a non-pipelined fashion is bounded by:
$$
Lat_{R_{op}}^{l,S} \ge \lfloor TC/UF \rfloor \times Lat_{R_{op}}^{L^{'}}
$$    
\end{theorem}

Note this bound requires to build the operation graph for the whole loop body. This is straightforward for inner loops and/or fully unrolled loop nests, but impractical if the loop body contains other loops. We therefore define a weaker, but more practical, bound:

%\FIXME{Need to precise that we use DSA. And cite Feautrier papier for transformation Affine -> DSA}

\begin{theorem}[Minimal latency of a partially unrolled loop with factor UF and complex loop bodies]
\label{th:minlatunrolled2}
Given a loop $l$ with trip count $TC_l$ and loop body $L$, and unroll factor $UF \le TC$. Given available resources $R_{op}$ and latencies $L(op) \ge 1$. Then the minimal latency of $l$ if executed in a non-pipelined fashion is bounded by:
$$
Lat_{R_{op}}^{l,S} \ge \lfloor TC/UF \rfloor * Lat_{R_{op}}^L
$$    
\end{theorem}

Consider the scenario of loop $L0$ within Listing \ref{lst:reference_bicg}, which has been unrolled by a factor denoted as $UF \leq TC_{L0}$ where $TC_{L0}$ is the trip count of $L0$. The latency for one iteration of $S0$ is denoted as $Lat^{S0}_{R_{o}} > 0$.
In the absence of pipelining, the lower bound of the latency of this sub-loop body is: $\lfloor TC_{L0} / UF \rfloor \times Lat^{S0}_{R_{o}}$.

% \FIXME{The above theorem covers the composition case of C operator, with children made of loops with a bound and statements. Perhaps worth rewriting the theorem (or a corrollary) to show this composition of children is valid? That would conclude the formulas with the pipelining below.}

% % Vitis allows to do a reduction with a tree reduction in logarithmic time with the option ``unsafe-math''. 
% The Vitis compiler, part of the Vitis development environment for FPGA design, offers an “unsafe-math” option that allows reductions to be performed using a tree-based reduction algorithm. This method enables the reduction operation to be executed in logarithmic time, improving performance in certain computational tasks by leveraging this optimization.

The Vitis HLS compiler enables reductions to be performed using a tree-based reduction algorithm, achieving logarithmic time complexity. This is possible by enabling the 
\textit{unsafe\_math\_optimizations} 
option, as described in the Xilinx documentation.

% \textcolor{red}{TODO: Should we decompose red and // operation ?}

\begin{theorem}[Minimal latency of a partially unrolled loop with factor UF for reduction loop with tree reduction]
\label{th:minlatunrolled3}
Given a reduction loop $l$ with trip count $TC_l$ and loop body $L$, and unroll factor $UF \le TC$. Given available resources $R_{op}$ and latencies $L(op) \ge 1$. Then the minimal latency of $l$, if executed in a non-pipelined fashion and the tree reduction is legal is bounded by:
$$
Lat_{R_{op}}^{l,S} \ge \lfloor TC/UF \rfloor \times Lat_{R_{op}}^L \times \lfloor \log_2(UF)  \rfloor
$$    
\end{theorem}

\subsubsection{Loop pipelining} 

% Loop pipelining amounts to overlapping multiple iterations of the loop, so that the next iteration can start prior to the completion of the preceding one. The initiation interval (II) measures in cycles the delay between the start of two consecutive iterations. 

Loop pipelining amounts to overlapping multiple iterations of the loop, so that the next iteration can start prior to the completion of the preceding one. The initiation interval (II) measures in cycles the delay between the start of two consecutive iterations. It is easy to prove our formula template accurately integrates the latency of pipelined loops with the $I$ operator.
We compute the minimal II in function of the dependencies of the pipelined loop and the iteration latency of the operations of the statements during the NLP generation. 
Let $RecMII$ and $ResMII$ be the recurrence constraints and the resource constraints of the pipelined loop, respectively. We have $II \geq \max{(ResMII, RecMII)}$. 
$RecMII = \max_i \lceil \frac{ delay(c_i)}{ distance(c_i)} \rceil$ with $delay(c_i)$ the total latency in dependency cycle $c_i$ and $distance(c_i)$ the total distance in dependency cycle $c_i$.
% As we cannot compute accurately $ResMII$ as we don't know how the resource will be used by the compiler 
We suppose that $ResMII=1$, as we do not know how the resource will be used by the compiler.
Hence, if the loop is a reduction loop then the $II \geq \frac{IL_{reduction}}{1}$ with $IL_{reduction}$ the iteration latency of the operation of reduction. For a kernel like the Listing \ref{lst:ex_II}
% $for (j = 0; j < N; j++) y[j] = y[j-2] + 3;$ 
the $II \geq \lceil \frac{IL_+}{2} \rceil$.

All dependencies are computed using PolyOpt, and the highest possible value for II is selected based on the dependencies, as defined by the formula.

\begin{lstlisting}[label={lst:ex_II},caption={Demonstration of a code snippet showcasing a scenario where a loop pipelined with a dependency of distance 2 results in an initiation interval ($II$) that satisfies $II \geq \lceil \frac{IL_+}{2} \rceil$. }]
for (j = 0; j < N; j++) 
    y[j] = y[j-2] + 3;
\end{lstlisting}
\vspace{1cm}

It follows a bound on the minimal latency of a pipelined loop:
\begin{theorem}[Minimal latency of a pipelined loop with known II]
\label{th:minlatpipeline}
Given a loop $l$ with trip count $TC_l$ and loop body $L$. Given available resources $R_{op}$ and latencies $L(op) \ge 1$. Then the minimal latency of $l$ if executed in a pipelined fashion is bounded by:
$$
Lat_{R_{op}}^{l,P} \ge Lat_{R_{op}}^L + II *( TC_l - 1)
$$    
\end{theorem}

% \begin{proof}[Proof Th \ref{th:minlatpipeline}]
% $Lat_{R_{op}}^L$ is the minimal latency to complete one iteration of $l$ by Theorem~\ref{th:latlbresources}. The initiation interval measures the number of elapsed cycles before the next iteration can start, it takes therefore at least $TC_l * II - 1$ to start $TC_l - 1$ iterations, irrespective of their completion time. Therefore the latency of the loop is at least the latency of one iteration to complete, and for all iterations to be started.
% \end{proof}

\subsubsection{Loop pipelining and unrolling}

A loop $l$ with trip count $TC_l$ can be pipelined and partially unrolled with $UF < TC_l$, in this case there is loop splitting where the trip count of the innermost loop equal to the unroll factor and the trip count of the outermost loop equal to $\frac{TC_l}{UF}$.

\begin{theorem}[Minimal latency of a pipelined loop with known II and partially unrolled]
\label{th:minlatpipeline_unrolled}
Given a loop $l$ with trip count $TC_l$, partially unrolled by an unroll factor $UF < TC_l$ and a loop body $L$. Given available resources $R_{op}$ and latencies $L(op) \ge 1$. Given $L^{'}$ the loop body obtained by replicating UF times the original loop body $L$.
Then the minimal latency of $l$ if executed in a pipelined fashion is bounded by:

$$
Lat_{R_{op}}^{l,P} \ge Lat_{R_{op}}^{L^{'}} 
+ II * ( \frac{TC_l}{UF} - 1)
$$    
\end{theorem}

% \begin{proof}[Proof Th \ref{th:minlatpipeline_unrolled}]

% By construction and Theorem \ref{th:minlatflattenednest} the latency $Lat_{R_{op}}^{L^{'}}$ is a lower bound of $L^{'}$. As the loop was split due to the partial unrolling, the trip count of the pipelined loop is $\frac{TC_l}{UF}$. Theorem \ref{th:minlatpipeline} gives us the lower bound of the latency for a loop with a trip count equal to $\frac{TC_l}{UF}$.

% \end{proof}

\subsubsection{Non-Parallel, Non-Pipelined Loops}

We continue with a trivial case: if the loop is not optimized by any directive (including any automatically inserted by the compilers), i.e., not parallelized nor pipelined, then every next iteration of the loop starts only after the end of the prior iteration.

\begin{definition}[Lower bound on latency of a non-parallel, non-pipelined loop under resources constraints]
\label{def:loopIteratedSequentially}
Given a loop $l$ with trip count $TC_l$ which is neither pipelined nor parallelized, that is, iteration $i+1$ starts after the full completion of iteration $i$, for all iterations. Given $Lat_{R_{op}}^L$ the minimal latency of its loop body. Then
$$
Lat_{R_{op}}^l \ge TC_l * Lat_{R_{op}}^L
$$

\end{definition}

\subsubsection{Coarse-Grained parallelization}

Coarse-grained parallelization is a performance enhancement technique involving the unrolling of a loop which iterates a loop body not fully unrolled i.e., containing at least a pipelined loop or a loop executed sequentially. It is therefore impossible to do a coarse-grained parallelization with a reduction loop because the $n$ sub loop body are dependent on each other.

It follows a bound on the minimal latency of a coarse-grained unrolled loop:
\begin{theorem}[Minimal latency of coarse-grained unrolled loop]
\label{th:coarse_grained}

Given a loop $l$, which is not a reduction loop, with trip count $TC_l$, an unroll factor $UF \le TC_l$ and  $L$ the loop body iterated by the loop $l$ with a latency lower bound $Lat^{L}_{R_{op}}$. 
Given available resources $R_{op}$ and latencies $L(op) \ge 1$. Given $L^{'}$ the loop body obtained by replicating $UF$ times the original loop body $L$. Then the minimal latency of $l$ if executed in a non-pipelined fashion is bounded by:
$$
Lat_{R_{op}}^{l,S} \ge \lfloor TC/UF \rfloor \times Lat_{R_{op}}^{L^{'}}
$$
\end{theorem}

\subsubsection{Program latency lower bound under resource constraints} 

We now focus on the latency lower bound of a program, under resource constraints. This bound takes into account the limitations imposed by available resources, which can significantly affect the achievable performance.
We assume here that the resources consumed are only consumed by the computing units and resource use by the computational unit of one operation can not be reused by the computational unit of another operation executing at the same time. We also assume that the compilers have implemented the pragma configuration given as input.

For DSPs we suppose we have a perfect reuse i.e., that the computation units for the same operation can be reused as soon as the computation unit is not in use. Under-estimating the resources used is fundamental to proving the latency lower bound, as otherwise another design that consumes less resources than predicted may be feasible, itself possibly leading to a better latency.

\begin{theorem}
\label{th:minres}
    Given a loop body $L$,  the set of set of  statements $\mathcal{S}_{seq}$ non executed in parallel, 
$\#L_{op}^s$ the number of operations $op$  for the statements $s$, $DSP_{op}$ the number of resources (DSPs) used for the operation $op$, $MCU_{op}^s$ the maximal number of computational units the statement $s$ can use in parallel at any given time,  and  the configuration of pragma $\vec{PV_i}$ for each loop.
The minimal number of resource (DSPs) consumed, $R_{used}^{min}$, by $L$ for the pragma configuration  is the sum, for each operation, of the maximum number of DSPs used in parallel by a statement.
This corresponds to:
$$R^{min}_{used} =  \sum_{op} \max_{\mathcal{S} \in \mathcal{S}_{seq}}( \sum_{s \in \mathcal{S}}
\#L_{op}^s
\times DSP_{op} 
\times MCU_{op}^{s})$$
\end{theorem}

% \begin{proof}[Proof Th \ref{th:minres}]
%     Considering perfect resource reuse, where all unused computational units can be reused, and assuming that the compilers have implemented the pragma configuration provided as input. 
% For each statement $s$, the maximum number of computational units used in parallel is determined. This means that each statement $s$ requires at least $\#L_{op}^s \times DSP_{op} \times MCU_{op}^s$ DSPs. 
% If a set of statements $\mathcal{S}$ are executed in parallel they cannot share the resource so the execution in parallel of the statement $s \in \mathcal{S}$ will require $( \sum_{s \in \mathcal{S}}
% \#L_{op}^s
% \times DSP_{op} 
% \times MCU_{op}^{s})$ DSPs.
% % 
% By considering the maximum across all statements, we can guarantee that at least one set of statement executed in parallel will require $\max_{s \in \mathcal{S}_{seq}}(\sum_{s \in \mathcal{S}} \#L_{op} \times DSP_{op} \times MCU_{op}^s)$ DSPs.
% Since there is no possibility of resource reuse between different operations, the summation of the resource consumed for each operation remains the minimum consumption of resources. In other words, the sum of the individual resource consumption for each operation represents the minimum amount of resources required.
% \end{proof}

Given a program and the available resource of DSP $DSP_{avail}$, if $R^{min}_{used} < DSP_{avail}$ the lower bound is valid and the program does not over-utilize the resources. 
In practice, \( MCU_{op}^s \) is determined by the unroll factor, which specifies how many times computational units need to be duplicated. However, if the initiation interval (II) of the pipelined loop is strictly greater than one, resource reuse becomes possible within the same statement, reducing the overall number of computational units required.

%%% LNP: I cannot parse this sentence.
%The latency lower bound of a kernel is therefore the latency lower bound of the configuration such that the minimal resource consumption is less than or equal to the resource available.

% \begin{theorem}[Latency Lower bound of a program]
% \label{th:minlatpipelineunderresourceconstraint}
% % Given a kernel, available resource $R_{op}$, the set of statements $\mathcal{S}$ and all the configuration of pragmas which are constitute of  the vectors $\Vec{PV_i}$ for each loop.
% Given a loop body $L$,  the set of statements $\mathcal{S}$,
% $\#L_{op}^s$ the number of operation $op$  for the statements $s$, $DSP_{op}$ the number of resources (DSPs) used for the operation $op$, $MCU_{op}^{s}$ the maximal number of computational unit the statement $s$ can use in parallel, and  the configuration of pragma  $\Vec{PV_i}$ for each loop.

% The configuration of pragma, which has the minimal lower bound and which respect $R_{used}^{min} = \sum_{op} \max_{s \in \mathcal{S}}(
% \#L_{op}^s
% \times DSP_{op} 
% \times MCU_{op}^{s}) \leq R_{op}$, is the lower bound of the kernel.

% \end{theorem}

% \begin{proof}[Proof Th \ref{th:minlatpipelineunderresourceconstraint}]

%     According to Theorem \ref{th:minres}, $R_{used}^{min}$ is the minimal number of DSP a configuration can use. Hence, all configuration which does not respect the constraint $R_{used}^{min}\leq R_{op}$ have a over-utilization of the resource and hence are invalid.

%     Hence, the minimal lower bound of each configuration of pragma is the lower bound of the kernel.
    
% \end{proof}

%   

\subsubsection{Memory transfer}

AMD/Xilinx Merlin manages automatically the memory transfer. The memory transfer and computation are not overlap (no dataflow) hence the latency is the sum of the latency of computation and communication. We assume that for each array the contents of the array are in the same DRAM bank.

\begin{theorem}[Lower bound of the memory transfer latency for an array]
\label{th:memlatonearray}

Given a loop body $L$,  the set of array $\mathcal{A}$,  an array $a \in \mathcal{A}$, and $Lat^{mem}_a$ the latency to transfer the array $a$ from off-chip to on-chip (inputs) and from on-chip to off-chip (outputs).
% $\forall a \in \mathcal{A}, LAT^{mem}_a \geq 
% (\mathds{1}_{a \in V_O^L} + \mathds{1}_{a \in V_I^L}) \times footprint_a / max\_burst\_size$.
\begin{equation*}
\left.
\begin{cases}
 \forall a \in \mathcal{A}, Lat^{mem}_a \geq 
 \frac{footprint_{a}}{max\_burst\_size}, & \text{ if $a$ is only read or only write i.e. } (a \in V_I^L \text{ and } a \notin V_O^L) \\
 & \text{ or } (a \notin V_I^L \text{ and } a \in V_O^L) \\
 \forall a \in \mathcal{A}, Lat^{mem}_a \geq 
2 \times \frac{footprint_{a}}{max\_burst\_size}, & \text{ if $a$ is read and write i.e. } a \in V_I^L \text{ and } a \in V_O^L \\
\end{cases}
\right.
\end{equation*}

\end{theorem}

\begin{theorem}[Lower bound of the memory transfer latency]
\label{th:memlat}

Given a loop body \( L \), the set of arrays \( \mathcal{A} \), the set of loops \( \mathcal{L} \), 
the booleans \( loop\_l\_cache\_array\_a \) to indicate if the array \( a \) is transferred under the loop \( l \),
and \( footprint_{a} \), the footprint of the array \( a \),
the latency for communication \( Lat^L_{communication} \) 
% assuming no reuse buffers are used, 
is bounded by:
% $\max_{a \in \mathcal{A}} (\mathds{1}_{a \in V_O^L} + \mathds{1}_{a \in V_I^L}) \times footprint_a / max\_burst\_size$.
\[Lat^L_{communication} \geq \sum_{l\in \mathcal{L}} \max_{a \in \mathcal{A}} (loop_l\_cache\_array\_a \times \frac{footprint_{a}}{max\_burst\_size} ) \]

\end{theorem}

\subsection{Summary}

By composing all the theorems, this allows us to end up with the final latency lower bound of the program which is presented in theorems \ref{th:comp_lb} for the computation and \ref{th:global_lb} for the computation and communication.

\begin{theorem}[Computation Latency Lower bound of a Program]
\label{th:comp_lb}

Given available resource $DSP_{avail}$, the properties vector $\Vec{PV}_i$ for each loop and a program which contains a loop body $L$.
The properties vector allows to give all the information concerning the trip counts and the II of the pipelined loops and to decompose the loop body $L$ with a set of loops $\mathcal{L}^{non \; reduction}_{L}$ potentially coarse-grained unrolled with $\forall l \in \mathcal{L}_{L}, UF_l$ and a set of reduction loops executed sequentially $\mathcal{L}^{reduction}_{L}$ which iterates a loop body $L_{pip}$. 
By recursion the loop body $L_{pip}$ contains a pipelined loop $l_{pip}$ which iterate a loop body $L_{fg}$ fully unrolled. The loop body $L_{fg}$ contains operations which can be done in parallel with a latency $Lat^{L_{par}}_{R_{op}}$ 
and operations which are reduction originally iterated by the loops $\mathcal{L}^{reduction}_{L_{fg}}$ with a latency $Lat_{L_{seq}}$. 

The computation latency lower bound of $L$, which respected $DSP^{min}_{ued} \leq DSP_{avail}$, executed with tree reduction is:

$$
Lat^{L}_{R_{op}} \geq \prod_{l\in \mathcal{L}^{par}_{L}} \frac{TC_l}{UF_l} \times 
\prod_{l\in \mathcal{L}^{reduction}_{L}} TC_l \times Lat^{L_{pip}}_{R_{op}}
$$

with 

$Lat^{L_{pip}}_{R_{op}} = (Lat^{L_{fg}}_{R_{op}} + II\times (\frac{TC_{l_{pip}}}{l_{pip}\_UF}-1) )$
and 
$Lat^{L_{fg}}_{R_{op}} = Lat_{L_{par}} +
Lat_{L_{seq}} \times \prod_{l\in \mathcal{L}^{reduction}_{L_{fg}}} \frac{TC_l}{UF_l} \times \log_2(UF_l)   $.

\end{theorem}

% \begin{proof}[Proof Th \ref{th:comp_lb}]

% % By composition and Theorems \ref{th:minlatunrolled1} and \ref{th:minlatunrolled2}, $Lat_{L_{fg}}$ is a lower bound of the fully unrolled sub-loop body of $L$, $L_{fg}$.

% % By composition and Theorems \ref{th:minlatpipeline} and \ref{th:minlatpipeline_unrolled}, $Lat_{L_{pip}}$ is a lower bound of $L_{pip}$.

% % By composition, Theorem \ref{th:coarse_grained} and Definition \ref{def:loopIteratedSequentially}, $Lat_{L}$ is a lower bound of $L$.

% Through composition and the application of Theorems \ref{th:minlatflattenednest}, \ref{th:minlatunrolled1} and  \ref{th:minlatunrolled2}, $Lat^{L_{fg}}_{R_{op}}$ serves as a computation latency lower bound for the fully unrolled sub-loop body of $L$, denoted as $L_{fg}$, where $Lat_{L_{par}} +
% Lat_{L_{seq}} \times \prod_{l\in \mathcal{L}^{reduction}_{L_{fg}}} \frac{TC_l}{UF_l} \times \log_2(UF_l)$ represent the critical path of $Lat^{L_{fg}}_{R_{op}}$.

% By employing composition alongside Theorems \ref{th:minlatpipeline} and \ref{th:minlatpipeline_unrolled}, $Lat^{L_{pip}}_{R_{op}}$ stands as a computation latency lower bound for $L_{pip}$.

% Utilizing composition, Theorem \ref{th:coarse_grained}, and Definition \ref{def:loopIteratedSequentially}, $Lat^{L}_{R_{op}}$ emerges as a computation latency lower bound for $L$.

% \end{proof}

\begin{theorem}[Latency lower bound of a program optimized with Merlin pragmas]
\label{th:global_lb}

Given available resource $DSP_{avail}$ and a program which contains a loop body $L$ with a computation latency $Lat^{L}_{computation}$ and a communication latency $Lat^{L}_{communication}$.

The lower bound for $L$ which respected $DSP^{min}_{ued} \leq DSP_{avail}$ and where the computation and communication can not be overlap is:

$$
Lat_{L} = Lat^{L}_{computation} + Lat^L_{communication}
$$

\end{theorem}

% \begin{proof}[Proof Th \ref{th:global_lb}]
% The AMD/Xilinx Merlin compiler does not overlap computation and communication. Hence the computation and communication is a sum of the latency of computation and latency of communication.
% By composition and Theorems \ref{th:comp_lb} and \ref{th:memlat}.

% \end{proof}

%% file: sources/implementation.tex
 \label{sec:implementation}

We now present our Design Space Exploration (DSE) approach. 
Our approach focuses on identifying designs with the most promising theoretical latency within the available design space. However, it may result in suboptimal designs if the selected pragmas are not applied during compilation. To address this potential issue and ensure high QoR, we conduct an additional exploration within a restricted subspace. 
Our DSE explores two additional parameters: the type of parallelism and the maximum array partitioning factor. Array partitioning is a technique commonly used in FPGA contexts to divide arrays or matrices into smaller sub-arrays, which can be stored in independent memory blocks known as Block RAMs (BRAMs). AMD/Xilinx HLS has a limit of 1,024 partitions per array. 
The array partitioning is calculated by taking the product of loops that iterate the same arrays on different dimensions (cf. Section \ref{sec:formulation}). So constraining the maximal array partitioning also constrains the maximal UF.
This NLP based DSE technique is presented in Algorithm \ref{alg:implementation}. 
The DSE starts without constraint on parallelism and array partitioning. Then we alternate constraints on parallelism while decreasing the maximum unrolling factor and array partitioning.

% \vspace{-0.36cm}
\begin{algorithm}[h]
 
\caption{\framework}\label{alg:implementation}
\KwData{$kernel$ \tcp*[f]{without Pragma}}
\KwData{$Space\_Array\_Partitioning$ \tcp*[f]{e.g., $\{\infty, 2048, 1024, 512, 256, 128, 64, 32, 16, 8,  1\}$}}
% \KwData{$nb\_workers$ \tcp{Number of threads}} 
\KwData{$timeout\_HLS$,  $timeout\_NLP$} 
\KwResult{$kernel$ \tcp*[f]{with Merlin Pragma}}

$nlp\_file \gets generate\_nlp\_file(kernel)$ , $min\_lat \gets \infty$\;
\For{$max\_array\_partitioning \in Space\_Array\_Partitioning$}{
    \For{$parallelism \in \{coarse+fine, fine\}$}{
        $current\_nlp\_file \gets $ change\_max\_array\_partitioning($copy(nlp\_file)$, $max\_array\_partitioning$)\;
        \If{$parallelism == fine$}{
            add\_constraint\_only\_fine\_grained\_parallelism( $current\_nlp\_file$)\;
        }
        $pragma\_configuration, lower\_bound \gets SOLVER(current\_nlp\_file, timeout\_NLP)$ \;
        \If{$lower\_bound < min\_lat$}{
            % $current\_kernel \gets copy(kernel)$\;
            $current\_kernel \gets$ introduce\_pragma($copy(kernel)$, $pragma\_configuration$)\;
            $hls\_lat, valid \gets MERLIN(kernel, timeout\_HLS)$ \;
            \If(\tcp*[f]{no over-utilization}){$valid$} { 
            
                $min\_lat \gets min(min\_lat, hls\_lat)$\;
            }
        }
    }
}
% % \vspace{-\baselineskip}
\end{algorithm}

% \vspace{-0.4cm}

% 

In order to reduce the maximum unroll factor and array partitioning, we modify the parameters specified in the NLP file. And we automatically add constraints (Eq \ref{eq:fine_grained}) to restrict parallelism to fine-grained levels, as described in Section \ref{sec:formulation}.
The choice to restrict the maximum array partitioning to the power of 2 
is to improve the speed of the DSE. Adding possibilities would permit exploring a larger space and potentially finding a design with a faster latency at the cost of a longer DSE.

% \begin{figure}[H]
%     \centering
%     \includegraphics[width=\textwidth]{figures/figure_NLP_DSE.018.jpeg}
%     \caption{NLP-DSE}
%     \label{fig:enter-label}
% \end{figure}

%% file: sources/evaluation.tex
\label{sec:evaluation}
% \REMARK{Section 6 states that array partitioning is restricted to power of 2 values to reduce the design space. Is the same done in AutoDSE to ensure a fair comparison?}
% \FIXME{Explain that AutoDSE by using Merlin does not restrict the array partitioning and explain why it is not a problem \textcolor{red}{DONE?}}
% \FIXME{Complete the evaluation with CNN and LLM}

We now present our experimental results using a set of polyhedral computation kernels. 
% We first discuss the experiment setup and evaluated benchmarks before experimental results are discussed.
%We first discuss the experiment setup and evaluated benchmarks before examining the experimental results.

\subsection{Setup}

% FIXME REMOVE SPACE: Delete two subtitles ?

% \subsubsection{Benchmark description \\}

We use kernels from Polybench/C 4.2.1 \cite{polybench-web}. 
The complexities and sizes of the problems are detailed in Table \ref{tab:polybench} located in the Appendix.
In addition, we add a layer of a Convolutional Neural Network (CNN) to demonstrate that our method works effectively on different types of kernels.
 A single-precision floating point is utilized as the default data type in computations to compare to AutoDSE \cite{autodse} and ScaleHLS \cite{scalehls}. Computations operate on medium and large datasets from PolyBench/C \cite{polybench-web} in order to  have kernels with large footprint and have a large enough space to 
 explore. 
Selecting medium and large problem sizes is crucial to accurately reflect the complexities encountered in various fields such as scientific simulations, data analytics, and artificial intelligence. These sizes pose challenges that mirror the practical limitations of memory transfer, where efficiently managing data movement becomes paramount due to its potential to bottleneck performance. In such contexts, the careful orchestration of memory transfers is essential to ensure optimal resource utilization and prevent computational inefficiencies. Additionally, the scale of these problems often exceeds on-chip memory capacity, necessitating strategies for effectively handling data footprints that surpass available memory, further emphasizing the need for meticulous memory management and optimization techniques.
 The problem size and loop order of CNN are J,I=256, P,Q=5 H,W=224.
 A description of each benchmark can be found in Tables \ref{tab:polybench} and \ref{tab:eval}. The \textit{ludcmp}, \textit{deriche} and \textit{nusinnov} kernels are not present as PolyOpt-HLS \cite{pouchet:fpga13} does not handle negative loop stride. \textit{Cholesky} and \textit{correlation} contains a \texttt{sqrt()} operation which we do not support currently. Finally, we removed \textit{FDTD-2D} because it exposed a bug in Merlin, and this generated a program where data dependencies are not fully preserved.
 
 A double-precision floating point is utilized as the default data type in computations to compare to HARP. We chose to use the problem size used by HARP (small and medium) in order to reuse their model. 
 The problem size and kernel use by HARP can be found on the Table \ref{tab:harp}.

% \subsubsection{Circuit Generation and NLP Solver \\}

 % CFLAGS="-I $XILINX_HLS/include" merlincc --attribute burst_total_size_threshold=36700160 --attribute burst_single_size_threshold=36700160 --kernel_frequency 250  -funsafe-math-optimizations --platform=vitis::/opt/xilinx/platforms/xilinx_u200_xdma_201830_2/xilinx_u200_xdma_201830_2.xpfm  -I $XILINX_HLS/lnx64/tools/gcc/lib/gcc/x86_64-unknown-linux-gnu/4.6.3/include/ -I $XILINX_HLS/include/ -I /opt/merlin/sources/merlin-compiler/trunk/source-opt/include/apint_include/ -c -o mykernel_merlincc_polyopt --report=estimate --attribute polyopt=on --keep_prj $@

We evaluate designs with AMD/Xilinx Merlin \cite{merlin}.
The synthesis is carried out with AMD/Xilinx Vitis 2021.1. We choose the option "-funsafe-math-optimizations" to enable commutative/associative reduction operators and implementation of reductions in logarithmic time. 
We change the default on-chip memory size of Merlin by the size of the device we use.
% with the options: "burst\_total\_size\_threshold" and "burst\_single\_size\_threshold".
As the target hardware platform, we run the Xilinx Alveo U200 device where the target frequency is 250 MHz.

% We use 2 Intel(R) Xeon(R) CPU E5-2680 v4 @ 2.40GHz for our experiments.

% \subsubsection{NLP Solver \\}

% perte de place totale ici..
We analyze  the kernels and automatically generate  each NLP problem with a version of PolyOpt-HLS \cite{pouchet:fpga13}. We modified and extended for our work. 
Employing the AMPL description language to solve the NLP problems, we ran the commercial BARON solver \cite{baron1, baron2} version 21.1.13.
% We use the commercial BARON solver \cite{baron1, baron2}, version 21.1.13, using the AMPL description language 
% % \cite{ampl} 
% to solve the NLP problems.
% should we cite AMPL ? no one cite AMPL
%
For our experiments, we utilize 2 Intel(R) Xeon(R) CPU E5-2680 v4 @ 2.40GHz and 252GB DDR4 memory.

\subsection{Experimental Evaluation}

\subsubsection{AutoDSE}
% Change position of this
We compare our method with AutoDSE \cite{autodse}, described in Section \ref{sec:motivation}, and
we automatically generate  the space of AutoDSE with the command \textit{ds\_generator}. We replace the UF and tile size by all the UF and tile size which divide the TC in order to have the same space. 
AutoDSE does not impose any constraints on parallelism or the maximum array partitioning. It employs an incremental exploration approach, enabling it to make compiler-specific pragma selections.

Table \ref{tab:eval} displays the space size of each design.
 The DSE is done in 4 parts with two threads for each (default parameter), with a timeout for the generation of the HLS report of 180 minutes, and a timeout of the DSE of 600 minutes (not always respected cf. Table \ref{tab:eval}). 
For our method we take the same parameters,
and we add a timeout for BARON of 30 minutes. The space given as input to \framework is $\{\infty, 2048, 1024, 512, 256, 128, 64, 32, 16, 8,  1\}$.

\subsubsection{HARP}

The evaluation vs. HARP is done with the same parameters as the evaluation vs. AutoDSE. We change the space given as input due to the small problem size and we choose $\{\infty, 1024, 750, 512, 256, 128, 64, 32, 16, 8,  1\}$.

We run HARP for one hour in order to have a similar DSE time as \framework. 
This enables the exploration of an average of 75,000 distinct pragma configurations for each kernel.
HARP's DSE method navigates the space by iteratively adjusting the pragma in a bottom-up manner. It synthesizes the top 10 designs discovered by the DSE, employing a timeout of 3 hours for the HLS compiler, similar to the approach used in the \framework framework.

% The DSE of HARP explore the space by incrementally changing the pragma in a bottom-up way and synthesize the top-20 designs found by the DSE with a timeout for the HLS compiler of 3h similarly to \framework

\subsection{Comparison with AutoDSE}

\label{sec:evaluation_autodse}

\begin{figure}[!htb]
% \begin{figure}[]
    % \vspace{-.15cm}
    \centering
  \subfloat[\label{fig:th_large} GF/s]{%
       \includegraphics[width=\textwidth]{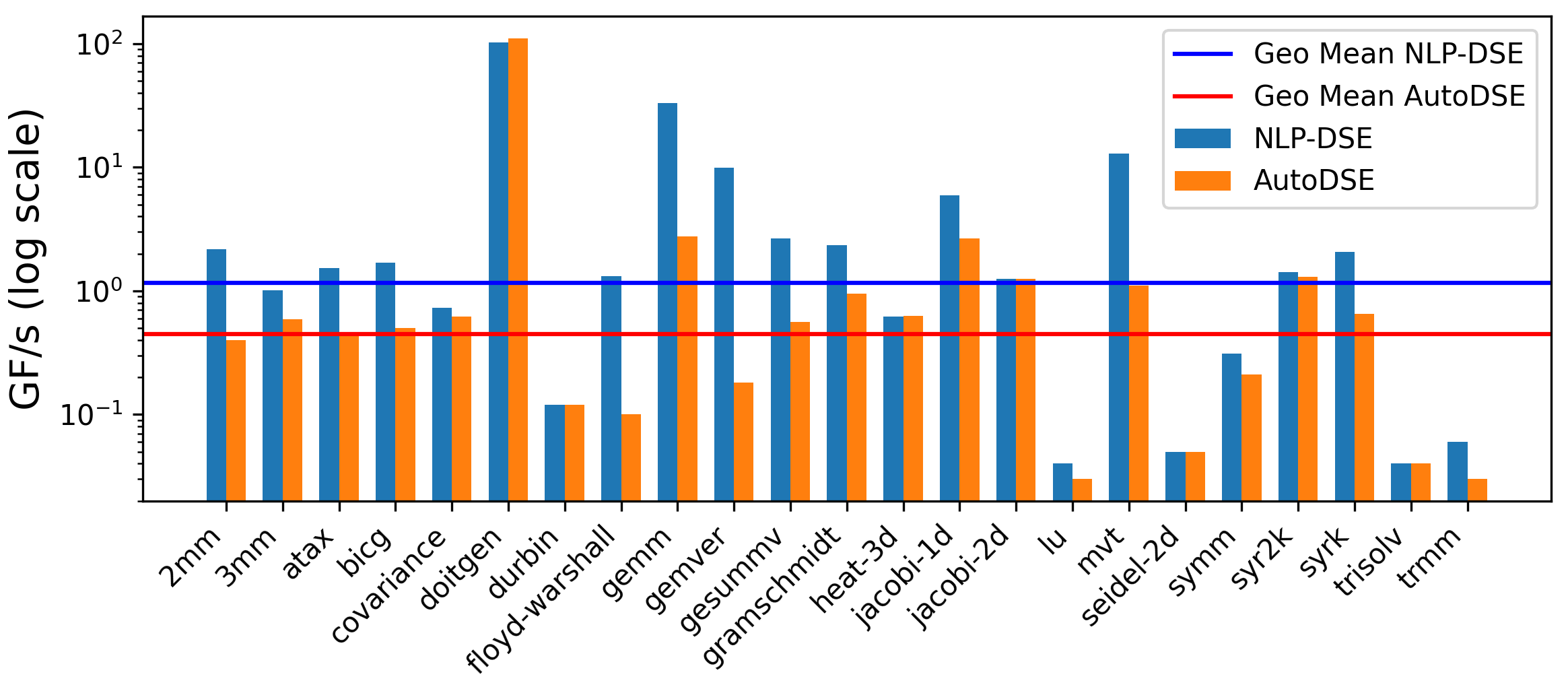}}
    \hfill
\vspace{-0.45cm}
  \centering
    \subfloat[\label{fig:time_large} DSE Time (min)]{%
        \includegraphics[width=\textwidth]{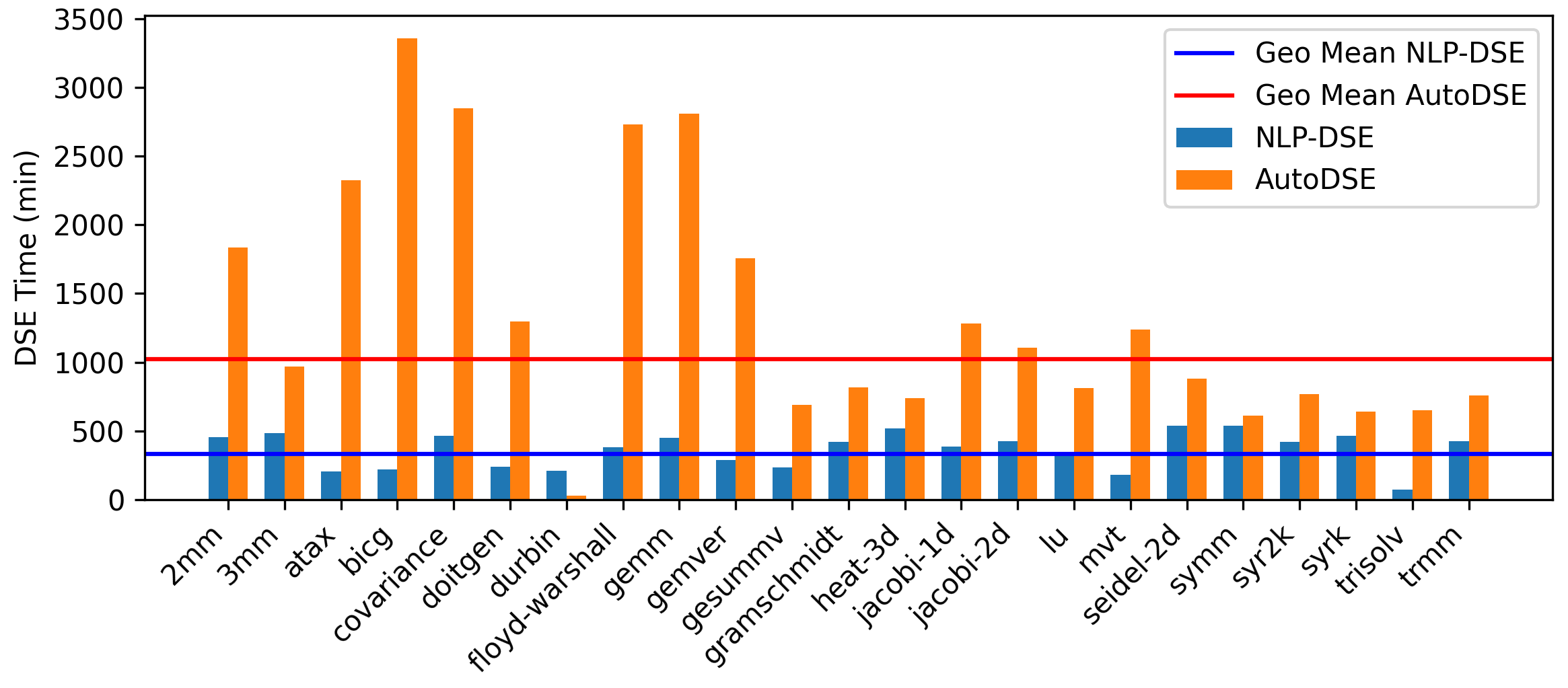}}
% \vspace{-.35cm}
  \caption{Comparison between the throughput (GF/s) and Design Space Exploration (DSE) time (min) of NLP-DSE and AutoDSE for large problem sizes in Polybench.}
  \label{fig:large} 
\end{figure}

\begin{figure}[!htb]
% \begin{figure}[]
    % \vspace{-.15cm}
    \centering
  \subfloat[\label{fig:th_medium} GF/s]{%
       \includegraphics[width=\textwidth]{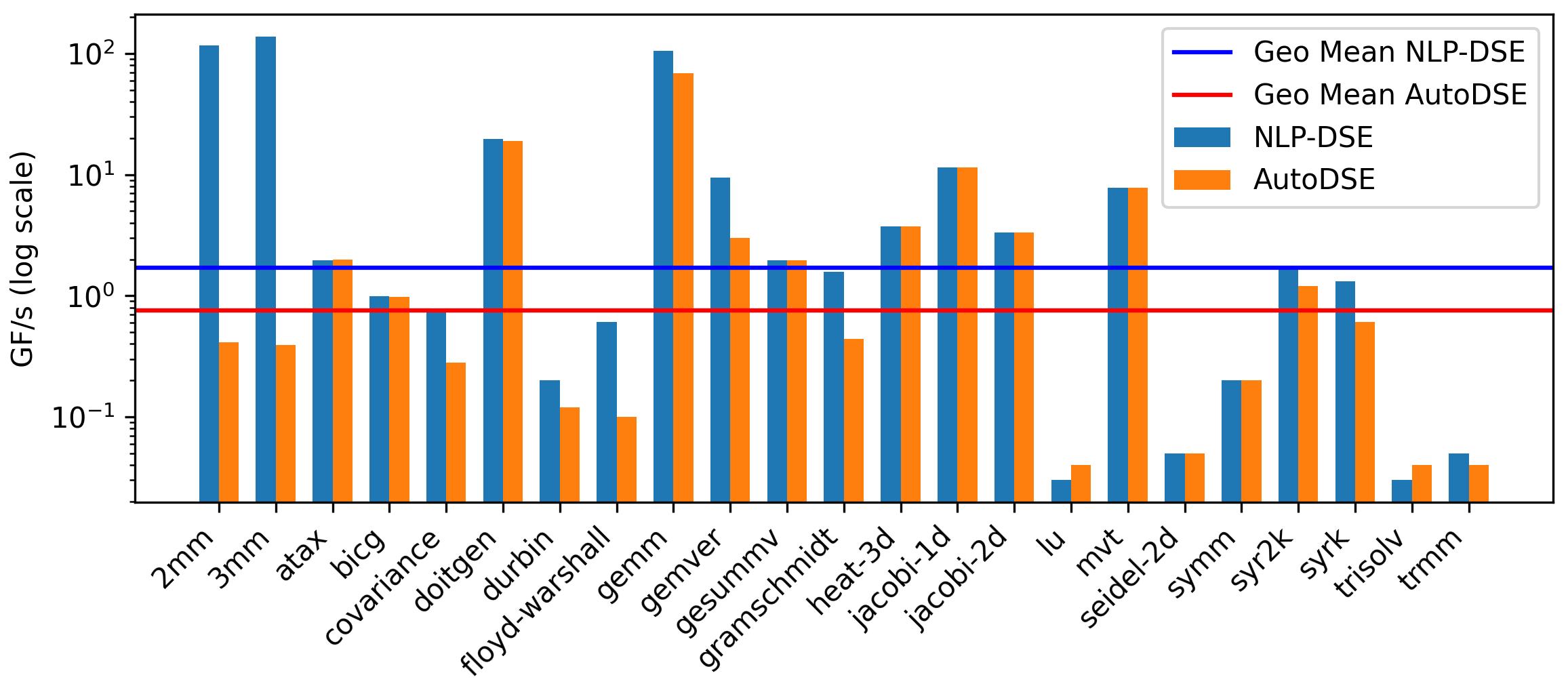}}
    \hfill
\vspace{-0.45cm}
  \centering
    \subfloat[\label{fig:time_medium} DSE Time (min)]{%
        \includegraphics[width=\textwidth]{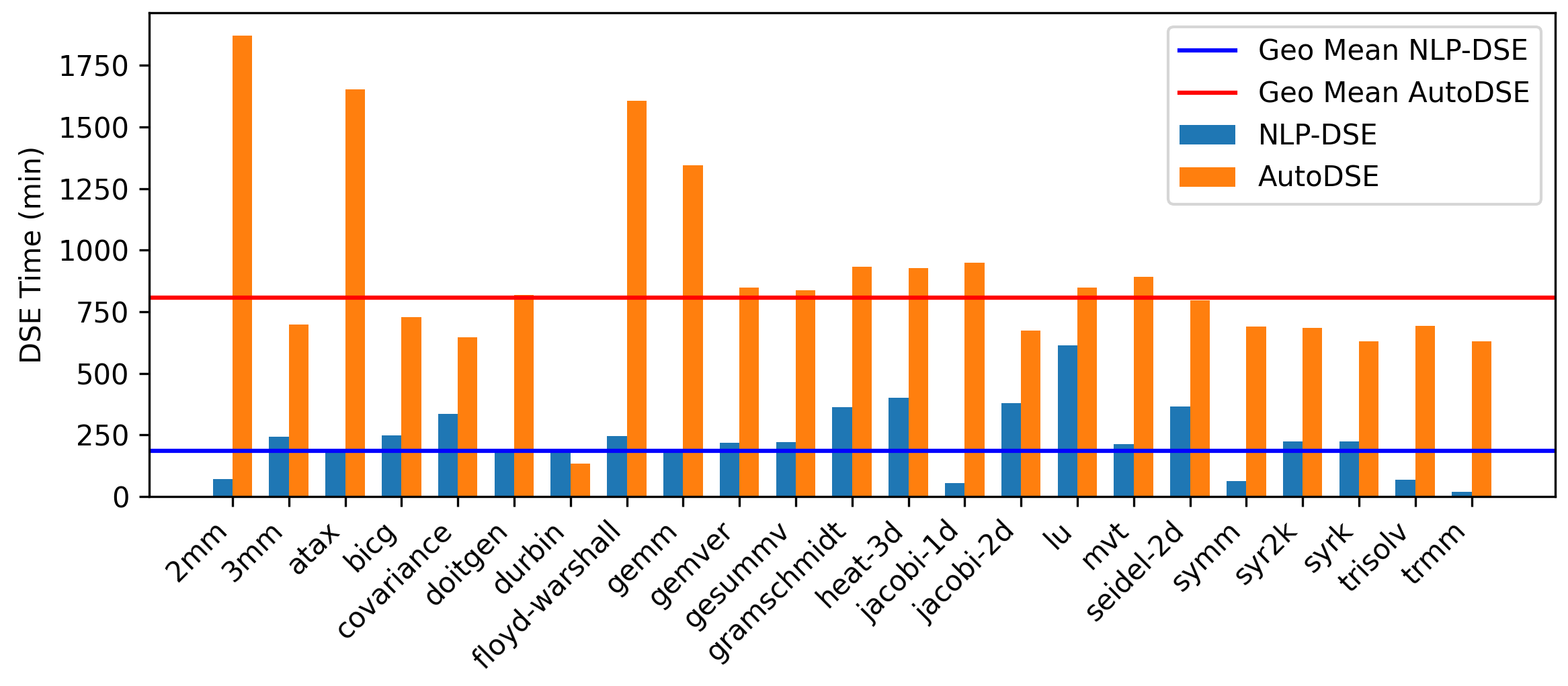}}
% \vspace{-.35cm}
  \caption{Comparison between the throughput (GF/s) and Design Space Exploration (DSE) time (min) of NLP-DSE and AutoDSE for medium problem sizes in Polybench.}
  \label{fig:medium} 
\end{figure}

Figures \ref{fig:large} and \ref{fig:medium} show the comparison with AutoDSE for Large and Medium problem size respectively.
% \begin{figure}[H]
Table \ref{tab:eval} shows the details of the comparison with AutoDSE. NL, ND, S, and Space S are respectively the number of loops, the number of polyhedral dependencies (WaR, WaW, RaW), problem size (L for Large and M for Medium) and space size. 
For each method we compute the throughput (GF/s) in GFLOPs per second, the total time of the DSE (T) in minutes, the number of designs explored (DE) and the number of designs timeout (DT). 
In addition, for AutoDSE we add the number of design that are early rejected/prune (ER) as AutoDSE prunes the design when AMD/Xilinx Merlin can not apply one of the pragmas, due to its analysis limitations.

To illustrate the performance achievable \emph{without a complete DSE}, the first  synthesizable  design found with \framework (FS) is displayed. Indeed, due to our under-estimation of resources, the theoretically best design produced by NLP solving may not be synthesizable. We report the improvements in DSE time (T) and throughput (GF/s).

The performance of the kernel evaluated  show significant improvements in both time and throughput. 
% 5.84	17.24
% 3.76	2.38
% 
The time of the DSE is 5.69x faster on average (3.70x for geo-mean)
and the throughput is 17.24x higher on average (2.38x for geo-mean) for the kernel evaluated.
For almost all (46/47) kernels and problem sizes the method identifies a design with a throughput similar to (+/- 2\%), or better than, AutoDSE. 
% Where there is a slowdown for some of the kernels and the problem size, it is negligible. It is 1.01 for 4 kernels, 1.02 for 2 kernels and 1.07 for Doitgen Large. 
% For this kernel, we have a performance difference because the design found with maximum array partitioning of 2048 is timeout and the best design is found with a maximum array partitioning of 1024. 
% 
% 
% We have a slowdown for Doitgen Large
% because, when our method explored for a design at the maximum array partitioning of 2048, it timed out and then explored with a maximum array partitioning of 1024 which is the best design for this method.
% However AutoDSE finds a design with a maximum array partitioning of 1280. By changing our parameter of maximum array partitioning at 1280 it comes up with the same configuration as AutoDSE. Thus it is possible to obtain designs with better performance but at the cost of a longer search.
% 
% 
We have a slight slowdown for \textit{Doitgen} Large
because \framework explores the design found by the NLP with a maximum array partitioning of 2048 which timeouts, and then 1024 which is the best design found.

\input{sources/tablev2.tex}
However AutoDSE finds a design with a maximum array partitioning of 1280. By changing the maximum array partitioning to 1280 we find the same configuration as AutoDSE. 
% Can be remove bellow
Thus it is possible to obtain designs with a better performance but at the cost of a longer search.
For all kernels and problem sizes, except \textit{Durbin}, \framework is faster than AutoDSE. AutoDSE prunes all configurations of \textit{Durbin} which explains the speed of AutoDSE for this kernel.

% vTODO

% analyze 2mm L vs M
% 2mm M nlp vs autodse
% Jacobi 2d L and M NLP vs AU 
% Covariance NLP vs AutoDSE

% TODO REFORMULATE

We can observe a difference of the performance for the same kernel in function of the problem size. If we take the examples of \textit{2mm} and \textit{3mm}, the difference has many factors. 
First as the footprint of the kernel becomes more important, it begins overusing the BRAMs. A large parallelism  requires a bigger array partitioning which considerably increases the number of BRAMs and uses more BRAMs than available.
Additionally, for large problems with high levels of parallelism, there are multiple instances of timeouts observed. 
Furthermore, the compilers applied 
% more efficiently the pragmas 
the pragmas more efficiently
for smaller problem sizes. We observe twice as many kernels where the pragmas are not applied as expected for the large problem size.

For \textit{AtAx} Large (Listing \ref{lst:reference1}), AutoDSE explores 166 designs of which 106 are early rejected and 30 timeout. AutoDSE starts by partially unrolling Loops 2 and 3 and will then attempt to do a coarse-grained parallelization on Loop 1 with all divisors, which is impossible due to dependencies. 
Although AutoDSE manages to prune/early reject the designs because Merlin cannot apply the pragmas, it still requires several minutes of compilation by Merlin for each unroll factor.
In parallel, AutoDSE tries to pipeline the outermost loops (and therefore unroll the innermost loops) which creates numerous timeouts.
Although the first two designs timeout due to too high level of parallelism, \framework allows us to find a configuration with the innermost loops unrolled with a UF=700. This allows us to find a design with a 3.46x higher throughput in 11.34x less time.

% For \textit{2mm}, \framework allows to find a configuration with fine-grained parallelization with each loop body optimized equivalently.
% Our DSE and AutoDSE uncovers the same configuration for both problem sizes when employing the kernel \textit{Jacobi-2d}.
% For \textit{Covariance}, there are numerous instances of design timeouts and pruned with AutoDSE. As for \textit{2mm}, the exploration of the unroll factor for the innermost loop in \textit{Covariance} is limited, and the space explored by AutoDSE within the time limit does not contain these better designs.
% 
% 
% 
% 
Our method experiences some timeouts for designs with high levels of parallelism. However, thanks to our DSE approach, we quickly identify optimized designs where each loop body has a similar level of parallelism. 
% In the case of Large problem sizes, there is an over-utilization of BRAMs for high parallel configurations. As a result, the throughput is reduced because in order to have valid design the parallelism has to be reduce.
% 
% 
 % C;est un peu nul dit comme ca....
For 20/47 cases, the first synthesizable design is equal to the best design of the DSE. This is because compilers can be conservative and not apply pragmas as expected. 
% In this case, it is another configuration than that found by the NLP which is applied
In this case, another configuration is applied than what was identified by the NLP, which explains the difference in performance.

% % \vspace{-0.3cm}

\subsection{Comparison with HARP}

\label{sec:eval_harp}

Utilizing the PolyBench problem size of HARP allows for direct reusability of the model, facilitating comparison and benchmarking against the framework HARP.
This also allows for the utilization of data that will enable achieving the best results with HARP.

Evaluating on other problem sizes would have required the creation of a database to at least fine-tune the model with the kernel and problem size in question.

\begin{figure}[!htb]
    \centering
    \includegraphics[width=\textwidth]{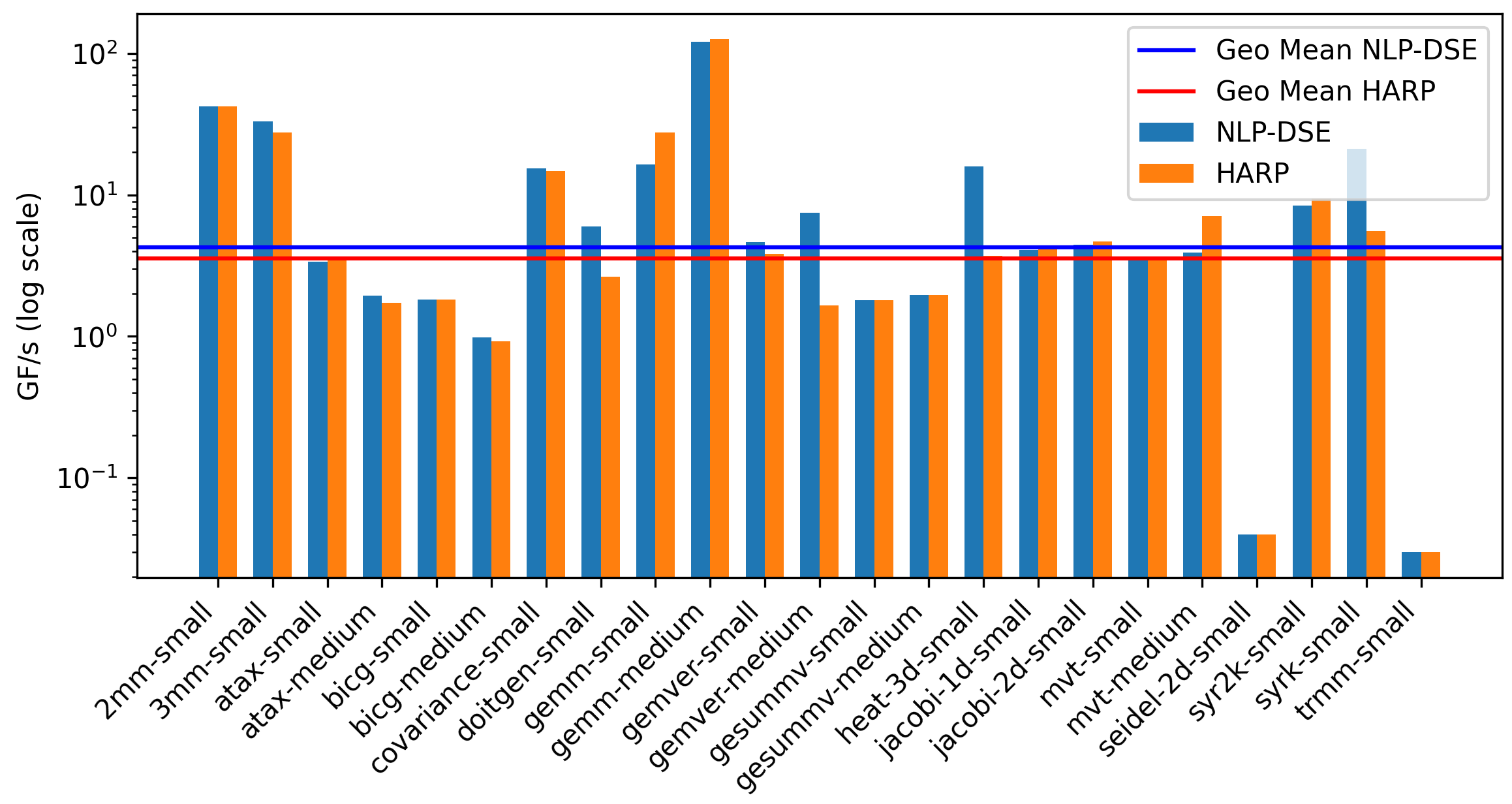}
    \caption{Comparison between the throughput (GF/s)  of NLP-DSE and HARP for small and medium problem sizes in Polybench.}
    \label{fig:harp}
\end{figure}

Figure \ref{fig:harp} shows the comparison with HARP for Small and Medium problem size.
Table \ref{tab:harp} in Appendix, shows the details of the comparison.

The throughput is 1.45x higher on average (1.21x for geo-mean) for the kernel evaluated in similar DSE time.
For 20/23 kernels the method identifies a design with a throughput similar to (+/- 10\%), or better than, HARP. 

% We observe a difference in performance improvement vs. the comparison to AutoDSE is due initially to the size of the space. HARP is capable of exploring 150,000 designs on average, which allows almost the entire space to be explored. From another point of view HARP is trained and/or fine-tuned with the specific kernel and problem size, which gives the model great knowledge of when the pragmas are not applied and the performance estimated, which gives it a advantage vs. AutoDSE which uses compiler as a black box.

We note a variation in the enhancement of performance compared to the evaluation in Section \ref{sec:evaluation_autodse} vs. AutoDSE, stemming from various factors. Primarily, the breadth of the exploration space plays a significant role. HARP has the capacity to traverse an average of 75,000 designs, enabling it to nearly exhaustively explore the entire space. Additionally, HARP is trained and/or fine-tuned with precise knowledge of the kernel and problem size, granting it deep insight into scenarios where pragmas are not applied and have enough training on these specific kernel to estimate the latency. This confers an advantage over AutoDSE, which treats the compiler as a black box.

We observe two significant slowdown with the kernel mvt for medium problem size and gemm for small problem size.

The kernel mvt does two matrix vector multiplications where the same matrix $A$ is read for the two statements in reverse order. 
Even though our model assumes that the same array $A$ can be used for both statements with the same partitioning based on unrolling, Merlin transfers the array $A$ twice—once for each statement.
However HARP find the design which allow the compiler to transfer only one times the array $A$. 
As we can see for the small size HARP is not able to find the configuration which allow Merlin to transfer $A$ one time and achieve to find a design with the same QoR as us. 
For our design the transfer of the two A take 97.2\% of the total latency. So transferring only one time allow to achieve at least the same throughput as HARP.

For Gemm small, HARP leverages Merlin’s double-buffering to transfer the output efficiently. Without this optimization, which was not included in our design space, HARP achieves a throughput of 14.64 GFLOPS/s.

With Gemver, we can attain a speedup by leveraging our capability to explore the entire space within a single optimization problem. The space of Gemver with medium size encompasses over $10^{11}$
  designs, making it impractical to thoroughly explore, even with HARP's estimation per design hovering around the millisecond range.

% For the kernel mvt with medium size we have a slowdown due to the fact that Merlin transfers two times the array $A$ which contains 159,900 double-precision float. 
% This kernel is already memory bound by transfering one time the array $A$. The time to transfer $A$ one time is 20,000 cycles with a bitwidth of 512 (maximum) so it is represent 40,000 cycles on and the total latency of the kernel is  40726 cycles. HARP was able to find the configuration which allow Merlin to transfer A only one time.

\subsection{Comparison with ScaleHLS}

We evaluated the throughput (GFLOPS/s) of \framework against ScaleHLS (S-HLS) \cite{scalehls} using medium-sized Polybench kernels with single-precision floating-point operations. The results are summarized in Table \ref{tab:scalehls}.

To ensure a fair comparison, we considered two scenarios based on the memory model:
\begin{itemize}
	\item \textit{Comparison 1: Including Memory Transfers}
Since ScaleHLS uses the Vivado flow, which assumes that data reside on-chip, we manually added memory transfers to the ScaleHLS designs. We optimized the transfers using the maximum burst size allowed by the problem (e.g., 128 bits for an array of size 210) and synthesized the design using the Vitis flow. To further improve efficiency, we overlapped off-chip to on-chip data loads and on-chip to off-chip writes wherever possible.
	\item \textit{Comparison 2: Excluding Memory Transfers}
For this scenario, we followed the methodology described in the ScaleHLS paper, where Vivado assumes all data are on-chip, eliminating memory transfer overhead. For a fair comparison, we extracted only the computation cycles from the Vitis flow report for our framework, isolating pure computation performance without memory overhead.
\end{itemize}

\input{sources/table_scale.tex}

Although ScaleHLS can perform code transformations that our framework does not currently implement, the design space we explore for pragma insertion is far more comprehensive and fully explored within seconds or minutes using our NLP solver. ScaleHLS relies on a QoR estimator to evaluate each pragma configuration, which is fast (in the order of milliseconds), but it cannot exhaustively explore a large design space, similar to the limitations of tools like HARP. Unlike ScaleHLS, our DSE is not focused solely on identifying a theoretical solution quickly. Instead, our NLP solver efficiently explores the entire design space to identify a theoretical optimal solution in seconds or minutes. 
Furthermore, our approach incorporates a DSE process to avoid cases where pragmas are not applied as intended during synthesis. In contrast, ScaleHLS does not validate whether the selected pragmas are correctly implemented by the compiler, potentially leading to discrepancies between expected and actual performance.

The results demonstrate that \framework delivers superior QoR. When memory transfers are included in the ScaleHLS results (Comparison 1), our framework achieves an average speedup of 11.63x across the evaluated kernels, with a geometric mean of 2.89x. When isolating computation latency (Comparison 2), \framework achieves an average speedup of 50.07x, with a geometric mean of 20.26x.

We did observe slowdowns in memory-bound kernels like bicg and mvt. In these cases, our manually optimized memory transfers were more efficient than those handled by Merlin. For instance, in the bicg kernel, Merlin repeatedly transfers two large arrays, consuming 99.2\% of the total latency. If these transfers were managed more efficiently, our framework could achieve a throughput of 1.96 GFLOPS/s. A similar inefficiency is observed in the mvt kernel.

Overall, \framework consistently outperforms ScaleHLS. Its memory management approach makes \framework more general and delivers better QoR by including on-chip memory transfers as part of the exploration space. Additionally, the theoretical design space is fully explored through the implementation of a cost model formulated as an NLP. A lightweight DSE is also employed to avoid cases where pragmas are not applied as expected by the compiler.

\subsection{Accuracy}

% \REMARK{Develop: The accuracy study is interesting in the Result-D section. It will be good if the authors can identify which pragma is not applied that causes high variation}

% \FIXME{Make a paragraph to explain the usual "perf gap" of Merlin \textcolor{red}{DONE}}

\begin{figure}[!htb]
% \begin{figure}[H]
% \begin{figure}[]
    % \vspace{-.15cm}
    \centering
  \subfloat[\label{fig:accuracy_all} For all explored designs]{%
       \includegraphics[width=\linewidth]{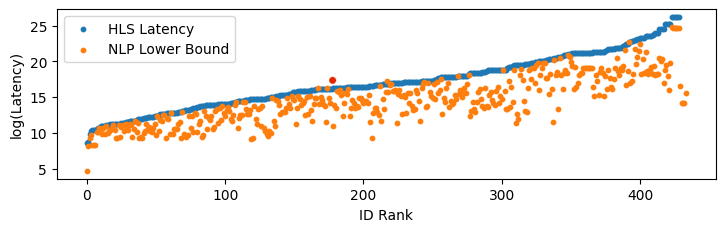}}
    \hfill
% \vspace{-0.45cm}
  \centering
    \subfloat[\label{fig:accuracy} For cases where pragmas were applied as expected]{%
        \includegraphics[width=\linewidth]{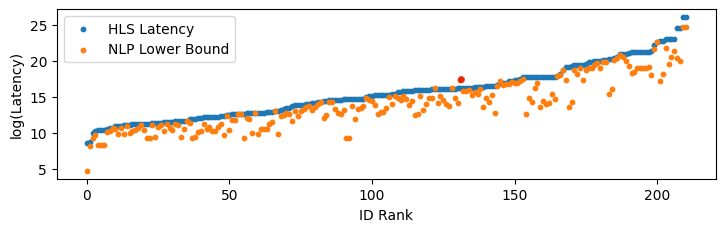}}
% \vspace{-.35cm}
  % \caption{Comparison of the latency between the design reported in the HLS report and the lower bound estimate provided by the nonlinear problem for all explored designs and specifically for cases where pragmas were applied as expected.}
  \caption{Comparison of the Latency Between the Design Reported in the HLS Report and the Lower Bound Estimate Provided by the Nonlinear Problem for All Explored Designs, Specifically for Cases Where Pragmas Were Applied as Expected, with ID Rank Representing the Order of Designs Sorted by HLS Latency.}
  \label{fig1} 
\end{figure}

% We  now briefly study the tightness of our performance model.
% 
The tightness of the lower bound estimation relies on the correct application of pragma directives such as pipeline and parallel. It also assumes that Merlin can efficiently transfer memory from off-chip to on-chip using 512-bit chunks. Finally, it assumes that Merlin optimally handles the transfer of memory from off-chip to on-chip.
% 
% Figures \ref{fig:accuracy_all} and \ref{fig:accuracy} compare the measured HLS latency for every synthesizable design explored during our DSE with its predicted latency per solving the NLP.
Figures \ref{fig:accuracy_all} and \ref{fig:accuracy} provide a comprehensive comparison between the measured HLS latency for every synthesizable design explored during our DSE process and the predicted latency obtained by solving the NLP.

The Y-axis represents log(latency) and the X-axis the rank of the design sorted by HLS latency. 
%The blues points represents the HLS latency from the report and the orange points the latency estimate with the NLP solver. 
For Fig~\ref{fig:accuracy} we exclude  designs when we detect that the pragma parallel and pipeline are not applied as defined by Merlin. We observe that about half of the designs have at least one pragma not applied, leading logically to a larger difference between measured and predicted latency. Generally speaking, for parallelization pragmas, Merlin is more restrictive for coarse-grained parallelization, in many cases these pragmas are not applied. 
Coarse-grained pragmas are typically not applied to kernels that do not have an outermost reduction loop and thus can theoretically have coarse-grained parallelization, which is present in most linear algebra kernels such as 2mm, 3mm, gemver, etc.
We also observe certain cases where the partitioning is not done correctly which does not allow a pipeline with II=1 when it is theoretically possible.
For Figure \ref{fig:accuracy} we observe a better overall accuracy, albeit imperfect. These differences are due in large part to how Merlin eventually implemented memory transfers, which we model optimistically. Internally, Merlin transforms the size of the arrays according to the program's unroll factors and in certain cases does not allow transfers with a bitwidth of 512 bits.

% \FIXME{What pragma is not applied and for which kernel}

% \begin{figure}[h]
%   \centering
%   \subfloat[\label{fig:accuracy_all} when some pragmas are not applied]{%
%     \includegraphics[width=0.49\linewidth]{figures/1.png}
%   }
%   \hfill
%   \subfloat[\label{fig:accuracy} when all pragmas are applied]{%
%     \includegraphics[width=0.49\linewidth]{figures/2.png}
%   }
%   \caption{Comparison of the HLS Latency vs. Lower Bound}
%   \label{fig1} 
% \end{figure}

% % \vspace{-0.4cm}
% \begin{figure}[h]
%   \centering
%   \subfloat[\label{fig:accuracy_all} when some pragmas are not applied]{%
%     \includegraphics[width=0.8\linewidth]{figures/1.png}
%   }
  
%   \subfloat[\label{fig:accuracy} when all pragmas are applied]{%
%     \includegraphics[width=0.8\linewidth]{figures/2.png}
%   }
  
%   \caption{Comparison of the HLS Latency vs. Lower Bound}
%   \label{fig1}
% \end{figure}
% % \vspace{-0.3cm}

% \begin{figure}[ht]
% \centering
% \begin{minipage}[b]{0.40\textwidth}
%   \centering
%   \includegraphics[width=\textwidth]{figures/2.png}
%   \caption{Comparison of the HLS Latency vs. Lower Bound}
%   \label{fig:accuracy}
% \end{minipage}
% \hfill
% \begin{minipage}[b]{0.40\textwidth}
%   \centering
%   \includegraphics[width=\textwidth]{figures/1.png}
%   \caption{Comparison of the HLS Latency vs. Lower Bound}
%   \label{fig:accuracy_all}
% \end{minipage}
% \end{figure}

% Our estimate is indeed a lower bound in the cases evaluated except for the cases where Vitis apply the automatic optimization loop flatten.
We observe in Fig.~\ref{fig:accuracy_all} and Fig.~\ref{fig:accuracy}  one configuration where the lower bound property is not maintained (shown in red). This corresponds to a configuration of the Heat-3d kernel, where the pragma \textbf{loop\_flatten} has been applied automatically, which changes the program structure.
Overall unless Vitis applies loop\_flatten automatically, which we do not model, our estimate is a lower bound for the cases evaluated.
Our model can easily implement the automatic flattened loop optimization:
We must multiply the TC of the loop pipeline by the TC of all the perfectly nesteed loops above pipeline loop (and remove them in the first products). 
Because this optimization is rarely applied, we prioritize having a tight lower bound.

% \vspace{-0.6cm}

Additionally, we evaluate the number of DSE steps needed to achieve the design with the best Quality of Results (QoR) of our DSE and the number of syntheses required before terminating the DSE due to finding a lower bound (LB) greater than the latency of an already synthesized kernel, sorting by latency estimation provided by the NLP in ascending order.

The results are presented in Table \ref{tab:top_x}. On average, it takes 8 steps of the DSE to discover the design with the best QoR and 15 steps to terminate the DSE.
We can observe that for some kernel we find the design with the best QoR at the first iteration of the DSE (which correspond to first shoot method in Table \ref{tab:eval}) but the DSE needs more step to stop  the guarantee that we cannot obtain better latency, which implies that the lower bound is not perfectly tight.

\input{sources/top_x}

\subsection{Scalability}

To mitigate prolonged solving times for specific kernels and problem sizes, we have implemented a 30-minute timeout constraint for the AMPL BARON solver. While this timeout does not guarantee achieving optimality, it ensures that the solver provides the best solution it has found within the time limit. In Table \ref{tab:scalability}, we present statistics regarding the number of problem timeouts (ND T/O) and problem non-timeouts (ND NT/O), along with the average time in seconds (Avg Time) for all problems and exclusively for those that did not time out. We can note that the 20 NLP problems for CNN finish in few seconds with an average of 3.71 seconds.

% % \vspace{-0.35cm}
% \begin{table}[]
% \begin{table}[htb]
% \begin{table}[H]
%     \centering
%     \begin{tabular}{@{}l rrrr@{}}
%     \toprule
%            % &                   &                   &  \multicolumn{2}{l}{\textbf{Average Time (s)}} \\
%            % % \cline{4-5}
%            % \cmidrule{4-5}
           
%       Size & ND T/O & ND NT/O & Avg Time  & Avg Time NT/O   \\
%        % &  Timeout &  Optimal &  &   \\
%       % \hline
%       \midrule
%        Medium & 7 & 469 & 55s & 29s \\
%        Large  &  119 & 361 & 479s & 43s  \\
%        All & 126 & 830 & 268s & 35s  \\
%        \bottomrule
%     \end{tabular}
%     \caption{Study of the scalability}
%     \label{tab:scalability}
% \end{table}
% % \vspace{-0.75cm}
% 

% \begin{table}[H]
\begin{table}[!htb]
\footnotesize
    \centering
    \begin{tabular}{@{}l rrrr@{}}
    \toprule
           % &                   &                   &  \multicolumn{2}{l}{\textbf{Average Time (s)}} \\
           % % \cline{4-5}
           % \cmidrule{4-5}
           
      Size & ND T/O & ND NT/O & Avg Time  & Avg Time NT/O   \\
       % &  Timeout &  Optimal &  &   \\
      % \hline
      \midrule
       Medium & 7 & 469 & 55s & 29s \\
       Large  &  119 & 361 & 479s & 43s  \\
       All & 126 & 830 & 268s & 35s  \\
       \bottomrule
    \end{tabular}
    \caption{Study of the scalability of the NLP solver across various sizes of Polybench and CNN. Comparison of the number of designs that timeout (ND T/O), the number of designs that do not timeout (ND NT/O), the average time to solve the problem (Avg Time), and the average time to solve the problem for non-timeout designs (Avg Time NT/O).}
    \label{tab:scalability}
\end{table}

We notice that 12 kernels exhibit at least one NLP problem that times out. 
To investigate scalability further, we conducted restarts for NLP problems that timed out at 30 minutes, extending the timeout to 30 hours. For 30 out of 126 problems (23.8\%), we found an optimal theoretical solution within an average time of 3.13 hours.
We observe that problems timing out after 30 hours often involve trip counts with numerous divisors, significantly expanding the space for the unroll factor. Consequently, non-linear conditions involving more than three unknown variables of unroll factors (UFs) become extremely time-consuming to resolve. By relaxing these constraints, we are able to find a solution in seconds but this can result in infeasible designs due to over-utilization of resources as these constraints are removed.
For 23.8\% of problems not timing out at 30 hours, we examined the disparity in objective function values provided by the solver when it times out at 30 minutes (representing the best solution found so far) versus when it discovers the optimal solution. For 25 out of 30 problems, the estimated latency is exactly the same. However, for the remaining 5 problems, the differences in the estimated latencies range from a mere 0.04\% up to 2,426\%. 

% \vspace{-0.1cm}

% \vspace{-0.4cm}

%% file: sources/tablev2.tex
% \newpage
% \newgeometry{left=2cm,top=0.8cm}
% \vspace{-2cm}
\begin{table}[]
% \begin{table*}[]
\footnotesize
\begin{adjustwidth}{-1in}{-1in}
% \begin{adjustheight}
    
\centering
% \begin{spacing}{0.95}
% \begin{tabular}{|l rrrr|r|rrrr|rrrrr|rr|}
\begin{tabular}{@{}l @{\hspace{-0.25cm}}
r
@{\hspace{+2.5mm}}
r
@{\hspace{+2.5mm}}
r
@{\hspace{+2.5mm}}
r
% @{\hspace{-0.05mm}}
r
@{\hspace{-1.6mm}}
rrrrr
@{\hspace{-1.6mm}}
rrr
@{\hspace{+2.5mm}}
r
@{\hspace{+2.5mm}}
r
% @{\hspace{-0.00002mm}}
@{\hspace{+2.5mm}}
rr
@{\hspace{-1.5mm}}
r
r@{}}
\toprule
 &  &  &  &  & FS & & \multicolumn{3}{l}{\textbf{\framework}} & & & \multicolumn{4}{l}{\textbf{AutoDSE}} & &  & \multicolumn{2}{l}{\textbf{Perf. Imp.}}   \\
% \cmidrule{6-17}
\cmidrule{6-6}
\cmidrule{8-11}
\cmidrule{13-17}
\cmidrule{19-20}
\textbf{Kernel} &  \textbf{NL} & \textbf{ND} & \textbf{S} & \textbf{Space S} & \textbf{GF/s} & &\textbf{GF/s} & \textbf{T} & \textbf{DE} & \textbf{DT} & & \textbf{GF/s} & \textbf{T} & \textbf{DE} & \textbf{DT} & \textbf{ER} & & \textbf{T} & \textbf{GF/s} \\

\midrule

%Data Mining
        cov. & 7 & 34 & M & 1.80E+11 & 0.08 &  & 0.75 & 336 & 21 & 8 &  & 0.28 & 645 & 161 & 15 & 115 &  & 1.92x & 2.64x \\ 
        cov. & 7 & 34 & L & 1.92E+13 & 0.39 &  & 0.73 & 466 & 21 & 9 &  & 0.62 & 2,849 & 209 & 68 & 118 &  & 6.11x & 1.16x \\ 

\midrule

%LA
        2mm & 6 & 13 & M & 1.37E+10 & 13.19 &  & 117.48 & 70 & 18 & 0 &  & 0.41 & 1,870 & 101 & 37 & 49 &  & 26.71x & 288x \\ 
        2mm & 6 & 13 & L & 1.15E+12 & 0.57 &  & 2.17 & 456 & 17 & 3 &  & 0.40 & 1,835 & 291 & 38 & 240 &  & 4.02x & 5.41x \\ 
        3mm & 9 & 19 & M & 1.20E+15 & 13.86 &  & 138.73 & 242 & 18 & 0 &  & 0.39 & 698 & 82 & 15 & 57 &  & 2.88x & 354x \\ 
        3mm & 9 & 19 & L & 6.18E+17 & 0.18 &  & 1.01 & 486 & 19 & 3 &  & 0.59 & 968 & 112 & 18 & 81 &  & 1.99x & 1.71x \\ 
        atAx & 4 & 12 & M & 1.40E+05 & 1.96 &  & 1.96 & 194 & 10 & 1 &  & 1.98 & 1,653 & 175 & 13 & 136 &  & 8.52x & 0.99x \\ 
        atAx & 4 & 12 & L & 1.60E+07 & 0.47 &  & 1.52 & 205 & 11 & 2 &  & 0.44 & 2,325 & 166 & 30 & 106 &  & 11.34x & 3.46x \\ 
        bicg & 3 & 10 & M & 1.90E+04 & 0.99 &  & 0.99 & 248 & 12 & 1 &  & 0.98 & 729 & 65 & 2 & 28 &  & 2.94x & 1.01x \\ 
        bicg & 3 & 10 & L & 4.44E+05 & 1.68 &  & 1.68 & 218 & 12 & 1 &  & 0.50 & 3,355 & 236 & 42 & 176 &  & 15.39x & 3.38x \\ 
    cnn & 6 & 2 & - & 6.43E+06 & 0.39 && 97.99 & 213 & 16 & 1 && 97.99 & 1,292 & 28 & 19 & 480 && 6.06x & 1.00x \\

        doitgen & 5 & 30 & M & 8.64E+06 & 19.75 &  & 19.75 & 193 & 13 & 0 &  & 18.95 & 819 & 296 & 14 & 248 &  & 4.24x & 1.04x \\ 
        doitgen & 5 & 30 & L & 3.63E+07 & 0.08 &  & 102.62 & 241 & 20 & 1 &  & 110.66 & 1,299 & 222 & 24 & 169 &  & 5.39x & 0.93x \\ 
        durbin & 4 & 55 & M & 1.08E+02 & 0.01 &  & 0.20 & 193 & 7 & 4 &  & 0.12 & 134 & 25 & 0 & 23 &  & 0.69x & 1.65x \\ 
        durbin & 4 & 55 & L & 9.00E+00 & 0.12 &  & 0.12 & 212 & 3 & 1 &  & 0.12 & 31 & 7 & 0 & 5 &  & 0.15x & 1.00x \\

        gemm & 4 & 6 & M & 2.30E+06 & 105.18 &  & 105.18 & 185 & 21 & 1 &  & 68.91 & 1,345 & 86 & 27 & 34 &  & 7.27x & 1.53x \\ 
        gemm & 4 & 6 & L & 1.47E+07 & 32.98 &  & 32.98 & 450 & 18 & 7 &  & 2.77 & 2,810 & 188 & 47 & 133 &  & 6.24x & 11.8x \\ 
        gemver & 7 & 13 & M & 7.72E+11 & 0.78 &  & 9.45 & 218 & 21 & 4 &  & 2.99 & 847 & 65 & 5 & 28 &  & 3.89x & 3.16x \\ 
        gemver & 7 & 13 & L & 1.28E+13 & 9.94 &  & 9.94 & 290 & 21 & 7 &  & 0.18 & 1,756 & 221 & 206 & 10 &  & 6.06x & 54.7x \\ 
        gesum. & 2 & 17 & M & 6.12E+02 & 1.97 &  & 1.97 & 220 & 14 & 3 &  & 1.97 & 836 & 80 & 8 & 47 &  & 3.80x & 1.00x \\ 
        gesum. & 2 & 17 & L & 6.33E+03 & 1.82 &  & 2.64 & 236 & 18 & 1 &  & 0.56 & 692 & 94 & 29 & 60 &  & 2.93x & 4.73x \\ 
        gram. & 6 & 34 & M & 1.75E+07 & 1.58 &  & 1.58 & 364 & 7 & 3 &  & 0.44 & 934 & 109 & 92 & 8 &  & 2.57x & 3.56x \\ 
        gram. & 6 & 34 & L & 1.22E+08 & 2.34 &  & 2.34 & 420 & 6 & 4 &  & 0.95 & 819 & 265 & 11 & 239 &  & 1.95x & 2.47x \\ 

        lu & 5 & 16 & M & 2.28E+03 & 0.03 &  & 0.03 & 614 & 19 & 11 &  & 0.04 & 849 & 193 & 1 & 159 &  & 1.38x & 0.98x \\ 
        lu & 5 & 16 & L & 3.99E+03 & 0.03 &  & 0.04 & 335 & 9 & 2 &  & 0.03 & 812 & 258 & 5 & 219 &  & 2.42x & 1.03x \\ 
        mvt & 4 & 6 & M & 1.38E+07 & 7.77 &  & 7.77 & 212 & 17 & 1 &  & 7.77 & 893 & 166 & 6 & 106 &  & 4.21x & 1.00x \\ 
        mvt & 4 & 6 & L & 7.41E+07 & 12.90 &  & 12.90 & 181 & 20 & 3 &  & 1.10 & 1,240 & 249 & 20 & 189 &  & 6.85x & 11.8x \\ 

        symm & 3 & 33 & M & 2.31E+04 & 0.04 &  & 0.20 & 63 & 5 & 0 &  & 0.20 & 691 & 142 & 35 & 89 &  & 10.97x & 1.00x \\ 
        symm & 3 & 33 & L & 6.64E+04 & 0.21 &  & 0.31 & 540 & 8 & 5 &  & 0.21 & 612 & 731 & 0 & 692 &  & 1.13x & 1.52x \\ 
        syr2k & 4 & 6 & M & 2.32E+04 & 0.07 &  & 1.74 & 224 & 16 & 2 &  & 1.20 & 685 & 230 & 17 & 203 &  & 3.06x & 1.45x \\ 
        syr2k & 4 & 6 & L & 6.67E+04 & 1.30 &  & 1.42 & 420 & 15 & 7 &  & 1.30 & 768 & 293 & 21 & 262 &  & 1.83x & 1.09x \\ 
        syrk & 4 & 6 & M & 2.32E+04 & 0.49 &  & 1.32 & 224 & 16 & 2 &  & 0.61 & 631 & 280 & 4 & 264 &  & 2.82x & 2.15x \\ 
        syrk & 4 & 6 & L & 6.67E+04 & 0.94 &  & 2.07 & 466 & 17 & 7 &  & 0.65 & 643 & 410 & 0 & 398 &  & 1.38x & 3.16x \\ 
        trisolv & 2 & 13 & M & 3.60E+02 & 0.03 &  & 0.03 & 69 & 12 & 0 &  & 0.04 & 694 & 69 & 33 & 23 &  & 10.06x & 0.98x \\ 
        trisolv & 2 & 13 & L & 6.30E+02 & 0.04 &  & 0.04 & 75 & 18 & 0 &  & 0.04 & 651 & 127 & 2 & 98 &  & 8.68x & 0.99x \\ 
        trmm & 3 & 8 & M & 2.31E+04 & 0.01 &  & 0.05 & 20 & 16 & 0 &  & 0.04 & 630 & 401 & 4 & 367 &  & 31.50x & 1.29x \\ 
        trmm & 3 & 8 & L & 6.64E+04 & 0.02 &  & 0.06 & 425 & 17 & 2 &  & 0.03 & 760 & 167 & 159 & 4 &  & 1.79x & 1.79x \\ 

% Medley
\midrule
        floyd-w & 3 & 21 & M & 8.65E+04 & 0.17 &  & 0.61 & 246 & 17 & 3 &  & 0.10 & 1,605 & 60 & 22 & 29 &  & 6.52x & 6.15x \\ 
        floyd-w & 3 & 21 & L & 3.29E+06 & 0.17 &  & 1.31 & 381 & 20 & 6 &  & 0.10 & 2,728 & 150 & 71 & 77 &  & 7.16x & 13.1x \\ 
% Stencil
\midrule
            heat-3d & 7 & 42 & M & 3.04E+07 & 0.23 &  & 3.75 & 402 & 17 & 7 &  & 3.75 & 928 & 75 & 33 & 35 &  & 2.31x & 1.00x \\ 
        heat-3d & 7 & 42 & L & 4.37E+07 & 0.13 &  & 0.62 & 520 & 13 & 6 &  & 0.63 & 740 & 109 & 70 & 35 &  & 1.42x & 0.99x \\ 
        jacobi-1d & 3 & 14 & M & 1.48E+03 & 11.43 &  & 11.43 & 55 & 4 & 0 &  & 11.53 & 948 & 126 & 3 & 74 &  & 17.24x & 0.99x \\ 
        jacobi-1d & 3 & 14 & L & 4.00E+04 & 5.95 &  & 5.95 & 386 & 9 & 6 &  & 2.65 & 1,283 & 173 & 21 & 123 &  & 3.32x & 2.25x \\ 
        jacobi-2d & 5 & 22 & M & 8.07E+06 & 0.20 &  & 3.32 & 379 & 12 & 5 &  & 3.32 & 674 & 158 & 26 & 98 &  & 1.78x & 1.00x \\ 
        jacobi-2d & 5 & 22 & L & 1.14E+07 & 0.26 &  & 1.25 & 427 & 9 & 6 &  & 1.25 & 1,106 & 231 & 39 & 171 &  & 2.59x & 1.00x \\ 
                seidel-2d & 3 & 27 & M & 4.26E+03 & 0.05 &  & 0.05 & 365 & 5 & 1 &  & 0.05 & 796 & 103 & 27 & 68 &  & 2.18x & 1.01x \\ 
        seidel-2d & 3 & 27 & L & 1.77E+05 & 0.05 &  & 0.05 & 540 & 13 & 7 &  & 0.05 & 880 & 91 & 33 & 48 &  & 1.63x & 1.00x \\ 
\midrule
% Average &  &  &  &  &  &  &  &  &  &  &  &  &  &  &  &  &  &  &  \\ 
% Geo. Mean &  &  &  &  &  &  &  &  &  &  &  &  &  &  &  &  &  &  &  \\ 
Average &  &  &  &  & 5.38 &  & 15.11 & 296 &  &  &  & 7.44 & 1,123 &  &  &  &  & 5.69x & 17.2x \\
Geo. Mean &  &  &  &  & 0.51 &  & 1.54 & 247 &  &  &  & 0.65 & 916 &  &  &  &  & 3.70x & 2.38x \\
\bottomrule

\end{tabular}
% \end{spacing}
% \vspace{0.4cm}
\end{adjustwidth}
\caption{
Comparison of DSE time and Throughput for \framework, \framework-FS, and AutoDSE across Polybench kernels at different problem sizes: NL, ND, S, and Space S denote the number of loops,  dependencies, problem size (L for Large and M for Medium), and space size, respectively. GF/s indicates throughput, T represents DSE time (min), DE signifies the number of explored designs, and DT denotes timeout designs. Additionally, ET reflects the count of designs early rejected/pruned by AutoDSE
}

\label{tab:eval}
% \vspace{-0.8cm}
% \end{adjustheight}

\end{table}

%% file: sources/table_scale.tex
% \begin{table}[H]
\begin{table}[!htb]
\footnotesize
\centering

\begin{tabular}{@{}l rrrrrrr@{}}
\toprule
  &  \multicolumn{3}{l}{With Memory} && \multicolumn{3}{l}{Without Memory}   \\

\cmidrule{2-4}
\cmidrule{6-8}

Kernel &  Th. S-HLS  & Th. NLP-DSE & Perf. Imp.  & & Th. S-HLS  & Th. NLP-DSE$^*$ & Perf. Imp.   \\

 &   Comparison 1 &  &  Comparison 1  & & Comparison 2 &  &  Comparison 2 \\

\midrule

2mm &23.61&117.48&4.98x&&27.91&247.33&8.86x \\
3mm&19.52&138.73&7.11x&&34.62&203.09&5.87x \\
Atax&1.43&1.96&1.37x&&4.85&181.70&37.43x \\
Bicg&1.65&0.99&0.60x&&9.03&287.07&31.79x \\
Gemm&38.48&105.18&2.73x&&44.83&197.29&4.40x \\
Gesummv&1.66&1.97&1.19x&&9.97&513.83&51.52x \\
Mvt&10.48&7.77&0.74x&&31.87&298.51&9.37x \\
Symm&0.10&0.20&2.09x&&0.07&0.20&3.09x \\
Syrk&0.44&1.32&2.99x&&0.01&1.33&134.21x \\
Syr2k&0.02&1.74&92.54x&&0.02&4.23&214.18x \\

\midrule
Average&&&11.63x&&&&50.07x \\
Geo-Mean&&&2.89x&&&&20.26x \\

\bottomrule

\end{tabular}

\caption{
Comparison of the throughput (GFLOPS/s) of \framework and ScaleHLS (S-HLS) for medium-sized Polybench kernels. In Case 1, we manually inserted memory transfers with the maximum burst size allowed by the problem size, then synthesized the designs using the Vitis flow. In Case 2, we followed the methodology from the ScaleHLS paper, running the ScaleHLS design with the Vivado flow, which assumes data is already on-chip. For NLP-DSE*, we extracted only the computation cycles from the Vitis report, excluding the cycles associated with off-chip memory transfers.
}

\label{tab:scalehls}
% \vspace{-0.7cm}
\end{table}

%% file: sources/top_x.tex
% \vspace{-0.4cm}
\begin{table}[!htb]
% \begin{table}[H]
\footnotesize
\begin{tabular}{
>{}l|
% \multicolumn{2}{>{\raggedright\arraybackslash}p{3cm}}
rr
% >{\raggedright\arraybackslash}p{3cm}     
% \multicolumn{2}{>{\raggedright\arraybackslash}p{3cm}}
|
rr
% >{\raggedright\arraybackslash}p{3cm} 
% >{\raggedright\arraybackslash}p{3cm} 
}
\toprule
Kernel & \multicolumn{2}{l}{To find the best QoR} & \multicolumn{2}{l}{To find a LB > than HLS result} \\
Problem Size&Large&Medium&Large&Medium \\
\midrule 
2mm&4&6&13&9 \\
3mm&7&2&11&7 \\
atAx&6&2&18&14 \\
bicg&2&2&16&18 \\
covariance&12&8&16&18 \\
doitgen&1&0&4&10\\
durbin&0&16&21&20\\
fdtd-2d&12&1&18&8\\
floyd-warshall&8&16&20&20\\
gemm&7&4&13&7\\
gemver&5&9&12&11\\
gesummv&0&4&14&16\\
gramschmidt&8&14&10&16\\
heat-3d&17&20&19&15\\
jacobi-1d&16&0&18&10\\
jacobi-2d&6&18&16&20\\
lu&18&0&20&20\\
mvt&1&0&4&6\\
seidel-2d&17&10&20&20\\
symm&12&10&16&21\\
syr2k&13&18&16&20\\
syrk&17&17&18&20\\
trisolv&0&0&17&20\\
trmm&16&4&20&17\\
% \textbf{Kernel} & \textbf{Size} & \textbf{Step} & \textbf{Size} & \textbf{Step} \\
% \midrule
% \textbf{2mm}& LARGE & 4 & MEDIUM& 6 \\
% \textbf{3mm}& LARGE & 7 & MEDIUM& 2 \\
% \textbf{atax} & LARGE & 6 & MEDIUM& 2 \\
% \textbf{bicg} & LARGE & 2 & MEDIUM& 2 \\
% \textbf{covariance} & LARGE & 12& MEDIUM& 8 \\
% \textbf{doitgen}& LARGE & 1 & MEDIUM& 0 \\
% \textbf{durbin} & LARGE & 0 & MEDIUM& 16\\
% \textbf{fdtd-2d}& LARGE & 12& MEDIUM& 1 \\
% \textbf{floyd-warshall} & LARGE & 8 & MEDIUM& 16\\
% \textbf{gemm} & LARGE & 7 & MEDIUM& 4 \\
% \textbf{gemver} & LARGE & 5 & MEDIUM& 9 \\
% \textbf{gesummv}& LARGE & 0 & MEDIUM& 4 \\
% \textbf{gramschmidt}& LARGE & 8 & MEDIUM& 14\\
% \textbf{heat-3d}& LARGE & 17& MEDIUM& 20\\
% \textbf{jacobi-1d}& LARGE & 16& MEDIUM& 0 \\
% \textbf{jacobi-2d}& LARGE & 6 & MEDIUM& 18\\
% \textbf{lu} & LARGE & 18& MEDIUM& 0 \\
% \textbf{mvt}& LARGE & 1 & MEDIUM& 0 \\
% \textbf{seidel-2d}& LARGE & 17& MEDIUM& 10\\
% \textbf{symm} & LARGE & 12& MEDIUM& 10\\
% \textbf{syr2k}& LARGE & 13& MEDIUM& 18\\
% \textbf{syrk} & LARGE & 17& MEDIUM& 17\\
% \textbf{trisolv}& LARGE & 0 & MEDIUM& 0 \\
% \textbf{trmm} & LARGE & 16& MEDIUM& 4 \\
% \midrule
% & Average && 8 Steps & \\ % TODO modify that look weird
\bottomrule
\end{tabular}
% \caption{Number of design evaluated to find the HLS design with the best QoR and the number of design to stop the DSE by founding a lower bound (LB) greater then the latency of a design synthetized with HLS compiler}
\caption{
The count of designs evaluated to identify the HLS design yielding the optimal Quality of Results (QoR), and the count at which the Design Space Exploration (DSE) ceases upon discovering a lower bound (LB) surpassing the latency of a design already synthesized with the HLS compiler}
\label{tab:top_x}
\end{table}
% \vspace{-0.6cm}

%% file: sources/example.tex
\label{sec:example}

In this section, we illustrate the significance of our method by contrasting it with AutoDSE and highlighting the advantages of our DSE approach. Our method excels in addressing domain-specific constraints, providing superior convergence in complex scenarios compared to AutoDSE. Furthermore, our DSE demonstrates adaptability and efficiency, proving to be robust in handling intricate design spaces, offering a more versatile and high-performing solution.

\subsection{2mm Medium}

\textit{2mm} serves as a linear algebra kernel, acting as a surrogate for transformer inference, such as Bert. The code snippet in Listing \ref{lst:2mm} illustrates the Medium-sized configuration with potential pragma options. The PIPE pragma can be either flattened or off (default), while PARA and TILE can be any divisor of the loop trip count, defaulting to 1.

\subsubsection{AutoDSE}

The optimal design identified by AutoDSE involves:
PARA\_L5 = 220, PIPE\_L3 = flatten, PARA\_L4 = 4
% \begin{itemize}
%     \item PARA\_L5 = 220
%     \item PIPE\_L3 = flatten
%     \item PARA\_L4 = 4
% \end{itemize}
 with all other parameters set to 1 or off. However, AutoDSE faces challenges in achieving a high Quality of Results (QoR) for 2mm due to two primary reasons:

For 2mm AutoDSE does not allow you to find a design with good QoR for two main reasons:
\begin{itemize}
    \item three out of four of workers initially over-utilize parallelism by flattening PIPE\_L0 
    % and $\setminus$ or 
    \textit{and/or}
    PIPE\_L1 (and hence unroll the innermost loops), causing timeouts in current High-Level Synthesis (HLS) tools. Even without considering timeouts, these designs exceed array partitioning limits, preventing the reading of all data in a single cycle as expected by the unrolling process. Moreover, these designs strain hardware resources, requiring backtracking, which extends the search duration.
% Moreover these designs over-utilized the resources (hardware constraint).
% This bottleneck method need a backtracking, but it would lead to an even longer search.

    \item one out of four mainly optimize a single loop body and is not able to optimize the second loop body. Moreover, even the loop body optimizer is not perfectly optimized and can be more parallelized.

\end{itemize}

\begin{figure}[!htb]
\begin{lstlisting}[label={lst:2mm},caption={Implementing the 2mm code with pragma directives for pipelining, tiling, and parallelization for each loop: $D=\alpha \times A \times B \times C + \beta \times D$.}]
#pragma ACCEL PIPELINE PIPE_L0
#pragma ACCEL TILE FACTOR=TILE_L0
#pragma ACCEL PARALLEL FACTOR=PARA_L0
Loop0: for (i1 = 0; i1 < 180; i1++) {
#pragma ACCEL PIPELINE PIPE_L2
#pragma ACCEL TILE FACTOR=TILE_L2
#pragma ACCEL PARALLEL FACTOR=PARA_L2
    Loop1: for (j1 = 0; j1 < 190; j1++) {
        S0: tmp[i1][j1] = 0.0;   
#pragma ACCEL PARALLEL FACTOR=PARA_L4
        Loop2: for (k1 = 0; k1 < 210; ++k1) {
            S1: tmp[i1][j1] += alpha * A[i1][k1] * B[k1][j1];
        }
    }
}
#pragma ACCEL PIPELINE PIPE_L1
#pragma ACCEL TILE FACTOR=TILE_L1
#pragma ACCEL PARALLEL FACTOR=PARA_L1
Loop3: for (i2 = 0; i2 < 180; i2++) {
#pragma ACCEL PIPELINE PIPE_L3
#pragma ACCEL TILE FACTOR=TILE_L3
#pragma ACCEL PARALLEL FACTOR=PARA_L3
    Loop4: for (j2 = 0; j2 < 220; j2++) {
        S2: D[i2][j2] *= beta;
#pragma ACCEL PARALLEL FACTOR=PARA_L5
        Loop5: for (k2 = 0; k2 < 190; ++k2) {
            S3: D[i2][j2] += tmp[i2][k2] * C[k2][j2];
        }
    }
}
\end{lstlisting}
\end{figure}

\subsubsection{\framework}

Now, we delve into how \framework overcomes these challenges to discover designs with superior QoR, emphasizing the usefulness of the DSE described in section \ref{sec:implementation}.

\begin{figure}[!htb]
    \centering
    \includegraphics[width=\textwidth]{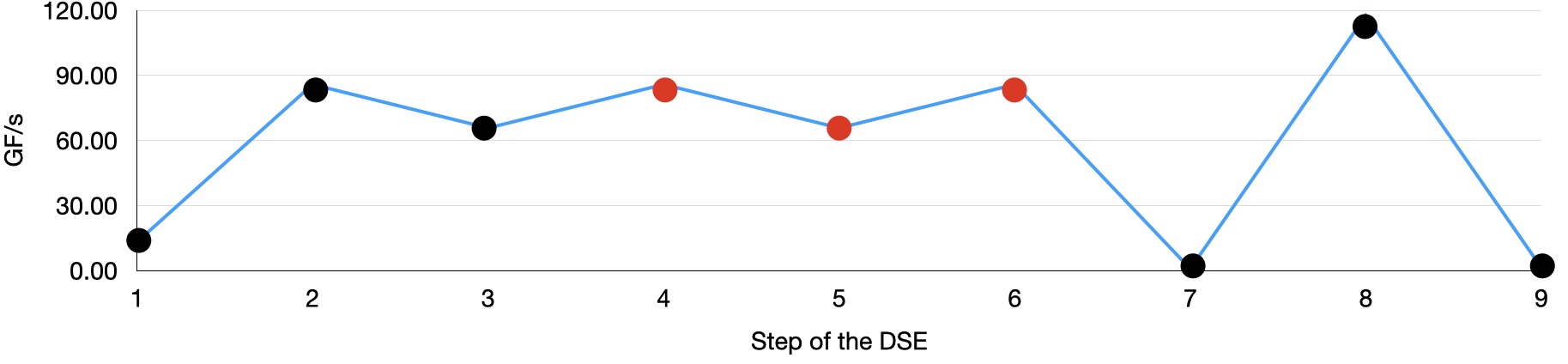}
    \caption{Representation of the throughput (GF/s) achieved for each design obtained at each stage of the NLP-DSE for the 2mm kernel.}
    \label{fig:ex_2mm_step}
\end{figure}

The initial \framework design features parameters (Step 1 in Figure \ref{fig:ex_2mm_step}):
PARA\_L0 = 3, PIPE\_L2 = flatten, PARA\_L4 = 210, PARA\_L1 = 6, PIPE\_L3 = flatten, PARA\_L5 = 190
% 
% \begin{itemize}
%     \item PARA\_L0 = 3
%     \item PIPE\_L2 = flatten
%     \item PARA\_L4 = 210
%     \item PARA\_L1 = 6
%     \item PIPE\_L3 = flatten
%     \item PARA\_L5 = 190
% \end{itemize}
 achieving 13 GFLOPS/s. However, the compiler fails to apply PARA\_L0 and PARA\_L1 pragmas, creating a performance gap.

The second design (Step 2 in Figure \ref{fig:ex_2mm_step}) uses the following parameters: PIPE\_L2 = flatten, PARA\_L2 = 2, PARA\_L4 = 210, PARA\_L3 = 5, PIPE\_L3 = flatten, and PARA\_L5 = 190. This configuration achieves a throughput of 85 GFLOPS/s. However, the unroll factor for PARA\_L3 does not allow the Xilinx Merlin compiler to efficiently transfer data from off-chip to on-chip, preventing the design from reaching the lower bound performance.

In Step 3, the design parameters are: PARA\_L0 = 3, PIPE\_L2 = flatten, PARA\_L4 = 210, PARA\_L1 = 3, PARA\_L3 = 2, PIPE\_L3 = flatten, and PARA\_L5 = 190. This configuration achieves a throughput of 65 GFLOPS/s, which is lower than the design found in Step 2, as the compiler did not apply the PARA\_L0 pragma as expected.
Steps 4, 5, and 6 (marked in red) identify the same configurations as Steps 2 and 3, so synthesis is not necessary as the results are already available.
For Step 7, the HLS compiler is unable to apply the configuration (PARA\_L0 = 2, PIPE\_L2 = flatten, PARA\_L4 = 210, PARA\_L1 = 6, PARA\_L3 = 44, PIPE\_L5 = flatten, PARA\_L5 = 5), leading to a throughput of only 0.82 GFLOPS/s.

The design with the best QoR in our search space is found at Step 8 of the DSE, with parameters: PIPE\_L2 = flatten, PARA\_L2 = 2, PARA\_L4 = 210, PARA\_L3 = 2, PIPE\_L3 = flatten, and PARA\_L5 = 190. However, due to Xilinx Merlin’s inability to optimally transfer data, the lower bound is not achieved, and the search continues.

In iteration 9, a design is found with the parameters PIPE\_L2 = flatten, PARA\_L4 = 210, PARA\_L1 = 6, PARA\_L3 = 22, PIPE\_L5 = flatten, and PARA\_L5 = 10. The pragma are not applied as expected, and the throughput is only 1.35 GFLOPS/s.
From iteration 10 the lower bound found by the NLP is greater than the latency (HLS report) of design 8, which makes it possible to stop the search because even if we reach the lower bound we will have a latency greater than what we have already obtained.
The Figure \ref{fig:ex_2mm_step} summarize the result achieved at each step of the DSE. 

% \begin{figure}[H]

We execute our Design Space Exploration (DSE) using 8 threads, allowing us to concurrently evaluate multiple designs in parallel. This approach anticipates that certain designs may not achieve the desired performance, and by running 8 designs simultaneously, we efficiently explore the design space. For this specific example, we perform a single iteration of the DSE.

%% file: sources/related.tex
% \FIXME{Redo}

\framework makes it possible to automatically introduce pragmas in order to obtain a design with a good QoR.
Many previous works using different DSE methods have the same objective.
% 
% Model-free DSEs \cite{autodse, s2fa, lorenzo} evaluate each candidate by generating the HLS report. The synthesis time or report generation can last several hours, which considerably reduces the explored design space.
%  Additionally, some DSEs, such as \cite{autodse}, 
%  may lead to local minima which prevent finding the optimal solution. 
These model-free Design Space Exploration (DSE) techniques, as exemplified by works such as \cite{autodse, s2fa, lorenzo}, employ a methodology where the compiler acts as a black-box, and they dynamically adapt their exploration strategies based on the outcomes of previous iterations.
In these approaches, each candidate design is evaluated by generating a High-Level Synthesis (HLS) report. However, the time required for the synthesis or report generation can extend over several hours, significantly limiting the breadth of the explored design space.
Moreover, it is worth noting that certain DSE methodologies, including those described in \cite{autodse}, might encounter challenges such as converging to local minima, which can impede the discovery of the globally optimal solution.

%  In order to avoid the synthesis time limitation,
%  model-based DSEs and AI-driven DSEs have been developed. They estimate the performance of each design using a cost model \cite{comba, lina, lin, 7372592}, Neural Network (NN) \cite{8060311, 7927161, 7551387, 6560643, 10.1145/3316781.3317754, schafer2012machine, 6930749},
% Graph Neural Networks (GNN) \cite{10323853, sohrabizadeh2022gnn, ironman} or decision trees (DT) \cite{ hls_predict, chimera,pyramide}.
%  These DSEs estimate the QoR of the design in a few milliseconds. However, the evaluation of a large number of designs can still take several hours and the accuracy may not match the HLS report. Hence, evaluating only the top-$n$ results with HLS may lead to sub-optimal solutions. 

To circumvent the constraints imposed by synthesis time, novel approaches in Design Space Exploration (DSE) have emerged, including model-based DSEs and AI-driven DSEs. These methods leverage sophisticated techniques such as cost modeling \cite{comba, lina, lin, 7372592}, Neural Networks (NN) \cite{8060311, 7927161, 7551387, 6560643, 10.1145/3316781.3317754, schafer2012machine, 6930749}, Graph Neural Networks (GNN) \cite{10323853, sohrabizadeh2022gnn, ironman}, or decision trees (DT) \cite{hls_predict, chimera, pyramide} to estimate the Quality of Results (QoR) of each design rapidly.
By utilizing these techniques, the evaluation time for a single design can be reduced to mere milliseconds. However, despite this acceleration, assessing a large number of designs still entails a significant time investment. Furthermore, while these rapid evaluations provide valuable insights, they may not perfectly align with the outcomes obtained from High-Level Synthesis (HLS) reports in terms of accuracy.
Consequently, relying solely on HLS validation for the top-$n$ results may lead to suboptimal solutions, as the rapid evaluation methods might not capture all pertinent design intricacies. Therefore, a more comprehensive approach that combines the strengths of both rapid evaluation techniques and traditional HLS validation is necessary to ensure optimal design outcomes.
 
 % Other works allow one-shot optimization with code transformations, e.g., \cite{scalehls, heterocl, pylog} but the space of hardware directives  and code transformations is limited. 
 % Some focus on a predefined micro-architecture, e.g., \cite{autosa, 10248016} but are limited in their application. But also specific application
 % such as DNN \cite{8587697, 10.1145/3400302.3415609}, stencil \cite{soda}, sparse linear algebra \cite{10.1145/3490422.3502368}, neural networks \cite{10.1145/3570928, 10.1145/3373087.3375321}, CNNs \cite{8778248}, etc. However, these methods are difficult to generalize outside their area.

In contrast, alternative methodologies offer one-shot optimization through code transformations and pragma insertion, as evidenced by works like \cite{scalehls, heterocl, pylog}. However, the efficacy of these approaches is constrained by the limited scope of available hardware directives and code transformations. While some endeavors concentrate on predefined micro-architectures, as seen in \cite{autosa, 10248016}, their applicability is restricted.
Although ScaleHLS offers valuable code transformations that are not implemented in our approach, our framework provides a more comprehensive design space for pragma insertion, which is exhaustively explored within seconds or minutes using our NLP solver. While ScaleHLS uses a QoR estimator for each pragma configuration—despite its quick estimation times (e.g., milliseconds)—it cannot explore large design spaces exhaustively. This limitation is similar to those observed with HARP.
In contrast to ScaleHLS, our design-space exploration (DSE) approach focuses on more than just rapidly identifying a theoretical solution. Our NLP solver thoroughly explores the entire design space in seconds or minutes to find the optimal theoretical solution. Additionally, our DSE approach ensures that pragmas are correctly applied by the compiler through synthesis, addressing a gap that ScaleHLS does not cover.
Furthermore, ScaleHLS employs the Vivado flow \cite{vivado_flow}, which assumes that data are on-chip. Thus, they consider a less complex space compared to ours, as we also account for the memory transfer from off-chip to on-chip.

Moreover, specialized applications such as Deep Neural Networks (DNN) \cite{8587697, 10.1145/3400302.3415609}, stencil computations \cite{soda}, sparse linear algebra operations \cite{10.1145/3490422.3502368}, and neural networks \cite{10.1145/3570928, 10.1145/3373087.3375321}, including Convolutional Neural Networks (CNNs) \cite{8778248, 10171329}, have garnered attention. Yet, these methods encounter challenges when extrapolated beyond their designated domains, rendering generalizations difficult.
Hence, while these approaches offer streamlined optimization strategies and tailored solutions for specific problem domains, their broader applicability beyond their respective niches remains a challenge.

 % 
% \framework is a hybrid approach: our cost model in the form of NLP makes it possible to consider very large design spaces in a few minutes. This makes it potentially even faster than existing cost models. Yet we rely on HLS to ensure accurate performance. 

% Recent advances in optimization solvers such as BARON \cite{baron1, baron2} have allowed NLP-based approaches to become a promising alternative to approximate ILP-based methods, as they can encode more complex and realistic performance models. Previous works \cite{pluto, pouchet:fpga13, 10.1145/3061639.3062195} encode a cost model as Linear Programming (LP) problem. However, it requires an approximation, such as approximating the volume of communication inter loops \cite{pluto,pouchet:fpga13}, or simplifying the space by exposing direct parallelization in the problem \cite{10.1145/3061639.3062195}. 
% There were no other NLP methods available for comparison, and LP methods were deemed less suitable due to their inability to handle non-linear constraints. There are multiple product terms  (e.g., products of UF of perfectly nested loops) in the objective function and constraints which are essential to have an accurate model, but cannot be easily approximated with linear models.

\framework presents a hybrid methodology, leveraging a NLP-based cost model to swiftly explore expansive design spaces within minutes, potentially outpacing existing models in speed. Nevertheless, to ensure precise performance evaluation, reliance on High-Level Synthesis (HLS) remains integral to our approach.
Recent advances in optimization solvers such as BARON \cite{baron1, baron2} have allowed NLP-based approaches to become a promising alternative to approximate ILP-based methods, as they can encode more complex and realistic performance models. Unlike prior works \cite{pluto, pouchet:fpga13, 10.1145/3061639.3062195} that frame the cost model as Linear Programming (LP) problems, NLP-based methods can handle more complex constraints without necessitating approximations, such as estimating communication volumes across loops \cite{pluto, pouchet:fpga13} or simplifying the space by exposing direct parallelization in the problem \cite{10.1145/3061639.3062195}.
It is noteworthy that the comparison landscape lacked other NLP-based methodologies, while Linear Programming methods were deemed less suitable due to their incapacity to manage nonlinear constraints effectively. The inclusion of multiple product terms within the objective function and constraints, such as the product of Unroll Factors (UF) for perfectly nested loops, underscores the necessity for accurate modeling, which is challenging to approximate with linear approximations.

The selection of tile sizes remains fundamental for the final QoR.
Similar to our approach, \cite{8056810} uses a cost model to select the tile size. Although their space is much more complete than ours as we are restricted to Merlin's transformations, their method does not allow the evaluation of the whole space.

Approaches that do not rely on precise static analysis, such as \cite{autodse, s2fa, lorenzo, sohrabizadeh2022gnn, ironman, ironman2}, can take as input any C/C++ kernel supported by the HLS compiler, thus expanding their applicability to a broader spectrum of programs. 
In contrast, \framework, akin to other model-based Design Space Exploration (DSE) approaches \cite{comba, lina, lin, 7372592}, is constrained to affine programs to ensure precise analysis and facilitate the modeling of latency and resource utilization. While this encompasses a significant subset of programs, including AI kernels, and aligns with the MLIR affine dialect, it does not match the versatility of other frameworks.
Consequently, there exists a tradeoff between accuracy and generality. Opting for a subset of programs provides more detailed information, thereby accelerating DSE and potentially improving the QoR.

%% file: sources/conclusion.tex
\label{sec:conclusion}

% Our work targets the automatic selection of pragma configurations for HLS with a framework that automatically inserts Merlin pragmas into loop-based programs. Our framework is guided by an analytical performance and resource model, which serves as a lower bound estimation for the achievable performance across all possible configurations. 
% By formulating this model as a Non-Linear Program (NLP), the theoretically optimal pragma configuration can be determined. Our framework facilitates efficient design-space exploration by leveraging the latency lower bound property, allowing for the rapid elimination of points in the search space using a lightweight DSE process.

In this paper, we presented \framework, a framework that automates the selection of pragma configurations for high-level synthesis (HLS) by inserting Merlin pragmas into loop-based programs. The core of our approach is an analytical performance and resource model, which serves as a lower bound estimation for achievable performance across all pragma configurations. By leveraging this model and formulating it as a Non-Linear Program (NLP), our framework can identify the theoretically optimal pragma configurations.

Our method significantly improves design-space exploration (DSE) efficiency by using the latency lower bound property to quickly eliminate suboptimal design points. This enables a lightweight DSE process that drastically reduces exploration time without sacrificing the quality of results (QoR). Compared to existing approaches like AutoDSE and HARP, \framework achieves equal or better QoR with fewer design points explored and in significantly less time.

Extensive evaluations on a variety of benchmarks have shown that \framework consistently delivers superior or equivalent results while maintaining fast exploration times. 
% By automating pragma insertion and optimizing design-space exploration, our framework paves the way for more efficient and effective HLS-based design, setting a new standard for DSE methodologies.

Now that our framework can automatically insert pragma configurations, the next step is to extend its capabilities to include code transformations. Transformations like loop permutation, tiling, and data reuse strategies can further optimize performance by restructuring the code to better exploit hardware resources. Integrating these transformations with pragma insertion will allow our framework to explore a broader range of design optimizations, potentially unlocking even greater improvements in both performance and resource utilization within the HLS process.

%% file: sources/appendix.tex
\section{Proofs for Latency and Resource Lower Bound}
\label{appendix:proof}
\subsection{Latency Lower Bound}

\begin{proof}[Proof Th \ref{th:latlbcp}]
Every operation $S_i$ is associated with at least one edge with a source in $<\{V_I,root\}>$, so there is a path between one of these nodes and every operation by construction in Def.~\ref{def:computationgraph}. For an operation to produce a useful output, there must be a path from its output to a node in $V_O$, otherwise the operation may be removed by dead code elimination. Therefore the shortest path $sp$ between a pair of nodes $(v_i, v_o) \in <\{V_I,root\},V_O>$ is the shortest sequence of operations in dependence that must be executed to produce $v_o$. As any operation takes at least one cycle to complete per Def.~\ref{th:latlbcp}, then this path must take at least $\#sp$ cycles to complete. As we take the largest of the shortest paths between all possible pairs  $(v_i, v_o)$ then $OG_{cp}^L$ is the length of the longest shortest path to produce any output $v_o$ from some input, via a sequence of producer-consumer operations. Therefore it must take at least $\#OG_{cp}^L$ cycles to execute this path.
\end{proof}

\begin{proof}[Proof Th \ref{th:latlbresources}]
Suppose $\forall o \in op,~R_{o} \ge \#L(o) $. Then there is equal or more resources available than work to execute, this is equivalent to the infinite resource hypothesis of Th.~\ref{th:latlbcp}, the minimal latency is $LO(\#OG_{cp}^L)$.

Suppose $\exists o \in op,~R_{o} < \#L(o)  $. Then there exists at least one unit in $R_{o}$ that is executing $\lceil \#L(o) /R_{o} \rceil$ operations. As every operation $op$ take at $L(op) \ge 1$ cycle to complete, this unit will execute in at least  $\lceil \#L(o) \times L_{o}/R_{o} \rceil$ cycles. If $\lceil \#L(o) \times L_{o}/R_{o} \rceil \ge LO(\#OG_{cp}^L)$, the computation cannot execute in less than $\lceil \#L(o) \times L_{o}/R_{o} \rceil$ cycles.
\end{proof}

\subsubsection{Loop Unrolling: full unroll}

\begin{proof}[Proof Co \ref{cor:fullflattenstraightlinecode}]
By construction the process of fully unrolling all the loops creates a straight-line code region without loop control, which therefore fits Def.~\ref{def:straightlinecode}.
\end{proof}

\begin{proof}[Proof Th \ref{th:minlatflattenednest}]
By Corollary~\ref{cor:fullflattenstraightlinecode}.
\end{proof}

\subsubsection{Loop Unrolling: partial unroll} 

\begin{proof}[Proof Th \ref{th:minlatunrolled1}]
By construction and Theorem~\ref{th:latlbresources}, $Lat_{R_{op}}^{L^{'}}$ is a lower bound on the latency of $L^{'}$, that is the sub-program made of $UF$ iterations of the loop. $\lfloor TC/UF \rfloor \le TC_l/UF$ is a lower bound on the number of iterations of the loop. As we assume a non-pipelined execution for the resulting outer loop, every iteration shall start after the completion of the preceding one, that is its iteration latency, itself bounded by $Lat_{R_{op}}^{L^{'}}$.
\end{proof}

\begin{proof}[Proof Th \ref{th:minlatunrolled2}]

% Key argument to prove: style of: lat(S1,forloop) >= max(lat(S1),lat(forloop) that is by adding a new instruction we dont reduce the cp length. A counter example is
% \{for i:  a = a+b\};a = 43

% if we consider a CDAG, adding the edge a = 43 gives a shorter path to a than any of the i iteration in the fully unrolled cdag. So we must add conditions such as writes to the same location have to be all executed, in the order of the original code (sensible condition in presence of data dependencies WAW).

% Proof flow:
% 1) Given $OG^{i}$ and $OG^j$ two CDAGs, for a pair of distinct iterations $i,j$ of loop $L$. If $V_O^i \cap V_O^j = \emptyset$, then the graph $OG^{i,j}$ made of the two iterations $i,j$ cannot have a smaller critical path length than $OG^{i}$ and $OG^j$: there is no edge crossing $OG^{i}$ and $OG^j$ in $OG^{i,j}$ since outputs are distincts, therfore $Lat(OG^{i,j} \ge \max(Lat(OG^i),Lat(OG^j))$.

% 2) Suppose now  $V_O^i \cap V_O^j \ne \emptyset$. Then iterations $i$ and $j$ produce at least one output in common. IDEA: DSA. If the program is in dynamic single assignent form, this is a contradiction (DSA: only 1 write per location). Need to adjust defs at start to be in DSA form? Mention automatic conversion from affine to DSA exists (Feautrier, 90s).

Given $OG^i$ and $OG^j$ two CDAGs, for a pair of distinct iterations $i,j$ of loop $l$. 

If $V_O^i \cap V_O^j = \emptyset$, then the graph $OG^{i,j}$ made of the two iterations $i,j$ cannot have a smaller critical path length than $OG^i$ and $OG^j$: there is no edge crossing $OG^i$ and $OG^j$ in $OG^{i,j}$ since outputs are distinct, therefore $Lat(OG^{i,j}) \ge \max(Lat(OG^i),Lat(OG^j))$.

If $V_O^i \cap V_O^j \ne \emptyset$. Then iterations $i$ and $j$ produce at least one output in common. As there is no useless operation, the graph $OG^{i,j}$ made of the two iterations $i,j$ can not be smaller than $OG^i$ or $OG^j$ and hence $Lat(OG^{i,j}) \ge \max(Lat(OG^i),Lat(OG^j))$.

\end{proof}

\begin{proof}[Proof Th \ref{th:minlatunrolled3}]
    
    By definition a reduction loop is a loop with a dependency distance of 1. Hence, at each iteration the same memory cell is read and write. 
Because of the dependency distance of 1, only one element can be added to the same memory cell.
However each data can be adding independently two by two and the result of this independent addition can also be adding two by two until we obtained one value. Hence the reduction can be done in $\log_2(UF)$ iterations with a tree reduction.
As the depth of the tree is $\log_2(UF)$ and each node at the same depth can be executed independently in $Lat_{R_{op}}^L$ cycles, the straight line code has a latency greater or equal to $Lat_{R_{op}}^L \times \lfloor \log_2(UF)  \rfloor$. And then similarly to Th. \ref{th:minlatunrolled2} and \ref{th:minlatunrolled3} the sequential execution of the loops without pragma repeat this process $\lfloor TC/UF \rfloor$ times.
    
\end{proof}

\subsubsection{Loop pipelining} 
\begin{proof}[Proof Th \ref{th:minlatpipeline}]
$Lat_{R_{op}}^L$ is the minimal latency to complete one iteration of $l$ by Theorem~\ref{th:latlbresources}. The initiation interval measures the number of elapsed cycles before the next iteration can start, it takes therefore at least $TC_l * II - 1$ to start $TC_l - 1$ iterations, irrespective of their completion time. Therefore the latency of the loop is at least the latency of one iteration to complete, and for all iterations to be started.
\end{proof}

\subsubsection{Loop pipelining and unrolling}
\begin{proof}[Proof Th \ref{th:minlatpipeline_unrolled}]

By construction and Theorem \ref{th:minlatflattenednest} the latency $Lat_{R_{op}}^{L^{'}}$ is a lower bound of $L^{'}$. As the loop was split due to the partial unrolling, the trip count of the pipelined loop is $\frac{TC_l}{UF}$. Theorem \ref{th:minlatpipeline} gives us the lower bound of the latency for a loop with a trip count equal to $\frac{TC_l}{UF}$.

\end{proof}

\subsubsection{Coarse-Grained parallelization}
\begin{proof}[Proof Th \ref{th:coarse_grained}]

By construction, Definition \ref{def:loopIteratedSequentially}, Theorem \ref{th:minlatpipeline} and definition of the loop body $L$, $Lat^{L}_{R_{op}}$ is a lower bound of the loop body $L$. As the loop is not a reduction loop there is no dependency between the loop bodies of $L$ for different iteration of $l$ and then the loop bodies can be executed in parallel. 
If $UF < TC_l$, then $\lfloor TC_l/UF \rfloor \leq TC_l / UF $ is a lower bound on the number of iterations of the loop. As we assume a non-pipelined execution for the resulting outer loop, every iteration shall start after the
completion of the preceding one, that is its iteration latency, itself bounded by 
$ Lat_{R_{op}}^{L^{'}} $.

\end{proof}

\subsubsection{Program latency lower bound under resource constraints} 
\begin{proof}[Proof Th \ref{th:minres}]
    Considering perfect resource reuse, where all unused computational units can be reused, and assuming that the compilers have implemented the pragma configuration provided as input. 
For each statement $s$, the maximum number of computational units used in parallel is determined. This means that each statement $s$ requires at least $\#L_{op}^s \times DSP_{op} \times MCU_{op}^s$ DSPs. 
If a set of statements $\mathcal{S}$ are executed in parallel they cannot share the resource so the execution in parallel of the statement $s \in \mathcal{S}$ will require $( \sum_{s \in \mathcal{S}}
\#L_{op}^s
\times DSP_{op} 
\times MCU_{op}^{s})$ DSPs.
By considering the maximum across all statements, we can guarantee that at least one set of statement executed in parallel will require $\max_{s \in \mathcal{S}_{seq}}(\sum_{s \in \mathcal{S}} \#L_{op} \times DSP_{op} \times MCU_{op}^s)$ DSPs.
Since there is no possibility of resource reuse between different operations, the summation of the resource consumed for each operation remains the minimum consumption of resources. In other words, the sum of the individual resource consumption for each operation represents the minimum amount of resources required.
\end{proof}

\subsubsection{Memory transfer}
\begin{proof}[Proof Th \ref{th:memlatonearray}]
    In order to transfer all the elements of the array $a$ we can use packing with a maximum packing allowed by the target device of $max\_burst\_size$, which means in practice the real burst size will be equal or less than $max\_burst\_size$.
    As all the elements of the array $a$ are in the same bank, the transfer is sequential. And as we as suppose all operation including memory transfers are done in at least one cycle, the minimum latency is $\frac{footprint_{a}}{max\_burst\_size}$ to transfer once the array $a$. As an array can be input, output or both we need to add the transfer from off-chip to on-chip for inputs i.e., $a \in V_I^L$ and from on-chip to off-chip for the outputs i.e., $a \in V_O^L$.
    
\end{proof}

\begin{proof}[Proof Th \ref{th:memlat}]
According to Theorem \ref{th:memlatonearray}, the lower bound for transferring one array \( a \) is given by \( \frac{footprint_{a}}{max\_burst\_size} \).
If the array \( a \) is transferred within a loop \( l \), we only need to transfer the footprint of the array \( a \) accessed within that loop multiplied by the trip count of the loops that iterate over this array and over the loop $l$, which corresponds to \( footprint_{a} \) in the best case because we need to transfer at least all the elements of \( a \) one time.
As the array can reside on different DRAM banks, the transfer from off-chip to on-chip can be performed in parallel if transferred consecutively, i.e., at the same level. However, at least one array will have a latency greater than or equal to \( \frac{footprint_{a}}{max\_burst\_size} \) so the latency of communication under the loop $l$ is bounded by $\max_{a \in \mathcal{A}} (loop_l\_cache\_array\_a \times \frac{footprint_{a}}{max\_burst\_size} )$.
Since communication cannot be overlapped if the transfers are not consecutive (i.e., not under the same loops), we sum the latency of communication for each loop.
\end{proof}

\subsection{Summary}

\begin{proof}[Proof Th \ref{th:comp_lb}]

% By composition and Theorems \ref{th:minlatunrolled1} and \ref{th:minlatunrolled2}, $Lat_{L_{fg}}$ is a lower bound of the fully unrolled sub-loop body of $L$, $L_{fg}$.

% By composition and Theorems \ref{th:minlatpipeline} and \ref{th:minlatpipeline_unrolled}, $Lat_{L_{pip}}$ is a lower bound of $L_{pip}$.

% By composition, Theorem \ref{th:coarse_grained} and Definition \ref{def:loopIteratedSequentially}, $Lat_{L}$ is a lower bound of $L$.

Through composition and the application of Theorems \ref{th:minlatflattenednest}, \ref{th:minlatunrolled1} and  \ref{th:minlatunrolled2}, $Lat^{L_{fg}}_{R_{op}}$ serves as a computation latency lower bound for the fully unrolled sub-loop body of $L$, denoted as $L_{fg}$, where $Lat_{L_{par}} +
Lat_{L_{seq}} \times \prod_{l\in \mathcal{L}^{reduction}_{L_{fg}}} \frac{TC_l}{UF_l} \times \log_2(UF_l)$ represent the critical path of $Lat^{L_{fg}}_{R_{op}}$.
By employing composition alongside Theorems \ref{th:minlatpipeline} and \ref{th:minlatpipeline_unrolled}, $Lat^{L_{pip}}_{R_{op}}$ stands as a computation latency lower bound for $L_{pip}$.
Utilizing composition, Theorem \ref{th:coarse_grained}, and Definition \ref{def:loopIteratedSequentially}, $Lat^{L}_{R_{op}}$ emerges as a computation latency lower bound for $L$.

\end{proof}

\begin{proof}[Proof Th \ref{th:global_lb}]

% The AMD/Xilinx Merlin compiler does not overlap computation and communication phases, resulting in total execution time being the sum of computation and communication latencies. By applying composition and Theorems \ref{th:comp_lb} and \ref{th:memlat}, a lower bound on latency can be derived.

The AMD/Xilinx Merlin compiler executes computation and communication sequentially, making total latency their sum. Applying composition and Theorems \ref{th:comp_lb} and \ref{th:memlat} derives a lower bound on latency.

% The AMD/Xilinx Merlin compiler does not support overlapping between computation and communication phases. Consequently, the overall execution time is determined by the sequential sum of computation latency and communication latency.
% % 
% % The separation of these phases simplifies the analysis of performance bounds. 
% By applying composition principles and leveraging Theorems \ref{th:comp_lb} and \ref{th:memlat}, it becomes possible to derive a lower bounds on latency. 

\end{proof}

\section{Evaluation vs. HARP}
\input{sources/table_harp}

\section{Problem Size}
% \FIXME{TODO}
\input{sources/table_polybench}

%% file: sources/table_harp.tex
\begin{table}[H]
\footnotesize
% \begin{adjustwidth}{-1in}{-1in}
\centering
% \begin{spacing}{0.95}
\begin{tabular}{@{}l lrrr@{}}
\toprule
\textbf{Kernel} & Problem Size & \textbf{GF/s NLP-DSE} & \textbf{GF/s Harp} & \textbf{Perf. Improvement} \\
\midrule
\textbf{2mm}& Small & 42.33& 42.33& 1.00 \\
\textbf{3mm}& Small & 33.04& 27.66& 1.19 \\
\textbf{atAx}& Small & 3.38 & 3.44 & 0.98 \\
\textbf{atAx}& Medium & 1.94 & 1.72 & 1.13 \\
\textbf{bicg}& Small & 1.82 & 1.83 & 0.99 \\
\textbf{bicg}& Medium & 0.98 & 0.92 & 1.07 \\
\textbf{covariance}& Small & 15.31& 14.76& 1.04 \\
\textbf{doitgen}& Small & 5.98 & 2.63 & 2.27 \\
\textbf{gemm}& Small & 16.41& 27.57& 0.60 \\
\textbf{gemm}& Medium & 120.63& 125.59 & 0.96 \\
\textbf{gemver}& Small & 4.63 & 3.84 & 1.21 \\
\textbf{gemver}& Medium & 7.47 & 1.66 & 4.51 \\
\textbf{gesummv} & Small& 1.80 & 1.80 & 1.00 \\
\textbf{gesummv}& Medium & 1.96 & 1.96 & 1.00 \\
\textbf{heat-3d}& Small& 15.85& 3.70 & 4.29 \\
\textbf{jacobi-1d}& Small& 4.07 & 4.24 & 0.96 \\
\textbf{jacobi-2d}& Small& 4.44 & 4.71 & 0.94 \\
\textbf{mvt}& Small& 3.62 & 3.62 & 1.00 \\
\textbf{mvt}& Medium & 3.93 & 7.07 & 0.56 \\
\textbf{seidel-2d}& Small& 0.04 & 0.04 & 1.01 \\
\textbf{syr2k}& Small& 8.40 & 9.49 & 0.88 \\
\textbf{syrk}& Small & 21.05& 5.54 & 3.80 \\
\textbf{trmm}& Small & 0.03 & 0.03 & 1.00 \\
\midrule
\textbf{Average} &&&&  1.45 \\
\textbf{Geo. Mean}&&&& 1.21 \\
\bottomrule
\end{tabular}
% \end{spacing}

\caption{Comparison between the throughput (GF/s) achieved by NLP-DSE and HARP, along with the performance improvement realized over HARP.}
\label{tab:harp}
% \end{adjustwidth}
\end{table}

%% file: sources/table_polybench.tex
% % \newpage
% % \newgeometry{left=2cm,top=0.8cm}
% % \vspace{-2cm}
% \begin{table}[]
% % \begin{table*}[]
% \begin{adjustwidth}{-1in}{-1in}
% % \begin{adjustheight}

% \centering
% \begin{spacing}{0.95}
% % \begin{tabular}{|l rrrr|r|rrrr|rrrrr|rr|}
% \begin{tabular}{@{}l @{\hspace{-1mm}}rrrrrrrrrr@{\hspace{-1.2mm}}rrrrrrrrr@{}}
% \toprule
%&&&&& FS & & \multicolumn{3}{l}{\textbf{\framework}} & & & \multicolumn{4}{l}{\textbf{AutoDSE}} & && \multicolumn{2}{l}{\textbf{Perf. Improvement}} \\

% \bottomrule

% \end{tabular}
% \end{spacing}
% \vspace{0.4cm}
% \caption{Polybench}

% \label{tab:eval}
% % \vspace{-0.8cm}
% % \end{adjustheight}
% \end{adjustwidth}
% \end{table}
% % \restoregeometry
% % % \vspace{-0.4cm}

\begin{table}[H]
\footnotesize
% \scriptsize
% \begin{tabular}{
% \begin{adjustwidth}{-1in}{-1in}
% >{}l lllll}
\begin{tabular}{
    >{}
    >{\raggedright\arraybackslash}p{1.9cm}
    >{\raggedright\arraybackslash}p{0.8cm}
    >{\raggedright\arraybackslash}p{0.8cm}
    >{\raggedright\arraybackslash}p{3cm} 
    >{\raggedright\arraybackslash}p{3cm} 
    >{\raggedright\arraybackslash}p{2.5cm} 
    % >{\raggedright\arraybackslash}p{3cm} 
}
\toprule
\textbf{Kernel} & \textbf{$\mathcal{O}(ops)$} & \textbf{$\mathcal{O}(Mem)$} & \textbf{Large}& \textbf{Medium} & \textbf{Small}  \\
\midrule
\textbf{2mm}& $\mathcal{O}(n^3)$& $\mathcal{O}(n^2) $ & NI 800, NJ 900, NK 1100, NL 1200& NI 180, NJ 190, NK 210, NL 220 & NI 40, NJ 50, NK 70, NL 80 \\
\textbf{3mm}& $\mathcal{O}(n^3) $ & $\mathcal{O}(n^2) $ & NI 800, NJ 900, NK 1000, NL 1100, NM 1200 & NI 180, NJ 190, NK 200, NL 210, NM 220 & NI 40, NJ 50, NK 60, NL 70, NM 80\\
\textbf{adi}& $\mathcal{O}(n^3) $ & $\mathcal{O}(n^2) $ & TSTEPS 500, N 1000 & TSTEPS 100, N 200 & TSTEPS 40, N 60\\
\textbf{atAx} & $\mathcal{O}(n^2)$& $\mathcal{O}(n^2) $ & M 1900, N 2100& M 390, N 410 & M 116, N 124\\
\textbf{bicg} & $\mathcal{O}(n^2)$& $\mathcal{O}(n^2)$& M 1900, N 2100& M 390, N 410 & M 116, N 124\\
\textbf{cholesky} & $\mathcal{O}(n^3) $ & $\mathcal{O}(n^2) $ & N 2000& N 400 & N 120\\
\textbf{correlation}& $\mathcal{O}(n^3)$& $\mathcal{O}(n^2) $ & M 1200, N 1400& M 240, N 260 & M 80, N 100\\
\textbf{covariance} & $\mathcal{O}(n^3) $ & $\mathcal{O}(n^2) $ & M 1200, N 1400& M 240, N 260 & M 80, N 100\\
\textbf{deriche}& $\mathcal{O}(n^2)$& $\mathcal{O}(n^2) $ & W 4096, H 2160& W 720, H 480 & W 192, H 128\\
\textbf{doitgen}& $\mathcal{O}(n^4)$& $\mathcal{O}(n^2) $ & NQ 140, NR 150, NP 160& NQ 40, NR 50, NP 60 & NQ 20, NR 25, NP 30 \\
\textbf{durbin} & $\mathcal{O}(n^2)$& $\mathcal{O}(n)$& N 2000& N 400 & N 120\\
\textbf{floyd-warshall} & $\mathcal{O}(n^3) $ & $\mathcal{O}(n^2) $ & N 2800& N 500 & N 180 \\
\textbf{ftdt-2d}& $\mathcal{O}(n^3)$& $\mathcal{O}(n^2)$& TMAX 500, NX 1000 NY 1200 & TMAX 100, NX 200, NY 240 & TMAX 40, NX 60
, NY 80  \\
\textbf{gemm} & $\mathcal{O}(n^3) $ & $\mathcal{O}(n^2) $ & NI 1000, NJ 1100, NK 1200 & NI 200, NJ 220, NK 240 & NI 60, NJ 70, NK 80\\
\textbf{gemver} & $\mathcal{O}(n^2) $ & $\mathcal{O}(n^2)$& N 2000& N 400 & N 120 \\
\textbf{gesummv}& $\mathcal{O}(n^2)$& $\mathcal{O}(n^2)$& N 1300& N 250 & N 90  \\
\textbf{gramschmidt}& $\mathcal{O}(n^3) $ & $\mathcal{O}(n^2)$& M 1000, N 1200& M 200, N 240 & M 60, N 80 \\
\textbf{heat-3d}& $\mathcal{O}(n^4) $ & $\mathcal{O}(n^3) $ & TSTEPS 500, N 120 & TSTEPS 100, N 40 & TSTEPS 40, N 20\\
\textbf{jacobi-1d}& $\mathcal{O}(n^2) $ & $\mathcal{O}(n)$& TSTEPS 500, N 2000& TSTEPS 100, N 400 & TSTEPS 40, N 120  \\
\textbf{jacobi-2d}& $\mathcal{O}(n^3)$& $\mathcal{O}(n^2) $ & TSTEPS 500, N 1300& TSTEPS 100, N 250 & TSTEPS 40, N 90 \\
\textbf{lu} & $\mathcal{O}(n^3)$& $\mathcal{O}(n^2) $ & N 2000& N 400 & N 120 \\
\textbf{ludcmp} & $\mathcal{O}(n^3) $ & $\mathcal{O}(n^2) $ & N 2000& N 400 & N 120 \\
\textbf{mvt}& $\mathcal{O}(n^2) $ & $\mathcal{O}(n^2) $ & N 2000& N 400 & N 120\\
\textbf{nussinov} & $\mathcal{O}(n^3)$& $\mathcal{O}(n^2)$& N 2500& N 500 & N 180 \\
\textbf{seidel-2d}& $\mathcal{O}(n^3) $ & $\mathcal{O}(n^2)$& TSTEPS 500, N 2000& TSTEPS 100, N 400 & TSTEPS 40, N 120  \\
\textbf{symm} & $\mathcal{O}(n^3)$& $\mathcal{O}(n^2) $ & M 1000, N 1200& M 200, N 240 & M 60, N 80 \\
\textbf{syr2k}& $\mathcal{O}(n^3) $ & $\mathcal{O}(n^2)$& M 1000, N 1200& M 200, N 240 & M 60, N 80 \\
\textbf{syrk} & $\mathcal{O}(n^3)$& $\mathcal{O}(n^2)$& M 1000, N 1200& M 200, N 240 & M 60, N 80 \\
\textbf{trisolv}& $\mathcal{O}(n^2)$& $\mathcal{O}(n^2)$& N 2000& N 400 & N 120 \\
\textbf{trmm} & $\mathcal{O}(n^3) $ & $\mathcal{O}(n^2) $ & M 1000, N 1200& M 200, N 240 & M 60, N 80 \\ 
\bottomrule
\end{tabular}
\caption{
Complexity analysis of the number of operations and memory requirements for Polybench's problem sizes categorized as large, medium, and small.}
\label{tab:polybench}
% \end{adjustwidth}
\end{table}